\documentclass[a4paper,11pt]{article}
\pdfoutput=1 

\usepackage{jheppub} 
                     
\usepackage{amsmath,amssymb,amsthm,amscd,graphicx}
\usepackage{psfrag}

\usepackage[english]{babel}
\usepackage[T2A]{fontenc}
\usepackage[utf8]{inputenc}
\usepackage{textcomp}
\usepackage{array}
\usepackage{hyperref}
\usepackage{tikz}
\usepackage{float}
\usetikzlibrary{decorations.markings}

\input epsf.sty

\usepackage{color}

\newcommand{\eqlb}[2]{\begin{equation} \label{#1} #2 \end{equation}}
\newcommand{\eq}[1]{\begin{equation} #1 \end{equation}}
\newcommand{\eqn}[1]{\begin{equation*} #1 \end{equation*}}

\newcommand{\rme}{\textrm{e}}
\newcommand{\rmd}{\textrm{d}}

\theoremstyle{definition}

\def\({\left(}
\def\){\right)}

\newcommand{\brc}[1]{\left(#1\right)}
\newcommand{\bsq}[1]{\left[#1\right]}
\newcommand{\bfi}[1]{\left\{ #1\right\}}

\newcommand{\abs}[1]{\left|#1\right|}

\newcommand{\at}[2]{\genfrac{}{}{0pt}{}{#1}{#2}}

\newcommand{\be}{\begin{equation}}
\newcommand{\ee}{\end{equation}}
\newcommand{\ba}{\begin{aligned}}
\newcommand{\ea}{\end{aligned}}
\newcommand{\ben}{\begin{eqnarray}\displaystyle}
\newcommand{\een}{\end{eqnarray}}

\newdimen\tableauside\tableauside=1.0ex
\newdimen\tableaurule\tableaurule=0.4pt
\newdimen\tableaustep
\def\phantomhrule#1{\hbox{\vbox to0pt{\hrule height\tableaurule width#1\vss}}}
\def\phantomvrule#1{\vbox{\hbox to0pt{\vrule width\tableaurule height#1\hss}}}
\def\sqr{\vbox{%
  \phantomhrule\tableaustep
  \hbox{\phantomvrule\tableaustep\kern\tableaustep\phantomvrule\tableaustep}%
  \hbox{\vbox{\phantomhrule\tableauside}\kern-\tableaurule}}}
\def\squares#1{\hbox{\count0=#1\noindent\loop\sqr
  \advance\count0 by-1 \ifnum\count0>0\repeat}}
\def\tableau#1{\vcenter{\offinterlineskip
  \tableaustep=\tableauside\advance\tableaustep by-\tableaurule
  \kern\normallineskip\hbox
    {\kern\normallineskip\vbox
      {\gettableau#1 0 }%
     \kern\normallineskip\kern\tableaurule}%
  \kern\normallineskip\kern\tableaurule}}
\def\gettableau#1{\ifnum#1=0\let\next=\null\else
\squares{#1}\let\next=\gettableau\fi\next}

\preprint{CERN-TH-2023-110}

\title{ Black hole perturbation theory and  multiple polylogarithms}
\author[a,b]{Gleb Aminov}
\author[c,d,e]{Paolo Arnaudo}
\author[c,d,e]{Giulio Bonelli}
\author[f,g]{Alba Grassi}
\author[c,d,e]{Alessandro Tanzini}

\affiliation[a]{ C.~N. Yang Institute for Theoretical Physics, State University of New York, Stony Brook, NY 11794-3840, USA}
\affiliation[b]{ Simons Center for Geometry and Physics, State University of New York, Stony Brook, NY 11794-3636, USA}
\affiliation[c]{International School of Advanced Studies (SISSA), via Bonomea 265, 34136 Trieste, Italy}
\affiliation[d]{INFN Sezione di Trieste, via Valerio 2, 34127 Trieste, Italy}
\affiliation[e]{Institute for Geometry and Physics, IGAP, via Beirut 2, 34151 Trieste, Italy}
\affiliation[f]{CERN, Theoretical Physics Department, CH-1211 Geneva 23, Switzerland}
\affiliation[g]{Section de Math\'ematiques, Universit\'e de Gen\`eve, 1211 Gen\`eve 4, Switzerland}
\abstract{We study black hole linear perturbation theory in a four-dimensional Schwarzschild (anti) de Sitter background. When dealing with a {\it positive} cosmological constant, the corresponding spectral problem is solved systematically via the Nekrasov-Shatashvili functions or, equivalently, classical Virasoro conformal blocks. However, this approach can be more complicated to implement for certain perturbations if the cosmological constant is {\it negative}. For these cases, we propose an alternative method to set up perturbation theory for both small and large black holes in an analytical manner. Our analysis reveals a new underlying recursive structure that involves multiple polylogarithms. We focus on gravitational, electromagnetic, and conformally coupled scalar perturbations subject to Dirichlet and Robin boundary conditions. The low-lying modes of the scalar sector of gravitational perturbations and its hydrodynamic limit are studied in detail.
}

\begin{document}
\maketitle

\flushbottom

\section{Introduction}

In this paper, we study analytical approaches to solve the differential equations describing black hole (BH) linear perturbation theory. 
The concrete problem consists in the study of Einstein equations with a cosmological constant, approximated around a particular BH solution to first order in perturbation theory.\footnote{Higher orders in perturbation theory can also be studied with the methods we develop here, but this is beyond the scope of this work.} Because of the symmetries of the Schwarzschild (anti) de Sitter solution, the linearised equations separate
and reduce to a second-order linear ODE of Fuchsian type; we refer to \cite{review} for a review and a list of references.  
In the particular cases we study in this work, the relevant equation has four or five regular singularities.
Fuchsian equations appear in many fields of theoretical and mathematical physics, and the relevance of the parametric analysis of their solutions and the corresponding connection coefficients goes well beyond the application to BH perturbation theory.
In this paper, we employ two distinct, complementary strategies to {\it analytically} study such equations.

First, by following modern developments in the context of the supersymmetric gauge theories, we
tackle such problems using the Nekrasov-Shatashvili (NS) functions \cite{nekrasov2010} (see Appendix A for their definition). These functions have been shown to be building blocks to compute quantum periods \cite{nekrasov2010,mirmor,mirmor2,Zenkevich2011,Bonelli:2011na},
eigenfunctions   \cite{ Gaiotto:2014ina,Kanno:2011fw,Jeong:2018qpc,Sciarappa:2017hds,Jeong:2017pai, Bonelli:2022ten,Piatek:2017fyn},
 Fredholm determinants \cite{ggm,ghn}, and connection coefficients \cite{Bonelli:2022ten}  for Fuchsian differential equations and their irregular limits.
These techniques were recently applied to studying spectral problems describing black hole perturbation theory. 
Initially introduced in \cite{Aminov:2020yma} for the study of quasinormal modes  (QNMs)  in  four-dimensional  asymptotically flat black holes, this approach has been generalized to various gravitational backgrounds and extends beyond the QNMs computation \cite{Bonelli:2021uvf,Bianchi:2021xpr,Bianchi:2021mft,Casals:2021ugr, Bianchi:2021yqs,Bianchi:2022wku,Dodelson:2022yvn,Consoli:2022eey,daCunha:2022ewy,Dodelson:2023vrw,Bianchi:2023sfs,Bhatta:2022wga,Fioravanti:2021dce,Imaizumi:2022dgj,Gregori:2022xks,Bianchi:2022qph, Bianchi:2023rlt,Giusto:2023awo}.\footnote{See also \cite{BarraganAmado:2021uyw, Amado:2020zsr,Novaes:2014lha,Amado:2021erf,Cavalcante:2021scq,Novaes:2018fry} for another approach based on Painlev\'e equations.}
Other interesting related results have been elaborated in
\cite{Bianchi:2023lrg,Chia:2023tle,DeLuca:2023mio,Charalambous:2023jgq,Fransen:2023eqj,Fioravanti:2023zgi,Berenstein:2022nlj,Kehagias:2022ndy,Hui:2022vov,Katagiri:2022vyz,Charalambous:2022rre,Fioravanti:2022bqf,Hui:2022sri,Oshita:2022pkc,Imaizumi:2022qbi,Xing:2022emg,Aguayo:2022ydj,Pereniguez:2021xcj,Hatsuda:2021gtn,Blake:2021hjj,Nakajima:2021yfz,Datta:2021hvm,daCunha:2021jkm,Yan:2020kkb}.

In this paper, we further extend this approach to the framework of four-dimensional BHs in the de Sitter (dS) background. More precisely, we compute the relevant connection formulae following the methodology developed in \cite{Bonelli:2022ten}, where exact connection formulae for the Heun equation were obtained from the classical limit of Virasoro conformal blocks, which is, in turn, related to NS functions.\footnote{For a discussion of these connection formulae from a mathematical viewpoint, see \cite{Lisovyy:2022flm}.}

The approach based on the NS functions can be applied in its simplest form only when 
the boundary conditions of the spectral problem are imposed at singular points of the differential equation. 
However, there are cases when this condition is not satisfied. For example, when considering gravitational or conformally coupled scalar perturbations of black holes in a four-dimensional anti-de Sitter (AdS) spacetime, see Sec.~\ref{section4}. 
Moreover, the NS functions expansion is potent but only in some regions of parameter space, and to explore other regions, it is sometimes necessary to resort to other analytic methods.   
This happens, for instance, in studying the so-called hydrodynamic limit.\footnote{Indeed, the natural expansion in this limit does not map to the instanton expansion in gauge theory. This obstacle may be overcome by using TBA techniques for the computation of the NS functions and the corresponding quantum periods, see \cite{Fioravanti:2021dce,Imaizumi:2022dgj,Gregori:2022xks}. However, we will not explore this path in this work.}
In these situations, we analyze the equation using an alternative "polylog" method where we reduce the relevant problem to recurrence relations which we solve in terms of multiple polylogarithms.  

This paper is structured as follows.

In Sec.~\ref{section2}, we briefly describe the NS and the polylog methods and comment on their implementation.

In Sec.~\ref{section3}, we study conformally coupled scalar, electromagnetic, and vector-type gravitational perturbations of four-dimensional Schwarzschild de Sitter black holes (SdS), where both methods are applicable. We use the Heun connection formulae to obtain the quantization condition in Sec.~\ref{section3.1}, which gives the quasinormal frequencies as series expansions in the radius of the black hole horizon $R_h$. In Sec.~\ref{section3.2}, we apply the polylog method and express the corresponding wave functions in terms of multiple polylogarithms in one variable. 
By gluing the relevant local solutions, we determine the frequencies of the QNMs, and the resulting series expansions in $R_h$ agree with the ones obtained in Sec.~\ref{section3.1}. In all computed orders, we find purely imaginary QNM frequencies in agreement with the earlier observations made by numerical computations \cite{KZ'22,Jansen:2017oag, Cardoso:2017soq}. 

In Sec.~\ref{section4}, we study the same class of perturbations of Sec.~\ref{section3} in the case of four-dimensional Schwarzschild anti-de Sitter black holes (SAdS) imposing Dirichlet boundary conditions at spatial infinity.  For these perturbations, one of the boundary conditions is imposed at a regular point of the equation. Hence the method based on the NS function is more complicated to implement, and we apply the polylog method instead.
As in the Schwarzschild de Sitter case, the two local solutions are described in terms of multiple polylogarithms in one variable.
We check our analytic results against  the numerical values of \cite{CKL2003}.\footnote{With the current version of {\tt Mathematica} (12.1 or higher), one can use the numerical implementation of Heun functions to get very accurate results for the QNM frequencies for most quantum numbers, see e.g~\cite{Hatsuda:2020sbn}.} Our results suggest that the leading order of the imaginary part of the QNM frequencies is $-c\, R_h^{2\ell+2}$ in the small $R_h$ approximation, where $\ell$ is the angular quantum number and $c$ is a real positive
constant depending on all the quantum numbers. This is consistent with earlier numerical results  obtained via the Breit-Wigner approach \cite{Berti:2009wx} and some earlier analytic studies in \cite{Cardoso:2006wa}.

In Sec.~\ref{section5}, we study the low-lying modes of the scalar sector of gravitational perturbations of Schwarzschild anti-de Sitter black holes in the big $R_h$ limit. In this Section, we use Robin boundary conditions at spatial infinity, which preserve the metric at the boundary and, as such, are more suited for holography. The corresponding differential equation has five regular singular points, and we use the polylog method to compute the relevant local solutions as Taylor expansions in $1/R_h$. To make this computation more efficient, we introduce three local regions. In the region near the horizon, the local solution is given in terms of multiple polylogarithms in several variables (all but the first argument are constants). The local solutions are described in terms of Laurent polynomials in the other two regions. The QNM frequencies are obtained by gluing the three local solutions, and the first two orders in $1/R_h$ agree with the results from \cite{michalogiorgakis2007low}. Theoretically, one can compute the QNM frequency up to any given order in $1/R_h$. However, due to the exponential growth of the number of polylogarithm functions that appear in each order in the perturbative expansion, we could determine the QNM frequency up to order $1/R_h^6$.  By taking the hydrodynamic limit, we can also reproduce the results from \cite{natsuume2008causal} and obtain four additional corrections in the expansion. 

Appendix \ref{appendixA} introduces the notations and conventions used in the main text for the NS functions. Appendix \ref{appendixB} reports relevant identities between classical polylogarithms and multiple polylogarithms and relations for multiple zeta values. In Appendix \ref{app:recurrence_rel}, we prove by induction on $K\in\mathbb{N}$ that the local solutions at order $R_h^K$ in Sec.~\ref{section3} and Sec.~\ref{section4} are given in terms of multiple polylogarithms in one variable of weight at most $K$. Appendix \ref{appendixD} presents the linear basis of multiple polylogarithms in several variables that describes the local solution near the horizon in Sec.~\ref{section5}. We also show how nontrivial identities arise between multiple polylogarithms at a fixed level. In Appendix \ref{appendixconnection}, we write the Heun connection formula relevant to the Schwarzschild anti-de Sitter case. We obtain the first order correction in $R_h$ of the QNM frequency with $n=0$, $\ell=s=1$, which matches the result obtained in Sec.~\ref{section4}, although the procedure is more involved and less efficient.

The interested reader can find the relevant {\tt Mathematica} files on {\url{https://github.com/GlebAminov/BH_PolyLog}}. More precisely
\begin{itemize}
\item In the folders "(A)dS General n,l,s", we list explicit results for some quantum numbers $(n,\ell,s)$ and also include executable files to make the computation at higher quantum numbers. This can be done just by specifying $(n,\ell,s)$ at the beginning of the file "$\rm Nf4_{-}{exe}$" in the subfolder "Executable files". 
In the subfolder "Heun wave functions", one can find the expansion to the first orders of the solutions to the differential equation in both left and right regions.
\item In the folder "Robin general m", we include the relevant orders for our computations of the solutions to the differential equation in the three regions in the files "$\rm WFL_{-}Robin.m$", "$\rm WFM_{-}Robin.m$", and "$\rm WFR_{-}Robin.m$". Moreover, in the file "$\rm Robin_{-}{QNM}$", we list the first seven orders of the expansion of the low-lying  QNM frequencies $\omega_k$ and their hydrodynamic limit in the file  "$\rm Robin_{-}{QNM_{-}hydro}$". 
\end{itemize}
More details are given in the {\tt README.md} and {\tt Comments.md} files on the  GitHub link above. Attached to the arXiv submission, the notebooks with the computed frequencies can also be found.

\vskip 0.5cm
\noindent {\large {\bf Acknowledgments}}
\vskip 0.5cm
We would like to thank 
L. Andersson, 
E. Barausse,
M. Bianchi,
J. Bluemlein,
V. Cardoso,
S. Cecotti,
O. Dias,
M. Dodelson,
F. Fucito,
S. Grozdanov,
Y. Hatsuda,
C. Iossa,
R. Karlsson, 
J.F. Morales,
R. Russo,
P. Vanhove,
S.T. Yau,
and
A. Zhiboedov
for interesting discussions.  

The research of G.B. is partly supported by the INFN Iniziativa Specifica ST\&FI and by the PRIN project ``Non-perturbative Aspects Of Gauge Theories And Strings''.
The research of P.A.\ and A.T.\ is partly supported by the INFN Iniziativa Specifica GAST.
The research of P.A.\, G.B and A.T.\ is partly supported
by the MIUR PRIN Grant 2020KR4KN2 
"String Theory as a bridge between Gauge Theories and Quantum Gravity".
The work of A.G.\ is partially supported by the Swiss National Science Foundation Grant No.185723 and the NCCR SwissMAP.
The work of A.T.\ is partly supported by the PRIN project ``Geometria delle variet\`a algebriche'' and InDAM GNFM.
P.A., G.B., A.G., and A.T. acknowledge funding from the EU project  
Caligola HORIZON-MSCA-2021-SE-01), Project ID: 101086123.

We acknowledge the Galileo Galilei Institute in Firenze and the participants of the  Workshop
"New horizons for (no-)horizon physics: from gauge to gravity and back" 
during which part of this research work was conducted.

\section{Methodology}\label{section2}
\subsection{The gauge theory approach}\label{sec:gaugetheory}

Following \cite{Aminov:2020yma,Bonelli:2021uvf,Bianchi:2021xpr,Bianchi:2021mft,Casals:2021ugr, Bianchi:2021yqs,Bianchi:2022wku,Dodelson:2022yvn,Consoli:2022eey,daCunha:2022ewy,Dodelson:2023vrw,Bianchi:2023sfs,Bhatta:2022wga}, our first approach to study black hole perturbation theory is to identify the differential equations coming from the gravitational side with the differential equations originating from Seiberg-Witten theory or, equivalently, satisfied by conformal blocks with a degenerate primary insertion; hence use the NS functions to solve the corresponding spectral problem.  In particular, we apply this method when dealing with the Heun equation, which is a second-order differential equation with four regular singularities, for which the connection formulae are given in \cite{Bonelli:2022ten}. The \emph{Heun equation} is given by
\begin{equation}\label{heuncanonical}
\begin{aligned}
&\left( \frac{d^2 }{dz^2}+\left( \frac{\gamma}{z}+\frac{\delta}{z-1}+\frac{\epsilon}{z-t} \right)\frac{d}{dz}+\frac{\alpha \beta z - q}{z(z-1)(z-t)} \right) \psi(z) = 0,\\
&\alpha+\beta+1=\gamma+\delta+\epsilon.
\end{aligned}
\end{equation}

To study the quasinormal modes equation, the gauge theory approach is useful when the boundary conditions are imposed at the singularities of the problem, like for the SdS case, analyzed in Sec.~\ref{section3}. Thanks to the power-like behaviour of the local solutions near the singular points, it is easy to identify the two local solutions selected by the two boundary conditions.
 These two solutions are then used to quantize the frequencies, taking into account the connection formula that relates them. In these cases the quantization condition is expressed in terms of the quantum periods of the underlying SW geometry, computed via NS functions (see formula \eqref{eq:Bperiod} and Appendix \ref{appendixA} for the conventions used).
If, as it happens in the SAdS case analyzed in Sec.~\ref{section4}, at least one of the boundary conditions is imposed in a regular point of the differential equation, the gauge theory approach is less effective: it is still possible to solve the problem with the connection formulae, but the quantization condition for the frequency will not be expressed in terms of NS functions only (see formula \eqref{quantizationtoexpand}).  

By introducing an appropriate change of variables, we can always transform the perturbation equation with four regular singularities in the Heun form and send the singularities in $z=0,1,t,\infty$. In all the cases in which the connection formulae are used, we will put us in a regime in which the complex modulus of $t$ is small, $|t|\ll 1$, and such that the relevant connection formula is among local solutions in $z=t$ and in $z=1$. The independent solutions of the Heun equation for $z\sim t$ are
\begin{equation}\label{Heunint}
\begin{aligned}
&\psi_-^{(t)} (z) =\mathrm{Heun}\biggl(\frac{t}{t-1},\frac{q-t\alpha\beta}{1-t},\alpha,\beta,\epsilon,\delta,\frac{z-t}{1-t}\biggr), \\ 
&\psi_+^{(t)} (z) =(z-t)^{1-\epsilon}\mathrm{Heun}\biggl(\frac{t}{t-1},\frac{q-(\beta-\gamma-\delta)(\alpha-\gamma-\delta)t-\gamma(\epsilon-1)}{1-t},\\
&\ \ \ \ \ \ \ \ \ \ \ \ \ \ \ \ \ \ \ \ \ \ \ \ \ \ \ \ \ \ \ \ \ \ \ \ -\alpha+\gamma+\delta,-\beta+\gamma+\delta,2-\epsilon,\delta,\frac{z-t}{1-t}\biggr),
\end{aligned}
\end{equation}
and the ones for $z\sim 1$ are
\begin{equation}\label{Heunin1}
\begin{aligned}
\psi_-^{(1)}(z)=&\biggl(\frac{z-t}{1-t}\biggr)^{-\alpha}\mathrm{Heun}\biggl(t,q+\alpha(\delta-\beta),\alpha,\delta+\gamma-\beta,\delta,\gamma,t\frac{1-z}{t-z}\biggr),\\
\psi_+^{(1)}(z)=&(z-1)^{1-\delta}\biggl(\frac{z-t}{1-t}\biggr)^{-\alpha-1+\delta}\mathrm{Heun}\biggl(t,q-(\delta-1)\gamma t-(\beta-1)(\alpha-\delta+1),\\
&\ \ \ \ \ \ \ \ \ \ \ \ \ \ \ \ \ \ \ \ \ \ \ \ \ \ \ \ \ \ \ \ \ \ \ \ \ \ \ \ -\beta+\gamma+1,\alpha-\delta+1,2-\delta,\gamma,t\frac{1-z}{t-z}\biggr).
\end{aligned}
\end{equation}
In terms of the connection matrices of hypergeometric functions
\begin{equation}
\mathcal{M}_{\theta \theta'}(a_1,a_2;a_3) = \frac{\Gamma(-2\theta'a_2)\Gamma(1+2\theta a_1)}{\Gamma\left(\frac{1}{2}+\theta a_1-\theta' a_2 + a_3\right) \Gamma\left(\frac{1}{2}+\theta a_1-\theta' a_2 - a_3\right)},
\end{equation}
where $\theta,\theta'=\pm$, the connection formula for small $t$ from $z\sim t$ to $z\sim 1$ is given by \cite{Bonelli:2022ten}
\begin{equation}
\begin{aligned}
&t^{-\frac{1}{2}+a_0\mp a_t}(1-t)^{-\frac{1}{2}+a_1}e^{\mp\frac{1}{2}\partial_{a_t}F(t)}\psi_{\pm}^{(t)}(z)=\\
&\left(\sum_{\sigma=\pm}\mathcal{M}_{\pm\sigma}(a_t,a;a_0)\mathcal{M}_{(-\sigma)-}(a,a_1;a_{\infty})t^{\sigma a}e^{-\frac{\sigma}{2}\partial_aF(t)}\right)(1-t)^{-\frac{1}{2}+a_t}e^{i\pi(a_1+a_t)}e^{\frac{1}{2}\partial_{a_1}F(t)}\psi_-^{(1)}(z)+\\
&\left(\sum_{\sigma=\pm}\mathcal{M}_{\pm\sigma}(a_t,a;a_0)\mathcal{M}_{(-\sigma)+}(a,a_1;a_{\infty})t^{\sigma a}e^{-\frac{\sigma}{2}\partial_aF(t)}\right)(1-t)^{-\frac{1}{2}+a_t}e^{i\pi(-a_1+a_t)}e^{-\frac{1}{2}\partial_{a_1}F(t)}\psi_+^{(1)}(z).
\end{aligned}
\end{equation}

Within the examples analyzed in this paper, the gauge theory approach is particularly effective in computing quasinormal modes at large $\ell$ in the SdS case (see subsection \ref{section:large_ell}).

\subsection{The multi polylog approach}
\label{sec:pert_method}

When the gauge theory approach proves less effective, we solve the QNM spectral problem order by order in some suitably chosen expansion parameter $\kappa$ and for fixed values of some quantum numbers.\footnote{Here "suitably chosen" means that the $0^{\rm th}$ order is solvable in terms of relatively simple functions (e.g.~rational functions or logarithms).
In addition, we would like good convergence properties for the expansions in $\kappa$ in the spectral problem.
We believe that this is the case, at least for the  example of Sec.~\ref{section3} where $\kappa$ is related to the instanton counting parameter $t$ of the underlying gauge theory (see \cite{Its:2014lga,Arnaudo:2022ivo,daCunha:2022ewy} for the study of convergence properties in gauge theory).
}
For instance, we can consider perturbation theory in $R_h$ or $R_h^{-1}$, $R_h$ being the radius of the black hole horizon. This is like doing  Hamiltonian perturbation theory,
and although the numerical implementation of this algorithm is well known (see e.g.~\cite{Cardoso:2001bb,Cardoso:2003vt,Horowitz:1999jd, review}), if we want an analytical answer, the calculation quickly becomes cumbersome. In particular, it is necessary to find suitable stratagems for higher orders.  For the situations we consider in this work, we find that higher orders can be determined systematically using the underlying structure that involves multiple polylogarithms (similar techniques are also used to compute Feynman integrals in QCD, see e.g.~\cite{Duhr:2019tlz,Panzer:2014caa,Frellesvig:2018lmm,c1,c2,c3} and references therein). In this section, we sketch the general idea while we give more details in concrete examples; see Sec.~\ref{section3}, Sec.~\ref{section4}, and Sec.~\ref{section5}. {\tt Mathematica} notebooks are also attached. 

The spectral problems we are interested in are two-point boundary value problems
associated with differential equations on the sphere with $n$ regular singularities.\footnote{The case of irregular singularities will appear elsewhere \cite{GlPa}.} More precisely, we will focus on the cases $n=4,5$. 
The boundary conditions are fixed at generic points $z=z_1$ and $z=z_2$, not necessarily coinciding with the position of the regular singularities.  

For the problems at hand, we can use the following Ansatz for the eigenfunctions in each region of the patch decomposition\footnote{We remark that the expansions in $z$ of the local solutions, performed around a given singularity of the problem, hold in an open disk centered in that singularity. The radius of this disk is equal to the distance between that singularity and the closest one.} of the $n$-punctured sphere
\be
\label{eq:psi_exp0}\psi\brc{z}=f_0\brc{z}+\sum_{K\geq 1} f_K(z) \kappa^K ~.\ee
Sometimes, it is useful to introduce additional regions 
with respect to the minimal patch decomposition to optimize the efficiency of perturbation theory. Different scalings in $\kappa$ of the regular singularities of the equation under the scale redefinitions $z\to \kappa^c z$ determine the possible number of regions. 
We assume that in the differential equation (and therefore in the position of the singularities too), only integer powers of $\kappa$ appear, therefore $c\in\mathbb{Z}$.
The position of the regular singularities can depend on the perturbation parameter $\kappa$. Therefore, in the perturbative expansion in $\kappa\sim 0$, the singularities will tend to cluster differently as we change the critical parameter $c$. These different clustering schemes define the different regions for the perturbative expansion and determine a finer structure in the patch decomposition, which also considers the different geometric situations describing the potential terms in the differential equation.
For a nontrivial example, see Sec.~\ref{section5}.
%
At each order in $\kappa$, $\psi\brc{z}$ is determined by a second-order equation
\be\label{eqp}\left(f_K(z)\right)''+ \varphi\brc{z} \left(f_K(z)\right)'+ \nu(z) f_K(z) +\eta_{K}(z)=0,\ee
which we solve by using the method of variation of parameters. 
The functions $\varphi$ and $\nu$ in \eqref{eqp} are 
known,\footnote{
The wave equation is understood to be Taylor expanded as
$\psi''+\sum_{H=0}^{+\infty}\kappa^H(\varphi_H\psi'+\nu_H\psi)=0$,
so that one finds explicitly 
$\eta_{K}=\sum_{L=1}^K(\varphi_Lf_{K-L}'+\nu_Lf_{K-L})$.
In the text $\varphi_0=\varphi$ and $\nu_0=\nu$.
} and the non-homogeneous part of the differential equation 
$ \eta_{K}(z)$ is fully determined by the solutions to the previous orders $f_{m}$ with $m\leq K-1$. Let $f_0, g_0$ be the two solutions of the homogeneous part of \eqref{eqp}.\footnote{These are the two solutions of  the leading order equation, $g_0$ being the one that does not satisfy the relevant boundary condition.} Then we  write the generic solution to   \eqref{eqp} as\footnote{The integrals appearing in \eqref{eq:psi_exp} are the indefinite ones. }
\be \label{eq:psi_exp}f_K(z)= b_{K} g_0(z)+c_{K} f_{0}(z) -g_0(z) \int^{z} f_{0}(z')\, \frac{\eta_{K}(z')}{W_0(z')}\, \rmd z' +f_{0}(z) \int^{z} g_0(z')\, \frac{ \eta_{K}(z')}{W_0(z')}\, \rmd z',\ee
where 
$W_0$ is the Wronskian of the two leading order solutions 
\eq{W_0\equiv f_{0}\brc{g_0}' - \brc{f_{0}}'g_0.}
In each region, the integration constants $c_K$'s can be absorbed into a normalization of the solution, and they can be fixed to zero without loss of generality.
Imposing the two boundary conditions and gluing the local solutions fixes the integration constants $b_K$ and gives the quantization of the frequency of the perturbation. If either $z_{1,2}$ has a non-trivial dependence on the parameter $\kappa$, 
the boundary condition is applied by expanding $\psi\brc{z_{1,2}}$ in powers of $\kappa$. In the following sections, we will be more detailed in describing how this expansion works case by case.

In principle, the relations \eqref{eq:psi_exp} allow us to compute the wave function up to any given order in $\kappa$. However, to implement this algorithm in practice, there is still a non-trivial step: explicitly compute the integrals in \eqref{eq:psi_exp}.
In all our examples, the leading order solutions are described in terms of rational or logarithmic functions, and their Wronskian is a rational function. Hence the wave function at order $\kappa^{K}$ is described in terms of multiple polylogarithms of weight $K$ and lower. In the cases analyzed in sections \ref{section3} and \ref{section4}, up to order $R_h^4$, one can avoid using multiple polylogarithms due to identities \eqref{eq:Id_MLtoL1}--\eqref{eq:Id_MLtoL2} presented in Appendix \ref{appendixB}. However, from order $R_h^5$ on, the expansions in multiple polylogarithms cannot be avoided to our knowledge. Here we would like to mention that a more general statement about multiple polylogarithms in several variables is well-known. According to Theorem D of \cite{Wojtkowiak'91}, not every multiple polylogarithm of weight $\geq 4$ can be expressed as a finite combination of classical polylogarithms. Since, in Sections \ref{section3} and \ref{section4}, we only deal with multiple polylogarithms in a single variable, we can push this bound to weight $5$.

\section{Perturbations of de Sitter black holes in four dimensions}\label{section3}

\subsection{Schwarzschild de Sitter black hole}\label{section3.1}

The metric describing the de Sitter Schwarzschild black hole in four dimensions ($\rm SdS_4$) is
\begin{equation}\label{metric}
\mathrm{d} s^2=- f(r) \mathrm{d} t^2+f(r)^{-1} \mathrm{d} r^2+r^2 \mathrm{d} \Omega_{2}^2
\end{equation}
with 
\begin{equation} f(r)=1-\frac{2 M}{r}-\frac{\Lambda}{3}r^2,\end{equation}
where $M$ is the mass of the black hole and $\Lambda>0$ is the cosmological constant. In what follows, we will fix $\Lambda=3$, and then we suppose $M$ to be in the range $0<M^2<1/27$ to have three real roots for the equation $rf(r)=0$, since otherwise we would have unphysical solutions. 
We will denote these roots by 
\begin{equation} R_h, \quad R_{\pm},  \end{equation}
where $R_h\in ]0,{1/\sqrt{3}}[$ is the smallest positive real root, and $R_\pm$ are real and given in terms of $R_h$ by
\begin{equation}
R_{\pm}=\frac{-R_h\pm\sqrt{4-3R_h^2}}{2}.
\end{equation} 
We will study a class of linear perturbations of the  $\rm SdS_4$ geometry with spin $s\in\{0,1,2\}$, encoded in the following radial equation (see  \cite{review} and reference therein) 
\begin{equation}\label{eq:RW_dS}
\left(\partial^2_r + \frac{f'(r)}{f(r)} \partial_r +\frac{\omega^2-V(r)}{f(r)^2}\right)\Phi(r)=0,
\end{equation}
where the potential is 
\begin{equation}\label{eq:pot}
V(r)= f(r)\left[\frac{\ell(\ell+1)}{r^2}+(1-s^2)\left(\frac{2M}{r^3}\right)\right].
\end{equation}
For $s=0$, this equation describes conformally coupled scalar perturbations; for $s=1$, electromagnetic perturbations; and for $s=2$, odd (Regge--Wheeler or vector-type) gravitational perturbations. 

The boundary conditions we impose on the wave function are the presence of only ingoing modes at the event horizon $R_h$ and the presence of only outgoing modes at the cosmological horizon $R_+$. These conditions can be made explicit by introducing the \emph{tortoise coordinate} $r_*$ defined by
\begin{equation}
\mathrm{d}r_*=\frac{\mathrm{d}r}{f(r)}.
\end{equation}
In terms of $r_*$, the behavior of $\Phi$ near $R_h,R_+$ is described by plane waves, so we ask that $\Phi$ behaves as $\exp(-i\omega r_*)$ for $r\sim R_h$ and as $\exp(i\omega r_*)$ for $r\sim R_+$. 
The latter radial equation apparently has five regular singular points located at $r=\{0,R_h,R_\pm,\infty\}$. However, as pointed out in \cite{suzuki1998}, under the change of variable
\begin{equation}
\label{eq:dS_z_trans}
z(r)=\frac{r(R_{+}-R_{-})}{R_{+}(r-R_{-})},
\end{equation}
and redefinition of the wave function
\begin{equation}\label{2.8}
\psi(z)=z^{-\gamma/2}(z-1)^{-\delta/2}(z-t)^{-\epsilon/2}\sqrt{f(r)}\, \frac{R_{-}(R_{+}-R_{-})}{R_{+}(r-R_{-})} \, \Phi(r),
\end{equation}
where
\begin{equation}
\begin{aligned}
&t=\frac{R_h(R_{-}-R_{+})}{R_{+}(R_{-}-R_h)},\\
&\gamma =1-2s,\\
&\delta = 1- \frac{2i\,\omega\,R_{+}}{(R_{+}-R_h)(R_{+}-R_{-})},\\
&\epsilon=1+\frac{2i \omega R_h}{1-3R_h^2},
\end{aligned}
\end{equation}
the singularity at infinity is removed, and the equation becomes a Heun equation \eqref{heuncanonical}
with
\begin{equation}
\begin{aligned}
&\alpha=1-s+\frac{2i\,\omega\,R_{-}}{(R_{-}-R_h)(R_{-}-R_{+})},\\
&\beta=1-s,\\
&q=\frac{\ell(\ell+1)}{R_{+}(R_{-}-R_h)} + \frac{(1-s)^2 R_h}{R_h-R_-} -
 \frac{s(1-s) R_{-}^2}{R_{+}(R_h-R_{-})}.
\end{aligned}
\end{equation}
In the $z$ coordinate, the horizon $r=R_h$ is mapped to $z=t$, 
the cosmological horizons $r=R_{\pm}$ are mapped to $z=1$ and $z=\infty$, respectively, while the origin, $r=0$, is mapped to $z=0$.

The boundary conditions described for $\Phi$ imply the following behaviors for the function $\psi$:
\begin{equation}\label{bc_SdS}
\begin{aligned}
&\psi(z)\sim 1\ \ \ \ \ \ \ \ \ \ \ \ \ \text{for}\ \ z\sim 1,\\
&\psi(z)\sim (z-t)^{1-\epsilon}\ \ \text{for}\ \ z\sim t.
\end{aligned}
\end{equation}

We now want to obtain the analytic formula from which the quasinormal modes can be computed in the limit where $t$ is small, $0<t\ll 1$, or, equivalently, $R_h$ is small, $R_h\ll 1$. For this purpose, we write the following dictionary for the gauge parameters in terms of Heun's parameters and gravitational quantities (see appendix \ref{appendixA} for the conventions used):
\begin{equation}
\begin{aligned}
&t=\frac{R_h(R_{-}-R_{+})}{R_{+}(R_{-}-R_h)}~,\\
a_0&=\frac{1-\gamma}{2}=s~,\\
a_1&=\frac{1-\delta}{2}=\frac{i\,\omega\,R_{+}}{(R_{+}-R_h)(R_{+}-R_{-})}~,\\
a_t&=\frac{1-\epsilon}{2}=-\frac{i \omega R_h}{1-3R_h^2}~,\\
a_{\infty}&=\frac{\alpha-\beta}{2}=\frac{i\,\omega\,R_{-}}{(R_{-}-R_h)(R_{-}-R_{+})}~,\\
u^{(0)}&=\frac{-2q+2t\alpha\beta+\gamma\epsilon-t(\gamma+\delta)\epsilon}{2(t-1)}~.
\end{aligned}
\end{equation}

\subsubsection{Connection Problem}\label{sec:connection}

The computation of quasinormal mode frequencies is obtained by imposing purely ingoing boundary conditions at the event horizon $z=t$ and purely outgoing at the positive cosmological horizon $z=1$.
The independent solutions of the Heun equation for $z\sim t$ are given in \eqref{Heunint}, and the ones for $z\sim 1$ are given in \eqref{Heunin1}.
Taking into account the boundary conditions \eqref{bc_SdS}, the connection coefficient between $\psi_+^{(t)}$ and $\psi_+^{(1)}$ has to be set equal to zero.

The connection formula \cite{Bonelli:2022ten}\footnote{For the computation of the traces of monodromies of the Heun equation, see also \cite{Jeong:2018qpc,Hollands:2017ahy}.} 
 for small $t$ from $z\sim t$ to $z\sim 1$ is given by
\begin{equation}
\begin{aligned}
&t^{-\frac{1}{2}+a_0-a_t}(1-t)^{-\frac{1}{2}+a_1}e^{-\frac{1}{2}\partial_{a_t}F(t)}\psi_+^{(t)}(z)=\\
&\left(\sum_{\sigma=\pm}\mathcal{M}_{+\sigma}(a_t,a;a_0)\mathcal{M}_{(-\sigma)-}(a,a_1;a_{\infty})t^{\sigma a}e^{-\frac{\sigma}{2}\partial_aF(t)}\right)(1-t)^{-\frac{1}{2}+a_t}e^{i\pi(a_1+a_t)}e^{\frac{1}{2}\partial_{a_1}F(t)}\psi_-^{(1)}(z)+\\
&\left(\sum_{\sigma=\pm}\mathcal{M}_{+\sigma}(a_t,a;a_0)\mathcal{M}_{(-\sigma)+}(a,a_1;a_{\infty})t^{\sigma a}e^{-\frac{\sigma}{2}\partial_aF(t)}\right)(1-t)^{-\frac{1}{2}+a_t}e^{i\pi(-a_1+a_t)}e^{-\frac{1}{2}\partial_{a_1}F(t)}\psi_+^{(1)}(z).
\end{aligned}
\end{equation}
This leads us to the quantization condition for the quasinormal modes in the form 
\begin{equation}
\begin{aligned}
&\sum_{\sigma=\pm}\mathcal{M}_{+\sigma}(a_t,a;a_0)\mathcal{M}_{(-\sigma)+}(a,a_1;a_{\infty})t^{\sigma a}e^{-\frac{\sigma}{2}\partial_aF(t)}=0,
\end{aligned}
\end{equation}
which can be rewritten as
\begin{equation}\label{eq:quantSdS4}\small
\begin{aligned}
\frac{\Gamma(1+2a)^2\Gamma(\frac{1}{2}-a+a_t+a_{0})\Gamma(\frac{1}{2}-a+a_t-a_{0})\Gamma(\frac{1}{2}-a-a_1-a_{\infty})\Gamma(\frac{1}{2}-a-a_1+a_{\infty})}{\Gamma(1-2a)^2\Gamma(\frac{1}{2}+a+a_t+a_{0})\Gamma(\frac{1}{2}+a+a_t-a_{0})\Gamma(\frac{1}{2}+a-a_1-a_{\infty})\Gamma(\frac{1}{2}+a-a_1+a_{\infty})}t^{-2a}e^{\partial_aF(t)}=1.
\end{aligned}
\end{equation} 
Note that this is nothing but (see Appendix \ref{appendixA})
\begin{equation}\label{eq:Bperiod}
\exp\left(\partial_aF_{\mathrm{full}}(t)\right)=1,
\end{equation}
where $F_{\mathrm{full}}(t)$ is the full NS free energy, since the ratio of Gamma functions in \eqref{eq:Bperiod} represents the 1-loop corrections.

\subsubsection{QNMs at large \texorpdfstring{$\ell$}{}} \label{section:large_ell}
The previous quantization condition gets simplified in the large $\ell$ limit, where we neglect non-perturbative effects in $\ell$ of the form $R_h^\ell$. This regime was studied for ${\rm AdS}_5$ black holes in \cite{Dodelson:2022eiz,Dodelson:2022yvn}, since in this limit, the quasinormal mode frequencies become real, and, via the AdS/CFT correspondence, they compute the dimensions of certain operators in the holographic conformal field theory, see \cite{Li:2020dqm,Esper:2023jeq,Karlsson:2021duj,Kulaxizi:2018dxo,Karlsson:2019qfi,Kulaxizi:2019tkd,Karlsson:2019dbd,Karlsson:2020ghx} and references therein. 
In the dS case, in this regime, the quasinormal mode frequencies are purely imaginary, and their interpretation from the point of view of holography is, at present, less clear (at least to us). 

In the leading order in $R_h$, $a\sim \pm\left(\ell+\frac{1}{2}\right)$. Choosing the plus sign, the quantization condition
\begin{equation}
\sum_{\sigma=\pm}\mathcal{M}_{+\sigma}(a_t,a;a_0)\mathcal{M}_{(-\sigma)+}(a,a_1;a_{\infty})t^{\sigma a}e^{-\frac{\sigma}{2}\partial_aF(t)}=0
\end{equation}
simplifies to
\begin{equation}
\mathcal{M}_{+-}(a_t,a;a_0)\mathcal{M}_{++}(a,a_1;a_{\infty})t^{-a}e^{\frac{1}{2}\partial_aF(t)}=0,
\end{equation}
since the other term is exponentially suppressed. This condition is satisfied if and only if
\begin{equation}
\frac{\Gamma(2a)\Gamma(1-2a_t)\Gamma(1+2a)\Gamma(-2a_1)}{\Gamma(\frac{1}{2}+ a+a_t+a_{0})\Gamma(\frac{1}{2}+ a+a_t-a_{0})\Gamma(\frac{1}{2}+ a-a_1-a_{\infty})\Gamma(\frac{1}{2}+ a-a_1+a_{\infty})}=0,
\end{equation}
which is solved at the poles of the Gamma functions in the denominator. Only the last one admits poles among the four Gamma functions in the denominator, consistently with our regime $R_h\ll 1$. These are given by condition 
\begin{equation}\label{bigellcond}
\frac{1}{2}+a-a_1+a_{\infty}=-n,\ \ \text{with}\ \ n\in\mathbb{Z}_{\ge 0}.
\end{equation}
Expanding the parameters in $R_h$ and writing $\omega$ as \begin{equation}
\label{dS_omega_exp}
 \omega=\sum_{k=0}^{\infty} \omega_k R_h^k,
\end{equation}
we obtain from this condition 
\begin{equation}
\begin{aligned}
&\omega_0=i (-\ell-n-1);\\
&\omega_1=0;\\
&\omega_2=-\frac{i}{8 \ell (\ell+1) (2 \ell+1)(2\ell-1)(2\ell+3)}\Bigl\{\ell^4 \left(60 n^2+60 n+22\right)+\ell^3 \bigl(120 n^2+48ns^2+122n+\\
&\ \ \ \ \ \ \ +24 s^2+45\bigr)+\ell^2 \left[8 n^2 \left(3 s^2+2\right)+n \left(96 s^2+19\right)+8 s^4+44 s^2+8\right]+\\
&\ \ \ \ \ \ \ +\ell \left[4 n^2 \left(6 s^2-11\right)+n \left(24 s^4-43\right)+20 s^4-4 s^2-15\right]+12 (n+1)^2 s^2 \left(s^2-2\right)\Bigr\};\\
&\omega_3=0;\\
&\qquad  \vdots
\end{aligned}
\end{equation}
Higher orders can also be computed systematically, but their expressions are cumbersome; hence we do not write them explicitly. 
Notice that in this limit, all the odd orders $\omega_{2k+1}$ seem to vanish.  Moreover, these formulas are correct for finite $\ell$ up to order $R_h^{2\ell+1}$, as will be shown in section \ref{section3.2}. We also note that we expect the series \eqref{dS_omega_exp} to be convergent\footnote{The convergence can be inferred from the convergence of the NS functions.} in $R_h$, the need for non-perturbative effects in $\ell$ can be inferred from the fact that at higher orders this series develops some unphysical poles in $\ell$. For instance, for $s=0$ we have an unphysical pole at $\ell=0$
\be \omega_4|_{s=0} =\frac{i (n+1)^4}{\ell}+\mathcal{O}\left(\ell^0\right)~.\ee

\subsection{Perturbation theory around \texorpdfstring{$\mathrm{dS}_4$}{} } \label{section3.2}

\subsubsection{QNMs in pure \texorpdfstring{$\mathrm{dS}_4$}{}}

The pure de Sitter case can be obtained by taking the limit $t \rightarrow 0$ or, equivalently, $R_h \rightarrow 0$. As the event horizon disappears in this limit, it is enough to consider only the region near the cosmological horizon $r=R_+$.
In this limit, the Heun equation becomes a Hypergeometric equation, whose solutions are
\begin{equation}
\label{pure_dS_sol}
z^{s-\ell-1}{}_2F_1(-\ell,-\ell-i\omega_0;-2\ell;z),\quad z^{\ell+s}{}_2F_1(\ell+1,\ell+1-i\omega_0;2\ell+2;z),
\end{equation}
where $\omega_0$ is the leading order term in the $R_h$ expansion of the frequency \eqref{dS_omega_exp}.
Since $\ell$ is a non-negative integer, the hypergeometric functions get truncated to polynomials as
\begin{equation}
\label{pure_dS_sol_sim}
\begin{aligned}
{}_2F_1\brc{-\ell,-\ell-i\omega_0;-2\ell;z}=&\sum_{k=0}^{\ell}\brc{-1}^k \binom{\ell}{k} \frac{\brc{-\ell-i \omega_0}_k}{\brc{-2\ell}_k} \, z^k,\\
{}_2F_1\brc{\ell+1,\ell+1-i\omega_0;2\ell+2;z}=& \brc{-1}^{\ell} \brc{\frac{z}{2}}^{-2\ell-1} 
\frac{\brc{\brc{2\ell+1}!!}^2}{2 \brc{2\ell+1}} \frac{\Gamma\brc{i \omega_0 - \ell}}{\Gamma\brc{i \omega_0 + \ell+1}}\times\\
\times & \sum_{k=0}^{\ell}\brc{-1}^k \binom{\ell}{k} \frac{\brc{-\ell-i \omega_0}_k}{\brc{-2\ell}_k} 
\brc{1-\brc{1-z}^{i \omega_0}\frac{\brc{-\ell+i \omega_0}_k}{\brc{-\ell-i \omega_0}_k}} z^k .
\end{aligned}
\end{equation}
The boundary conditions require that
 the radial part of the gravitational perturbation $\Phi\brc{r}$ is well-defined as $r\rightarrow 0$. Using the dictionary for the wave function \eqref{2.8}, we rewrite the latter requirement in terms of 
$\psi\brc{z}$:
\begin{equation}
z^{\gamma/2} \psi\brc{z} = z^{-s+1/2} \psi\brc{z} \sim 1 \quad \text{for} \quad z\sim 0.
\end{equation}
Thus, we have to pick a regular solution at $z\sim 0$ and consider an additional factor of $z^{-s+1/2}$.
Looking at the first solution from \eqref{pure_dS_sol}, we can see that $z^{-\ell-1/2}{}_2F_1(-\ell,-\ell-i\omega_0;-2\ell;z)$ is not regular at 
$z \sim 0$ for any allowed value of $\ell$. 
Indeed, the other combination gives the solution, which is regular at $z\sim 0$:
\begin{equation}
z^{\ell+1/2}{}_2F_1\brc{\ell+1,\ell+1-i\omega_0;2\ell+2;z}\sim z^{\ell+1/2}\sim 0.
\end{equation}
In addition, the boundary conditions at the cosmological horizon require 
the eigenfunction to be  regular with a well-defined Taylor expansion at $z=1$.
This is possible only if 
$i\omega_0 \in\mathbb{Z}_{\ge 0}$ (due to the term $\brc{1-z}^{i\omega_0}$ in \eqref{pure_dS_sol_sim}). Moreover, to avoid the poles in the Gamma functions in \eqref{pure_dS_sol_sim}:
\begin{equation}
\frac{\Gamma\brc{i \omega_0 - \ell}}{\Gamma\brc{i \omega_0 + \ell+1}}= \prod_{k=-\ell}^{\ell}\brc{i \omega_0 - k}^{-1},
\end{equation}
we must exclude all the values of $i\omega_0$ that are  smaller or equal to $\ell$ (these poles indicate that the second expression in \eqref{pure_dS_sol_sim} have to be rewritten in terms of $\log\brc{z-1}$ for $i\omega_0 = \ell,\ell-1,\dots,-\ell+1,-\ell$). This gives the well-known  quantization condition for the QNM frequencies of the pure $\mathrm{dS}_4$:
\begin{equation}
i\omega_0=\ell+n+1,\ \ \text{with}\ n\in\mathbb{Z}_{\ge 0}.
\end{equation}
The corresponding eigenfunction is 
\be
\label{eq:fL_dS}
f^L_0(z)=z^{\ell+s}\sum_{k=0}^{n}\brc{-1}^k \binom{n}{k}
\frac{(\ell+1)_k}{\brc{2 \ell+2}_k} \, z^k~.
\ee
 We also note that the discarded solution is
 \be \label{eq:fL_dS2}
g^L_{0}(z)=z^{s-\ell-1} \brc{1-z}^{\ell+n+1}\sum_{k=0}^{\ell}\brc{-1}^k \binom{\ell}{k} \frac{\brc{n+1}_k}{\brc{-2\ell}_k} \, z^k~.
\ee
The Wronskian between $f^L_0$ and $g^L_0$ is 
\begin{equation}
W^L_0(z)=-(2\ell+1)z^{2s-2}(1-z)^{\ell+n}.
\end{equation}

\subsubsection{Left Region}
\label{sec:dS_Left}

Here we call the region near the cosmological horizon $r=R_+$ left region due to the analogy with the corresponding quantum mechanical problem on the complex plane. The local variable in this region is $z$, and the leading order solutions in $R_h$ (and so in $t$) of the Heun equation \eqref{heuncanonical} are given in \eqref{eq:fL_dS}, \eqref{eq:fL_dS2}.
Expanding in small $R_h$ the solution and the frequencies we get for the outgoing solution $\psi_{-}^{(1)}$ at the cosmological horizon 
\begin{equation}
\psi_{-}^{(1)}\brc{z} = \frac{\ell! \brc{2 \ell +n+1}!}{\brc{2\ell+1}!\brc{\ell+n}!} \, f_0\brc{z} +
 \frac{\brc{-2}^{\ell} i \omega_1}{\brc{i \omega_1 +\ell+n+1}} \frac{n! \brc{2\ell-1}!!}{\brc{\ell+n}!} \, g_0(z) + \mathcal{O}\brc{R_h},
\end{equation}
where $\omega_1$ is a coefficient in the $R_h$ expansion of the frequency \eqref{dS_omega_exp}.  Since $g_0(z)$ blows up as $z\rightarrow 0$, it should not be present in the leading order of the wave function in the left region. Hence, we require $\omega_1=0$. On the other hand, the incoming wave solution at the cosmological horizon is
\begin{equation}
\label{eq:incw_left_dS}
\psi_{+}^{(1)}\brc{z} \sim \brc{z-1}^{i \omega}
\brc{1 + i \omega \log\brc{z-1} R_h} +\mathcal{O}\brc{R_h^2}.
\end{equation}
After we fix $\omega_1=0$ and proceed with the general method described in Sec.~\ref{sec:pert_method}, the logarithm function $\log\brc{z-1}$ appears in higher orders in $R_h$ (and $t$).
The only source of this function is the incoming wave solution (\ref{eq:incw_left_dS}), and we will be canceling any contributions of $\log\brc{z-1}$ by fixing the coefficients $b_K$ in the perturbative expansion of the wave function (\ref{eq:psi_exp0}), 
 (\ref{eq:psi_exp}).

After establishing the boundary condition, we compute the integrals in (\ref{eq:psi_exp}).
As we show in Appendix \ref{app:recurrence_rel}, these integrals are described in terms of the multiple polylogarithms in a single variable:

\eqlb{eq:PolyLog_def_dS}{\text{Li}_{s_1,\dots,s_k}\brc{z}=\sum_{n_1>n_2>\dots>n_k\geq 1}^{\infty}
\frac{z^{n_1}}{n_1^{s_1}\dots n_{k}^{s_k}}.}
The latter admits for $s_1\geq 2$:
\eq{\label{eq:propertypolylog}z \, \frac{\rmd }{\rmd\, z} \text{Li}_{s_1,\dots,s_k}\brc{z} = \text{Li}_{s_1-1,\dots,s_k}\brc{z}}
and for $s_1=1$, $k\geq 2$:
\eq{\label{eq:propertypolylog2}\brc{1-z} \frac{\rmd }{\rmd\, z} \text{Li}_{1,s_2,\dots,s_k}\brc{z} = \text{Li}_{s_2,\dots,s_k}\brc{z}.}
The \emph{weight} of the multiple polylogarithm $\text{Li}_{s_1,\dots,s_k}\brc{z}$ is $s_1+\dots +s_k$, and the \emph{level} is $k$. At each order $t^{K+1}$, both integrands in (\ref{eq:psi_exp}) are linear 
combinations of the following terms with maximum weight $K$:
\eqlb{eq:dS_vLog_Left}{\frac{\sum_{m=0}^{r_1} \alpha_m\, z^m}{z^{i_1} \brc{z-1}^{j_1}} \log\brc{z}^{p_1},\quad 
\frac{\sum_{m=0}^{r_2} \beta_m\, z^m}{z^{i_2} \brc{z-1}^{j_2}} \text{Li}_{s_1,\dots,s_k}\brc{1-z},}
where $r_{1,2}$, $i_{1,2}$, $j_{1,2}$, $p_{1}$ are some non-negative integers, and $0\leq p_1\leq K$, $s_1+\dots +s_k \leq K$.
After taking the integrals, the only new functions that appear are multiple polylogarithms of maximum weight $K+1$. Moreover, both integrals in (\ref{eq:psi_exp}) are linear combinations of terms similar to (\ref{eq:dS_vLog_Left}):
\eqlb{eq:dS_vLog_Left2}{\frac{\sum_{m=0}^{r_1+1} \gamma_m\, z^m}{z^{i_1-1} \brc{z-1}^{j_1-1}} \, \log\brc{z}^{p_1},\quad
\frac{\sum_{m=0}^{r_2+1} \delta_m\, z^m}{z^{i_2-1} \brc{z-1}^{j_2-1}} \, \text{Li}_{s_1,\dots,s_k}\brc{1-z}}
and terms containing new combinations of logarithms and multiple polylogarithms that were not present in (\ref{eq:dS_vLog_Left}):
\eqlb{eq:dS_vLog2_Left}{\log\brc{z-1},\quad  \log\brc{z}^{K+1},\quad \text{Li}_{\hat{s}_1,\dots,\hat{s}_{\hat{k}}}\brc{1-z},}
where the maximum weight is $K+1$:
\eq{\hat{s}_1+\dots +\hat{s}_{\hat{k}} \leq K+1.}
One of the differences between \eqref{eq:dS_vLog_Left} and \eqref{eq:dS_vLog_Left2} is that $r_{1,2}$, $i_{1,2}$, and $j_{1,2}$ are shifted by $1$ or $-1$. These shifts are specific to the left region of the $\mathrm{SdS}_4$ case (and even then may be subjected to reevaluation for some values of quantum numbers $n$, $\ell$, and $s$ that we did not consider).  In the right region, the shifts are different but can be determined on the case by case basis (the details can be found in the attached {\tt Mathematica} files, where we distinguish two regimes with $\ell \leq n$ and $\ell> n$). Even though the optimal choice of shifts depends on the case at hand, there is a choice of big enough shifts applicable to all quantum numbers for both regions.

To summarize, we reduced the problem of solving the initial ODE in a given order in $t$ to a system of linear equations on the coefficients in front of the functions from (\ref{eq:dS_vLog2_Left}) and $\gamma_m$, $\delta_m$.\footnote{Here we simplified the index structure of $\gamma_m$ and $\delta_m$, the full list of indices should be $\gamma_m(p_1)$ and  $\delta_m(p_1, s_1,\dots,s_k)$.} The resulting corrections $f^L_K\brc{z}$ to the wave function in the left region are linear combinations of the following functions:
\begin{equation}
\sum_{m=-k_1}^{l_1} \zeta^L_m\, z^m \, \log\brc{z}^{p_1} ,\quad 
\sum_{m=-k_2}^{l_2} \xi^L_m\, z^m  \, \text{Li}_{s_1,\dots,s_k}\brc{1-z},
\end{equation}
where $k_{1,2}$, $l_{1,2}$, $p_{1}$ are some non-negative integers, $0\leq p_1\leq K$, $s_1+\dots +s_k \leq K$, and $\zeta^L_m,\xi^L_m$ are $z$-independent quantities.

\subsubsection{Right Region}
\label{sec:Right_dS}

\begin{figure}
\begin{center}
\includegraphics[width=1\linewidth]{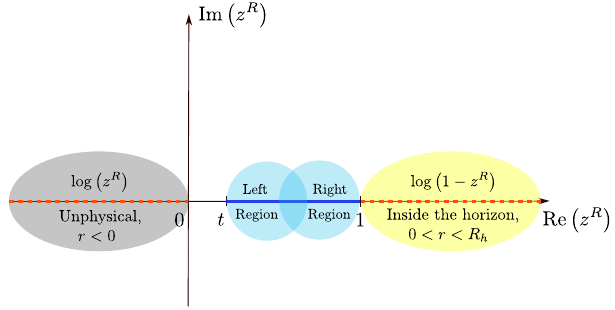}
\caption{Branch cuts (dashed red  lines) on the complex $z^R$ plane for de Sitter black holes.}
\label{fig:dS_zR}
\end{center}
\end{figure}

The right region is near the event horizon $r=R_h$, or $z=t$. We introduce the local variable $z^R=t/z$ so that the horizon is at $z^R=1$. In the $z^R$ variable, the equation \eqref{heuncanonical} reads
\begin{equation}
\frac{\rmd^2 \psi(z^R)}{(\rmd z^R)^2}+\left( \frac{2-\gamma}{z^R}+\frac{\delta t}{z^R(z^R-t)}+\frac{\epsilon}{z^R(z^R-1)} \right)\frac{\rmd\psi(z^R)}{\rmd z^R}+\frac{\alpha \beta  t - q z^R}{(z^R)^2(z^R-1)(z^R-t)}\psi(z^R) = 0.
\end{equation}
In the remaining part of this subsection, we will mostly omit the R index on the $z$ variable (except for the cases where it could be confusing). We take as leading order solutions in $R_h$ (and so in $t$) of this equation 
\begin{equation}
\label{eq:dS_HeunR}
\begin{aligned}
f_0^R(z)=&z^{-\ell-s}{}_2F_1(-\ell-s,-\ell+s;-2\ell;z)=\\
=&z^{-\ell-s}\sum_{k=0}^{\ell-s} \frac{\brc{s-\ell}_k \brc{-\ell-s}_k}{\brc{-2\ell}_k} \,\frac{z^k}{k!},\\
g_{0}^R(z)=&
z^{-s}\left\{\sum_{m=-s}^{\ell-1}a_{s\ell m}z^{-m}+\log(1-z)\sum_{m=s}^{\ell}b_{s\ell m}z^{-m}\right\},
\end{aligned}
\end{equation}
with
\begin{equation}
\begin{aligned}
b_{s\ell m}=&\frac{(-1)^{\ell+m+1}}{(m+s)!(m-s)!}\frac{(2\ell+1)!}{(\ell+s)!(\ell-s)!}\frac{(\ell+m)!}{(\ell-m)!},\\
a_{s\ell m}=&-b_{s\ell m}(H_{\ell+s}+H_{\ell-s}-H_{m+s}-H_{m-s}).
\end{aligned}
\end{equation}
The Wronskian between $f^R_0$ and $g^R_0$ is
\begin{equation}
W^R_0(z)=\frac{2\ell+1}{z^{2s}(z-1)}.
\end{equation}
Here we would like to comment on the choice of the logarithm function $\log\brc{1-z^R}$ in the solution $g_0^R$. The other possible choice of the logarithm could be, for example, $\log\brc{z^R-1}$. This choice dictates what functions will appear in higher orders in $t$ and affects the $R_h$ expansion of the frequency $\omega$. Throughout the paper, we work with the principal value of the complex logarithm, and thus the change in the argument affects the position of the branch cut on the complex $z$ plane. Our wave function $\psi\brc{z}$ can be viewed as an analytic continuation of the physical solution on half of the real line $r\geq 0$. In the de Sitter case, the coordinate transformation $z\brc{r}$ is \eqref{eq:dS_z_trans} with real parameters $R_{\pm}$. Since we want the solution to be continuous across the real slice $R_h <r< R_{+}$, the branch cut should not cross the interval $t<z^R< 1$, where $t$ is small and positive. This leaves us with $\log\brc{1-z^R}$, and the branch cut runs from $z^R=1$ to $z^R=+\infty$. The other logarithm function that appears in higher orders in $t$ is $\log\brc{z^R}$, and the corresponding branch cut runs from $z^R=0$ to $z^R=-\infty$ also avoiding the interval $t<z^R< 1$ (see Figure \ref{fig:dS_zR}).

The boundary condition near the horizon requires us to keep the solution corresponding to the incoming wave and discard the one corresponding to the outgoing wave. According to \eqref{Heunint}, the two solutions behave like
\begin{equation}
\label{rwv_out_inc}
\begin{aligned}
\text{outgoing wave}:\quad &\psi_{-}^{(t)}\brc{z^R} \sim 1, &  z^R\sim 1 ,\\
\text{incoming wave}:\quad &\psi_{+}^{(t)}\brc{z^R} \sim \brc{1-z^R}^{1-\epsilon}, &  z^R\sim 1 .
\end{aligned}
\end{equation}
Since $1-\epsilon=\mathcal{O}(R_h)$, both waves in the rhs of \eqref{rwv_out_inc} have Taylor expansions in $R_h$ that start with $1$. One must also consider the higher orders in $R_h$ to distinguish the two expansions.
The incoming wave solution has a particular dependence on the logarithm function $\log\brc{1-z}$ in each order in $R_h$ (or $t$):
\begin{equation}
\label{rwf_inc_exp}
\psi_{+}^{(t)}\brc{z} \sim  1 -2 i \omega_0 \log\brc{1-z} R_h +\mathcal{O}\brc{R_h^2},
\quad z\sim 1.
\end{equation}
In the leading order in $R_h$ both $\psi_{-}^{(t)}$ and $\psi_{+}^{(t)}$  are given by the same function $f_{0}^{R}\brc{z}$. Since the other function $g_0^{R}\brc{z}$ contains the logarithm, it enters $\psi_{+}^{(t)}$ in the higher orders in $R_h$. The constants $b_K$ from \eqref{eq:psi_exp} are fixed by matching with the logarithmic behavior of the incoming wave solution \eqref{rwf_inc_exp} in each order in $R_h$. 

The integrals in \eqref{eq:psi_exp} are again described in terms of the multiple polylogarithms in a single variable (see Appendix \ref{app:recurrence_rel}). We construct the linear basis of functions for each integral in the way it was done in the previous section for the left region. The only difference is that we need to add powers of the second logarithm function $\log\brc{1-z}$ to formulas \eqref{eq:dS_vLog_Left}, \eqref{eq:dS_vLog_Left2} and \eqref{eq:dS_vLog2_Left}. 
In particular, the second integrand from \eqref{eq:psi_exp} at order $t^{K}$ of the form 
\eqlb{eq:Int2R_dS}{g^R_0(z)\, \frac{ \eta^R_{K}(z)}{W^R_0(z)}}
will have a maximum weight $K$ because the logarithm function $\log\brc{1-z}$ is present in the leading order solution $g^R_0\brc{z}$. The resulting integral, however, will be of the same weight $K$ due to the pole structure in \eqref{eq:Int2R_dS}. Eventually, the corrections $f^R_{K}\brc{z}$ to the wave function in the right region are linear combinations of the following functions of maximum weight $K$:
\begin{equation}
\label{basisrightdS}
\begin{aligned}
&\sum_{m=-k_1}^{l_1} \zeta^R_m\, z^m \, \log\brc{1-z}^{p_1} \log\brc{z}^{p_2},\\
&\sum_{m=-k_2}^{l_2} \xi^R_m\, z^m  \, \log\brc{1-z}^{p_3} \text{Li}_{s_1,\dots,s_k}\brc{1-z},
\end{aligned}
\end{equation}
where $k_{1,2}$, $l_{1,2}$, $p_{1,2,3}$ are some non-negative integers, and $0\leq p_1+p_2\leq K$, $p_3+s_1+\dots +s_k \leq K$.

\subsubsection{Results for QNM frequencies}
The final step in the procedure described in Section \ref{sec:pert_method} is to glue the local solutions by requiring that the wave function and its first derivative are continuous at the intersection of the two regions. There is a certain freedom in choosing the intersection point as long as it lies in the region of convergence of both local solutions. 
We choose the point $z=t^{1/2}$, which is the same as $z^R=t^{1/2}$. 
Note that the expansions of $\psi^{L,R}\brc{z^{L,R}}$ are given as series expansions around $z^{L,R}=1$ up to orders $t^{m_{L,R}}$:
\begin{equation}
\label{eq:dS_cont_exp}
\begin{aligned}
\psi^{L}\brc{z}= & f^L_0\brc{z}+\sum_{K=1}^{m_L} f^L_K(z) t^K + O\brc{t^{m_L+1}}, \\
\psi^{R}\brc{z^R}= & f^R_0\brc{z^R}+\sum_{K=1}^{m_R} f^R_K(z^R) t^K  + O\brc{t^{m_R+1}}.
\end{aligned}
\end{equation}
What happens when we take $z^{L,R}\sim t^{1/2}$ and expand for a small $t$? Some terms $f^L_K(z) t^K$ in $\psi^{L}\brc{z}$ will contribute to orders lower than $t^K$. This could lead to a reshuffling, where, for example, $f^L_1(z) t$ becomes the leading order contribution at $z\sim t^{1/2}$. This happens when $\ell\geq 1$, as seen from \eqref{eq:fL_dS}:
\eqlb{eq:disord_dS}{f^L_0\brc{t^{1/2}} \sim t^{\brc{s+\ell}/2},\quad 
f^L_1\brc{t^{1/2}} t \sim t^{\brc{s-\ell+1}/2}.}
However, since we are within the radius of convergence of $\psi^{L}\brc{z}$, this reshuffling involves only a finite number of terms. For all values of quantum numbers we have considered, the reshuffling is superficial and goes away after the frequency is set to one of the quasinormal modes.

The continuity condition 
\eqlb{eq:cont_short}{\partial_z \left.\log\brc{\frac{\psi^L(z)}{\psi^R(t/z)}}\right|_{z=t^{1/2}}=0}
can be equivalently stated as 
\eqlb{eq:bclong_dS}{\psi^{L}\brc{t^{1/2}}=C\brc{t} \psi^{R}\brc{t^{1/2}},\quad 
\partial_z\psi^{L}\brc{z}\bigg|_{z=t^{1/2}}=C\brc{t} \partial_z\psi^{R}\brc{t/z}\bigg|_{z=t^{1/2}},}
where $C(t)$ is a normalization factor. The advantage of \eqref{eq:bclong_dS} is that we can use one of the equations to understand which orders in $t$ we can trust when expanding \eqref{eq:dS_cont_exp} at $z^{L,R}=t^{1/2}$,  then use the other one to fix the frequencies.   
 
Using {\tt Mathematica}, we compute the local solutions up to orders $m_L=10$ and $m_R=7$. This allows us to determine the $R_h$ expansion of the frequency up to order $R_h^{9}$ or less depending on the value of $\ell$. In all computed orders, we find the real part of the quasinormal modes is zero, which agrees with the earlier observations made by numerical computations \cite{KZ'22,Jansen:2017oag, Cardoso:2017soq}.
The results for the imaginary part of the quasinormal mode frequencies $\omega_{n,\ell,s}$, starting from $n=0$, are
\begin{equation}\small
\begin{aligned}
\text{Im}\brc{ \omega_{0,0,0}}=&\ -1-\frac{5}{8}R_h^2-3R_h^3-\left[ \frac{1287}{128}+2 \log \left(2 R_h\right)\right]R_h^4+\left[\pi ^2-\frac{119}{4}-15 \log(2 R_h)\right]R_h^5+\\
&+\left[\frac{25}{3} \pi ^2 -\frac{102\,621}{1024} -\frac{271}{4}\log\brc{2 R_h} -5 \log ^2(2 R_h)+6\,\zeta (3)\right]R_h^6+\mathcal{O}\brc{R_h^7},\\
\text{Im}\brc{\omega_{0,1,1}}=&\ -2-\frac{7}{12}\,R_h^2+\frac{7123}{1728}\, R_h^4+8\,R_h^5+\left[\frac{2\,757\,809}{124\,416}+\frac{32}{3} \log\brc{2 R_h}\right]R_h^6-\\
&-\frac{4}{27}\left[13+72\, \pi ^2-468 \log (2R_h)\right]R_h^7+\mathcal{O}\brc{R_h^8},\\
\text{Im}\brc{ \omega_{0,2,2}}=&\ -3-\frac{27}{40}R_h^2+\frac{51\,423}{16\,000}R_h^4-\frac{72\,333\,747}{3\,200\,000}R_h^6-\frac{72}{5}R_h^7+\left[\frac{60\,278\,884\,503}{512\,000\,000}-\right.\\
&-\left.\frac{144}{5} \log \left(2 R_h\right)\right]R_h^8
+\frac{9}{50}\left[625+240\, \pi^2-1008 \log \left(2 R_h\right)\right]R_h^9+\mathcal{O}\brc{R_h^{10}}.
\end{aligned}
\end{equation}
Let us also report the results for $n=1$:
\begin{equation}\small
\begin{aligned}
\text{Im}\brc{ \omega_{1,0,0}}=& -2-\frac{17}{4}R_h^2-24\,R_h^3 - \bsq{\frac{9791}{64}+32 \log\brc{2 R_h}}R_h^4+ \left[32\,\pi^2 -654 -384 \log\brc{2 R_h}\right] R_h^5+\\ 
&+\left[\frac{1408}{3} \pi^2 - \frac{1\,770\,481}{512} - 3276 \log\left(2 R_h\right) - 256  \log^2\left(2 R_h\right) +384\, \zeta\brc{3}\right]R_h^6 + \mathcal{O}\brc{R_h^7},\\
\text{Im}\brc{ \omega_{1,1,1}}=& -3-\frac{21}{8}R_h^2+\frac{4137}{128}R_h^4+72\, R_h^5+\left[ \frac{249\,879}{1024} + 144 \log \left(2 R_h\right)\right]R_h^6+\\
&+\left[303- 216\,\pi^2 + 1188 \log\left(2 R_h\right)\right] R_h^7 + \mathcal{O}\brc{R_h^8},\\
\text{Im}\brc{ \omega_{1,2,2}}=& -4-\frac{71}{30}R_h^2+\frac{1\,910\,399}{108\,000}R_h^4-\frac{44\,927\,058\,551}{194\,400\,000}R_h^6-\frac{768}{5}R_h^7+\left[\frac{685\,871\,572\,615\,439}{279\,936\,000\,000} - \right. \\
&\left.- \frac{2048}{5}\log \left(2 R_h \right)\right]R_h^8 + \frac{64}{225}\left[2880 \,\pi ^2-53 -10\,656 \log\left(2 R_h\right)\right]R_h^9+\mathcal{O}\brc{R_h^{10}}.
\end{aligned}
\end{equation}
Some of the results presented above were shortened for the reader's convenience. The full expressions and more expansions of frequencies for other choices of $\ell$ and $s$ can be found in the attached {\tt Mathematica} files.  The irrational numbers entering these QNM frequencies are $\log(2)$ and multiple zeta values.

\section{Perturbations of anti-de Sitter black holes in four dimensions}\label{section4}

The metric describing the $\rm AdS_4$ Schwarzschild  black hole  is given by \eqref{metric}, with $\Lambda<0$. 
We denote the roots of $rf(r)=0$ by 
\be R_h, \quad R_{\pm},  \ee
where, for $\Lambda<0$ , $R_\pm$ are complex conjugate and given by
\begin{equation}
R_{\pm}=\frac{-R_h\pm i\sqrt{3R_h^2-\frac{12}{\Lambda}}}{2},
\end{equation}
in terms of the BH horizon $R_h\in \mathbb{R}_{>0}$.  We will fix $\Lambda=-3$ and study the same perturbations of the Schwarzschild de Sitter case, described by equation \eqref{eq:RW_dS}.
According to $\mathrm{AdS}_4/\mathrm{CFT}_3$ holography, the conformally coupled scalar field is dual to scalar operators of conformal dimension $\Delta=1$ or $\Delta=2$, from the relation $\mu^2=\Delta(\Delta-3)$.
The main difference with the $\mathrm{SdS}_4$ case lies in the boundary conditions we impose on the solution. Indeed, we will still require the presence of only ingoing modes near the horizon, but we will impose the vanishing Dirichlet boundary condition at the AdS boundary.\footnote{In the context of AdS/CFT, these are not always the more physically relevant boundary conditions. Alternatively, one often considers Robin boundary conditions, which we discuss in Sec.~\ref{section5}.}

With the following change of variables 
\begin{equation}
\label{eq:z_trans_AdS}
z(r)=\frac{r(R_{-}-R_{+})}{R_{-}(r-R_{+})},
\end{equation}
and redefinition of the wave function
\begin{equation}
\psi(z)=z^{-\gamma/2}(z-1)^{-\delta/2}(z-t)^{-\epsilon/2}\sqrt{f(r)}\, \frac{R_{+}(R_{-}-R_{+})}{R_{-}(r-R_{+})} \, \Phi(r),
\end{equation}
with 
\begin{equation}\label{dictionaryAdS1}
\begin{aligned}
&t=\frac{R_h(R_{+}-R_{-})}{R_{-}(R_{+}-R_h)},\\
&\gamma =1-2s,\\
&\delta = 1- \frac{2i\,\omega\,R_{-}}{(R_{-}-R_h)(R_{-}-R_{+})},\\
&\epsilon=1-\frac{2i \omega R_h}{1+3R_h^2},
\end{aligned}
\end{equation}
the singularity at infinity is removed, and the equation \eqref{eq:RW_dS} becomes a Heun equation \eqref{heuncanonical} with 
\begin{equation}\label{dictionaryAdS2}
\begin{aligned}
&\alpha=1-s+\frac{2i\,\omega\,R_{+}}{(R_{+}-R_h)(R_{+}-R_{-})},\\
&\beta=1-s,\\
&q=\frac{\ell(\ell+1)}{R_{-}(R_{h}-R_+)} + \frac{(1-s)^2 R_h}{R_h-R_+} - \frac{s(1-s) R_{+}^2}{R_{-}(R_h-R_{+})}.
\end{aligned}
\end{equation} 
In these coordinates, the horizon is at $z=t$ while the boundary is at
\be z_{\infty}=1-\frac{R_+}{R_{-}}.\ee 
We will also consider the small black hole limit, $R_h\ll 1$. 

The Dirichlet boundary conditions in terms of the $\psi$ function are given by 
\begin{equation}
\label{eq:BC_AdS}
\begin{aligned}
&\psi(z)\sim 1\ \ \text{for}\ z \sim t,\\
&\psi(z_{\infty})=0.
\end{aligned}
\end{equation} 
Notice that the  AdS boundary ($z=z_{\infty}$) is not a singular point of the perturbation equation. This makes the approach based on the Seiberg-Witten theory less effective. One can write the quantization condition using the connection formulae between Heun functions, but in this case, an expansion of the Heun functions in $R_h$ is needed. We will report some results in this direction in appendix \ref{appendixconnection}.

\subsection{QNMs in pure \texorpdfstring{$\mathrm{AdS}_4$}{} }

The pure $\mathrm{AdS}_4$ case can be recovered in the limit $t\to 0$ or, equivalently, $R_h\to 0$. 
In this limit, the $z$ variable is given by
\begin{equation}
z=\frac{2r}{r-i},
\end{equation}
and the AdS boundary is at $z=2$.
The leading order solutions in $t$ of the Heun equation \eqref{heuncanonical} are given by
\begin{equation}
\label{pure_AdS_sol}
z^{s-\ell-1}{}_2F_1(-\ell,-\ell+\omega_0;-2\ell;z),\quad z^{\ell+s}{}_2F_1(\ell+1,\ell+1+\omega_0;2\ell+2;z),
\end{equation}
where $\omega_0$ is the leading order term in the $R_h$ expansion of the frequency \eqref{dS_omega_exp}.
As in the de Sitter case, these hypergeometric functions reduce to \eqref{pure_dS_sol_sim}, where we replace $-i\omega_0$ by $\omega_0$.

The first boundary condition from \eqref{eq:BC_AdS} tells us that the wave function $\psi\brc{z}$ is regular at $z=0$. This singles out the second solution from \eqref{pure_AdS_sol}.
Then, the second boundary condition at $z=2$ requires the following expression to vanish:
\begin{equation}\label{quantcondads}
{}_2F_1(\ell+1,\ell+1+\omega_0;2\ell+2;2)=4^{-\ell-1}\frac{(2\ell+1)!}{\ell!} \frac{\Gamma\left(\frac{-\omega_0-\ell}{2}\right)}{\Gamma\left(\frac{-\omega_0+\ell+2}{2}\right)}\left[1+\brc{-1}^{\ell-\omega_0+1}\right],
\end{equation}
which gives the quantization condition for the QNM frequencies of the pure $\mathrm{AdS}_4$
\begin{equation}
\label{eq:pure_AdS_omega}
\omega_0=\ell+2n+2, \quad n\in \mathbb{Z}_{\ge 0}\ \ \text{or}\ \ \omega_0=-\ell-2n-2, \quad n\in \mathbb{Z}_{\ge 0}.
\end{equation}
Here we have two branches of frequencies, positive and negative, and one is related to another by the complex conjugation of the radial part of the perturbation $\Phi\brc{r}$.

In the following subsections, we will perturb around the pure AdS case to obtain the corrections for the Schwarzschild anti-de Sitter small black holes. Following the same logic as in the de Sitter case, we will divide the space into two regions: left ($L$) and right ($R$). The left region describes the physical space near the AdS boundary with $r\rightarrow \infty$, and the right one is the space near the horizon $r=R_h$. After having determined the expansion of the solution $\psi(z)$ in each region up to certain orders in the expansion parameter $t$, we require that the function $\psi\brc{z}$ and its first derivative are continuous in a point in the intersection of two regions, which we can fix at $z=t^{1/2}$ (other values of $z$ are possible as long as they lie inside the convergence radius of the two solutions).

\subsection{Left Region}
\label{sec:AdS_Left}
The local coordinate in the left region is $z$, and the AdS boundary is at $z_\infty$, which has the following expansion in $R_h$:
\begin{equation}
z_{\infty}=\frac{3R_h^2+4+i\,R_h \sqrt{3R_h^2+4}}{2R_h^2+2}=2+iR_h-\frac{R_h^2}{2}+\mathcal{O}\left(R_h^3\right).
\end{equation}
The wave function in the left region $\psi^L\brc{z}$ satisfies the same Heun equation \eqref{heuncanonical}. The form of the leading order solutions depends on which branch of frequencies we choose in \eqref{eq:pure_AdS_omega}.
For the negative branch $\omega_0=-\ell-2n-2$, we have
\begin{equation}
\begin{aligned}
f^L_0\brc{z}=&z^{\ell+s}\sum_{m=0}^{2n+1} \brc{-1}^m \brc{\at{2\,n+1}{m}} \frac{\brc{\ell+1}_m}{\brc{2\,\ell+2}_m}\, z^m,\\
g^L_{0}\brc{z}=&z^{s-\ell-1} \sum_{m=0}^{\ell} \brc{-1}^m \brc{\at{\ell}{m}} \frac{\brc{-2\,\ell-2\,n-2}_m}{\brc{-2\,\ell}_m}\, z^m,
\end{aligned}
\end{equation}
and for the positive branch 
$\omega_0=\ell+2n+2$:
\begin{equation}
\begin{aligned}
f^L_0\brc{z}=&\frac{z^{\ell+s}}{\brc{1-z}^{2n+\ell+2}} \sum_{m=0}^{2n+1} \brc{-1}^m \brc{\at{2\,n+1}{m}} \frac{\brc{\ell+1}_m}{\brc{2\,\ell+2}_m}\, z^m,\\
g^L_{0}\brc{z}=&\frac{z^{s-\ell-1}}{\brc{1-z}^{2n+\ell+2}} \sum_{m=0}^{\ell} \brc{-1}^m \brc{\at{\ell}{m}} \frac{\brc{-2\,\ell-2\,n-2}_m}{\brc{-2\,\ell}_m}\, z^m.
\end{aligned}
\end{equation}
For both branches, the Wronskian can be written in terms of $\omega_0$ as
\begin{equation}
W_0^L(z)=-(2\ell+1)z^{2s-2}(1-z)^{-\omega_0-1}.
\end{equation}
We will apply the perturbative method described in Section \ref{sec:pert_method} to both positive and negative values of $\omega_0$, but the final result is straightforward. The only difference between the two branches is the sign of the real part of the frequency expansion \eqref{dS_omega_exp}, which again corresponds to complex conjugation of $\Phi\brc{r}$.

The boundary condition in the left region is simply $\psi^L\brc{z_{\infty}}=0$. Since $f^L_0\brc{2}=0$ and $g^L_{0}\brc{2}\neq 0$, we get the following perturbative expansion for the wave function in the left region:
\be\label{eq:psiL_AdS}
\psi^L\brc{z}=f^L_0\brc{z}+\sum_{K\geq 1} f_K^L(z) t^K ,\ee
where $f_K^L(z)$ are given by \eqref{eq:psi_exp}. The constants $b_K$ in \eqref{eq:psi_exp} are fixed by expanding $\psi^L\brc{z_{\infty}}$ in powers of $t$ and requiring the coefficients in this expansion to vanish.

\begin{figure}
\begin{center}
\includegraphics[width=1\linewidth]{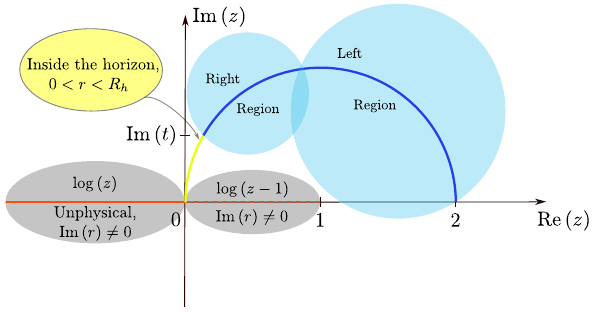}
\caption{Branch cuts (red  lines) on the complex $z$ plane for anti-de Sitter black holes.}
\label{fig:AdS_z}
\end{center}
\end{figure}

As we explain in Appendix \ref{app:recurrence_rel}, the integrals in \eqref{eq:psi_exp} are described in terms of the multiple polylogarithms in a single variable \eqref{eq:PolyLog_def_dS}. 
Since the weights of the multiple polylogarithms appearing at order $t^K$  are less or equal to $K$, we can construct a linear basis of functions in which the integrals in \eqref{eq:psi_exp} can be expanded. We take the same steps \eqref{eq:dS_vLog_Left}--\eqref{eq:dS_vLog2_Left} as we did in the $\mathrm{SdS}_4$ case to do this. The only difference is that we add the second logarithm function $\log\brc{z-1}$ to \eqref{eq:dS_vLog_Left}. To be more precise, the integrands in \eqref{eq:psi_exp} at order $t^{K+1}$ are given by the linear combination of the following functions:
\begin{equation}
\begin{aligned}
&\frac{\sum_{m=0}^{r_1} \alpha_m\, z^m}{z^{i_1} \brc{z-1}^{j_1}} \, \log\brc{z-1}^{p_1} \log\brc{z}^{p_2},\\
&\frac{\sum_{m=0}^{r_2} \beta_m\, z^m}{z^{i_2} \brc{z-1}^{j_2}} \, \log\brc{z-1}^{p_3} \text{Li}_{s_1,\dots,s_k}\brc{1-z},
\end{aligned}
\end{equation}
where $r_{1,2}$, $i_{1,2}$, $j_{1,2}$, $p_{1,2,3}$ are some non-negative integers and $p_1+p_2\leq K$, $p_3+s_1+\dots +s_k \leq K$.
The reasoning behind our choice of the branches of the logarithm functions $\log\brc{z}$ and $\log\brc{z-1}$ is the same as in Section \ref{sec:Right_dS}. We want the wave function $\psi\brc{z}$ to be continuous across the real slice $R_h<r<+\infty$. In the $\mathrm{SAdS}_4$ case, the coordinate transformation $z\brc{r}$ is given by \eqref{eq:z_trans_AdS} with complex parameters $R_{\pm}$. Taking into account that $r$ and $R_h$ are real, we have from \eqref{eq:z_trans_AdS}:
\eq{\brc{\text{Re}\brc{z}-1}^2 + \text{Im}\brc{z}^2=1.}
Thus, the real slice is approximately half the circle with the center in $z=1$ on the complex $z$ plane (see Figure \ref{fig:AdS_z}). It starts at $z=t$ and ends at $z=z_{\infty}$. Simple analysis shows that $\text{Im}\brc{t}>0$ and $\text{Im}\brc{z_{\infty}}>0$ when $R_h>0$. This justifies our choice of logarithm functions since both branch cuts do not cross the real slice. On the other hand, if one picks $\log\brc{1-z}$ instead of $\log\brc{z-1}$, the corresponding branch cut would touch the real slice at the point $z=2$ when evaluating $\psi^L\brc{z_{\infty}}$. This, in turn, would lead to incorrect results for QNM frequencies.

\subsection{Right region}
In the right region, we introduce local coordinate
\eq{z^R=\frac{t}{z}.}
The horizon is now situated at $z^R = 1$. The wave function in the right region $\psi^R\brc{z^R}$ satisfies the following equation in terms of $z^R$:
\eqlb{eq:Heun_R_AdS}{ \frac{\rmd^2 \psi^R}{\brc{\rmd z^R}^2}+\left( \frac{2-\gamma}{z^R}+\frac{\delta\, t}{z^R\brc{z^R-t}}+
    \frac{\epsilon}{z^R \brc{z^R-1}} \right)\frac{\rmd\,\psi^R}{\rmd z^R}+\frac{\alpha \beta \, t - q\,z^R}{\brc{z^R}^2\brc{z^R-1}\brc{z^R-t}}\, \psi^R = 0.}
Suppressing the R index on $z^R$, the two leading order solutions are given by
\begin{equation}
\begin{aligned}
\label{eq:fR_AdS}
f^R_{0}\brc{z}=& z^{-\ell-s} \sum_{m=0}^{\ell+s} \brc{-1}^m \binom{\ell+s}{m} \frac{\brc{s-\ell}_m}{\brc{-2\,\ell}_m}\, z^m,\\
g^R_{0}\brc{z}=& z^{-s}\Biggl\{\sum_{m=-s}^{\ell-1} a_{s\ell m}\, z^{-m} + \log\brc{1-z} \sum_{m=s}^{\ell} b_{s\ell m}\, z^{-m}\Biggr\},
\end{aligned}
\end{equation}
where the constants $a_{s\ell m}$, $b_{s\ell m}$ can be determined for any $\ell\geq s\geq 0$ as
\eqn{a_{s\ell m}=-b_{s\ell m} \brc{H_{\ell+s}+H_{\ell-s}-H_{m+s} -H_{m-s}},}
\eqn{b_{s\ell m}= \frac{\brc{-1}^{\ell+m+1}}{\brc{m+s}!\brc{m-s}!}\, \frac{\brc{2\,\ell+1}!}{\brc{\ell+s}!\brc{\ell-s}!} \frac{\brc{\ell+m}!}{\brc{\ell-m}!}.}
The expressions in \eqref{eq:fR_AdS} are independent of which branch of frequencies we choose in \eqref{eq:pure_AdS_omega} because the leading order of \eqref{eq:Heun_R_AdS} does not contain $\omega_0$.
The Wronskian of $f^R_{0}$ and $g^R_{0}$ is given by
\eq{W_0^{R}(z)=\frac{2\,\ell+1}{z^{2 s}(z-1)}.}
The boundary condition in the right region tells us that  $\psi^R$ is regular at $z^R=1$. Thus, we can write the following perturbative expansion:
\be\psi^R\brc{z}=f^R_0\brc{z}+\sum_{K\geq 1} f_K^R(z) t^K ,\ee
where $f_K^R(z)$ are computed using \eqref{eq:psi_exp}. Unlike in the left region, the choice of the logarithm function in $g^R_{0}\brc{z}$ is unimportant. This is due to the boundary condition that requires canceling contributions of $\log\brc{1-z}$ in each order $t^K$. The resulting corrections $f^R_{K}\brc{z}$ are linear combinations of the following functions of maximum weight $K$:
\begin{equation}
\sum_{m=-k_1}^{l_1} \zeta^R_m\, z^m \,  \log\brc{z}^{p_1},\quad
\sum_{m=-k_2}^{l_2} \xi^R_m\, z^m \, \text{Li}_{s_1,\dots,s_k}\brc{1-z},
\end{equation}
where $k_{1,2}$, $l_{1,2}$, $p_{1}$ are some non-negative integers, and $0\leq p_1\leq K$, $s_1+\dots +s_k \leq K$.

\subsection{Results for QNM frequencies}\label{resultsAdSsmall}
To determine the QNM frequencies, we use the continuity condition in the form  \eqref{eq:bclong_dS}:
\eqlb{eq:bclong_AdS}{\psi^{L}\brc{t^{1/2}}=C\brc{t} \psi^{R}\brc{t^{1/2}},\quad 
\partial_z\psi^{L}\brc{z}\bigg|_{z=t^{1/2}}=C\brc{t} \partial_z\psi^{R}\brc{t/z}\bigg|_{z=t^{1/2}},}
where $\psi^{L,R}\brc{z^{L,R}}$ are computed up to orders $m_{L,R}$ in $t$ around $z^{L,R}=1$:
\begin{equation}
\begin{aligned}
\psi^{L}\brc{z}= & f^L_0\brc{z}+\sum_{K=1}^{m_L} f^L_K(z) t^K + O\brc{t^{m_L+1}}, \\
\psi^{R}\brc{z^R}= & f^R_0\brc{z^R}+\sum_{K=1}^{m_R} f^R_K(z^R) t^K  + O\brc{t^{m_R+1}}.
\end{aligned}
\end{equation}
Similarly to the $\mathrm{SdS}_4$ case, the reshuffling of terms \eqref{eq:disord_dS} occurs in $\psi^{L}\brc{z}$ when we take $z\sim t^{1/2}$. For all values of quantum numbers we have considered, this reshuffling is superficial and goes away after the frequency is set to one of the quasinormal modes.

Using {\tt Mathematica}, we compute the local solutions up to orders $m_L=7$ and $m_R=8$ (sometimes even up to $m_L=9$ and $m_R=10$). This allows us to determine the $R_h$ expansion of the frequency up to order $R_h^{7}$ or less depending on the value of $\ell$. In all computed cases, the imaginary part does not appear before order $2\ell+2$ in $R_h$:
\begin{equation}\label{4.29}
    \text{Im}\brc{ \omega_{n,\ell,s}} \sim R_h^{2\ell+2}.
\end{equation}
As mentioned, the results computed for negative and positive branches of $\omega_0$ only differ by the sign in the real part of the frequency expansion.  
Below are the results for the real and imaginary parts of the quasinormal mode frequencies $\omega_{n,\ell,s}$ corresponding to the positive branch, starting from $n=0$:
\begin{equation*}\small
\begin{aligned}
\text{Re}\brc{ \omega_{0,0,0}}=& 2 -\frac{4}{\pi}\, R_h-\brc{\frac{1}{4}+\frac{24}{\pi ^2}} R_h^2 - \left(\frac{4\,\pi}{3}-\frac{94}{3\,\pi} -\frac{16}{\pi} \,\log\brc{4\,R_h}+\frac{208}{\pi^3} -\frac{112}{\pi^3} \, \zeta\brc{3}\right) R_h^3 +\mathcal{O}\brc{R_h^4},\\
\text{Im}\brc{ \omega_{0,0,0}}=& -\frac{8}{\pi}\, R_h^2-\brc{8+\frac{16}{\pi ^2}} R_h^3 - \left(\frac{40\,\pi}{3}-\frac{65}{\pi} -\frac{128}{\pi} \,\log\brc{2\,R_h}+\frac{192}{\pi^3}-\frac{448}{\pi^3} \, \zeta\brc{3}\right) R_h^4 +\mathcal{O}\brc{R_h^5},\\
\text{Re}\brc{ \omega_{0,1,1}}=& 3 -\frac{4}{\pi}\, R_h+\brc{\frac{27}{8}-\frac{140}{3\,\pi ^2}} R_h^2 - \left(3\,\pi-\frac{601}{12\,\pi} -\frac{18}{\pi} \,\log\brc{2}+\frac{2020}{3\,\pi^3} -\frac{168}{\pi^3} \, \zeta\brc{3}\right) R_h^3+\\ &+\mathcal{O}\brc{R_h^4},\\
\text{Im}\brc{\omega_{0,1,1}}=& -\frac{16}{\pi}\, R_h^4-\brc{24+\frac{96}{\pi ^2}} R_h^5 - \left(60\,\pi+\frac{579}{\pi} -\frac{264}{\pi} \, \log\brc{2\,R_h}+\frac{11\,536}{9\,\pi^3}-\frac{1344}{\pi^3} \, \zeta\brc{3}\right) R_h^6+\\ &+\mathcal{O}\brc{R_h^7},\\
\text{Re}\brc{ \omega_{0,2,2}}=& 4 -\frac{64}{15\,\pi}\, R_h+\brc{\frac{37}{6}-\frac{80\,896} {1125\,\pi ^2}} R_h^2 - \left(\frac{256\,\pi}{45} -\frac{1\,536\,256}{10\,125\,\pi} -\frac{512}{45\, \pi} \,\log\brc{2} + \frac{120\,946\,688} {84\,375\,\pi^3} -\right.\\
&\left. -\frac{57\,344} {225\,\pi^3} \, \zeta\brc{3}\right) R_h^3 +\mathcal{O}\brc{R_h^4},\\
\text{Im}\brc{\omega_{0,2,2}}=& -\frac{128}{5\,\pi}\, R_h^6-\brc{\frac{256}{5}+\frac{6144}{25\,\pi ^2}} R_h^7 +\mathcal{O}\brc{R_h^8}.
\end{aligned}
\end{equation*}
For $n=1$ we have:
\begin{equation*}\small
\begin{aligned}
\text{Re}\brc{ \omega_{1,0,0}}=& 4 -\frac{40}{3\,\pi}\, R_h+ \brc{\frac{25}{6}-\frac{5200}{27 \, \pi^2}} R_h^2 - \left(\frac{160\,\pi}{9}-\frac{45\,064 }{81\,\pi} -\frac{800} {9\,\pi} \, \log\brc{2}-\frac{128} {\pi} \, \log\brc{R_h}+ \right.\\
& \left. +\frac{1\,200\,800}{243\,\pi^3} - \frac{22\,400}{9\,\pi^3} \, \zeta\brc{3}\right) R_h^3 +\mathcal{O}\brc{R_h^4},\\
\text{Im}\brc{ \omega_{1,0,0}}=& -\frac{32}{\pi}\, R_h^2-\brc{64+\frac{2240}{9\,\pi ^2}} R_h^3 - \left(\frac{640\,\pi}{3}-\frac{4252}{3\,\pi} -\frac{1920}{\pi} \,\log\brc{2\,R_h} + \frac{101\,120}{9\,\pi^3} - \right.\\
&\left.-\frac{35\,840}{3\,\pi^3} \, \zeta\brc{3}\right) R_h^4 +\mathcal{O}\brc{R_h^5},\\
\text{Re}\brc{ \omega_{1,1,1}}=& 5 -\frac{172}{15\,\pi}\, R_h+\brc{\frac{2071}{120}-\frac{791\,372}{3375\,\pi ^2}} R_h^2 - \left(\frac{215\,\pi}{9}-\frac{27\,888\,631}{40\,500\,\pi} + \frac{40\,678}{225\,\pi} \,\log\brc{2} + \right.\\
&\left.+\frac{5\,269\,420\,724}{759\,375\,\pi^3} -\frac{103\,544}{45\,\pi^3} \, \zeta\brc{3}\right) R_h^3 +\mathcal{O}\brc{R_h^4},
\end{aligned}
\end{equation*}
\begin{equation*}\small
\begin{aligned}
\text{Im}\brc{\omega_{1,1,1}}=& -\frac{400}{3\,\pi}\, R_h^4-\brc{\frac{1000}{3}+\frac{39\,904}{27\,\pi^2}} R_h^5 - \left(\frac{12\,500\,\pi}{9}+\frac{328\,711}{27\,\pi} -\frac{49\,880}{9\,\pi} \, \log\brc{2\,R_h}+\right.\\
&\left. +\frac{14\,315\,216}{243\,\pi^3} -\frac{481\,600}{9\,\pi^3} \, \zeta\brc{3} \right) R_h^6 + \mathcal{O}\brc{R_h^7},\\
\text{Re}\brc{ \omega_{1,2,2}}=& 6 -\frac{384}{35\,\pi}\, R_h+\brc{\frac{675}{28}-\frac{12\,163\,072} {42\,875\,\pi ^2}} R_h^2 - \left(\frac{1152\,\pi}{35} -\frac{49\,433\,312}{42\,875\,\pi} +\frac{13\,824}{49\, \pi} \,\log\brc{2} + \right.\\
&\left. + \frac{1\,544\,254\,324\,736} {157\,565\,625\,\pi^3} -\frac{442\,368} {175\,\pi^3}\,\zeta\brc{3} \right) R_h^3 +\mathcal{O}\brc{R_h^4},\\
\text{Im}\brc{\omega_{1,2,2}}=& -\frac{1792}{5\,\pi}\, R_h^6-\brc{\frac{5376}{5}+\frac{385\,024}{75\,\pi ^2}} R_h^7 +\mathcal{O}\brc{R_h^8}.
\end{aligned}
\end{equation*}
Some of the results presented above were shortened for the reader's convenience. The full expressions and more expansions of frequencies for other choices of $n$, $\ell$, and $s$ can be found in the attached {\tt Mathematica} files. From these, one can see that the irrational numbers entering these QNM frequencies are $\log(2)$, $\pi$, and Euler sums.

Analytically computing $f_1^L\brc{z}$ from \eqref{eq:psiL_AdS}, we can also determine the subleading term in the QNM frequency expansion with $n=0$ and $\ell\ge 1$:
\begin{equation}\label{omega1}
\omega_{0,\ell,s}=\ell+2 - \frac{2^{2\ell+2}}{\pi} \,\frac{2\,\ell +s^2}{\ell\brc{\ell+1}}\,\frac{\brc{\brc{\ell+1}!}^2}{\brc{2\,\ell+2}!} \, R_h+\mathcal{O}\brc{R_h^2}.
\end{equation}
For small enough values of $R_h$, our results agree with the numerical ones obtained earlier in \cite{CKL2003}. Since the frequency expansions in higher orders in $R_h$ include multiple zeta values \eqref{MZV_def}, we use different identities of the form \eqref{eq:MZV_Id_1}--\eqref{eq:MZV_Id_2} to compute the corresponding numerical values. Tables \ref{tab:AdS_num1}--\ref{tab:AdS_num3} present the numerical results from the frequency expansions truncated at $R_h^7$ (in the scalar case with $n=l=0$, the expansion was computed up to order $R_h^6$ and truncated at the same order). In these tables, bold digits are the ones that are stable and agree with the numerical results obtained directly from the Heun function and the continuity condition \eqref{eq:cont_short}.
The digit is considered stable if it does not change when higher orders of $R_h$ are added to the expansion of the frequency. For example, below are the numerical results from electromagnetic frequency expansion with $n=0$, $\ell=1$ truncated at different powers of $R_h=1/20$: 
\begin{equation}
\begin{aligned}
R_h:\quad & \omega_{0,1,1} = 2.936338022763,\\
R_h^2:\quad & \omega_{0,1,1} = 2.932954718005,\\
R_h^3:\quad & \omega_{0,1,1} = 2.932365431000,\\
R_h^4:\quad & \omega_{0,1,1} = 2.932257833944 -0.000031830989\,i,\\
R_h^5:\quad & \omega_{0,1,1} = 2.932232789345 -0.000042370624\,i,\\
R_h^6:\quad & \omega_{0,1,1} = 2.932227305824 -0.000051050731\,i,\\
R_h^7:\quad & \omega_{0,1,1} = \mathbf{2.93222}\mathit{6938543} - \mathbf{0.00005}\mathit{3055262}\,i.\\
\end{aligned}
\end{equation}

\begin{table}
\centering
\begin{tabular}{| c | c  c| } 
  \hline			
  $R_h$ & $\text{Re}\brc{\omega_{0,0,0}}$ & $-\text{Im}\brc{\omega_{0,0,0}}$ \\
  \hline

  $1/16$ & $\mathbf{1.9095}\mathit{9612832}$ & $\mathbf{0.0136}\mathit{6850348}$ \\
  $1/18$ & $\mathbf{1.92054}\mathit{810947}$ & $\mathbf{0.0104}\mathit{3093333}$ \\
  $1/20$ & $\mathbf{1.92919}\mathit{836511}$ & $\mathbf{0.00820}\mathit{901816}$ \\
  $1/50$ & $\mathbf{1.9733862}\mathit{8700}$ & $\mathbf{0.0011184}\mathit{9414}$ \\
  $1/100$ & $\mathbf{1.986986250}\mathit{43}$ & $\mathbf{0.000265980}\mathit{52}$ \\
  \hline
\end{tabular}
\caption{Numerical results from conformally coupled scalar QNM frequency expansion with $n=0$, $\ell=0$.}
\label{tab:AdS_num1}
\end{table}

\begin{table}
\centering
\begin{tabular}{| c | c  c| } 
  \hline			
  $R_h$ & $\text{Re}\brc{\omega_{0,1,1}}$ & $-\text{Im}\brc{\omega_{0,1,1}}$ \\
  \hline
  
  $1/16$ & $\mathbf{2.9136}\mathit{28697405}$ & $\mathbf{0.0001}\mathit{51017506}$ \\
  $1/18$ & $\mathbf{2.92406}\mathit{3021823}$ & $\mathbf{0.00008}\mathit{6542953}$ \\
  $1/20$ & $\mathbf{2.93222}\mathit{6938543}$ & $\mathbf{0.00005}\mathit{3055262}$ \\
  $1/50$ & $\mathbf{2.973953080}\mathit{307}$ & $\mathbf{0.000000967}\mathit{146}$ \\
  $1/100$ & $\mathbf{2.9871273749}\mathit{10}$ & $\mathbf{0.0000000550}\mathit{27}$ \\
  \hline
\end{tabular}
\caption{Numerical results from electromagnetic QNM frequency expansion with $n=0$, $\ell=1$.}
\label{tab:AdS_num2}
\end{table}

\begin{table}
\centering
\begin{tabular}{| c | c  c| } 
  \hline			
  $R_h$ & $\text{Re}\brc{\omega_{0,2,2}}$ & $-\text{Im}\brc{\omega_{0,2,2}}$ \\
  \hline
  
  $1/15$ & $\mathbf{3.90327}\mathit{7526809}$ & $\mathbf{0.00000}\mathit{1160789}$ \\
  $1/18$ & $\mathbf{3.920419}\mathit{438200}$ & $\mathbf{0.000000}\mathit{363885}$ \\
  $1/20$ & $\mathbf{3.928811}\mathit{737917}$ & $\mathbf{0.000000}\mathit{186778}$ \\
  $1/50$ & $\mathbf{3.972361286}\mathit{120}$ & $\mathbf{0.000000000}\mathit{619}$ \\
  $1/100$ & $\mathbf{3.9863033746}\mathit{08}$ & $\mathbf{0.0000000000}\mathit{09}$ \\
  \hline
\end{tabular}
\caption{Numerical results from odd gravitational QNM frequency expansion with $n=0$, $\ell=2$.}
\label{tab:AdS_num3}
\end{table}

\section{Scalar Sector of Gravitational Perturbations - The low-lying modes}\label{section5}

Following \cite{kodama2003master}, one can consider a subdivision of gravitational perturbations in different sectors (scalar, vector, or tensor), whose distinction comes from the expansions in scalar, vector, or tensor spherical harmonics on the $S^2$ component of $\mathrm{AdS}_4$. In Sec.~\ref{section4} we considered the vector sector of gravitational perturbations ($s=2$).  We will now focus on the scalar sector and impose a new boundary condition at the $\mathrm{AdS}$ boundary, namely a Robin boundary condition \cite{michalogiorgakis2007low, friess:2006kw,siopsis:2007wn, Dias:2013sdc, Cardoso:2013pza,Grozdanov:2019uhi}, see also \cite{Kinoshita:2023iad} for very recent developments. This choice of boundary condition is motivated by the AdS/CFT correspondence, and it ensures that the perturbations do not deform the metric on the boundary of AdS.  

From the point of view of the dual CFT, these boundary conditions are related to  double-trace deformations, see for instance \cite{witten:2001ua, Gubser:2002vv, hartman:2006dy} and references therein.
In particular, we will analyze the so-called \emph{low-lying quasinormal frequencies}, which, according to AdS/CFT duality, are related to hydrodynamic modes of the $3d$ thermal CFT on the boundary \cite{policastro2002ads,Bhattacharyya:2007vjd,policastro2003ads,son2002minkowski,natsuume2008causal,Grozdanov:2019uhi,Grozdanov:2019kge,Grozdanov:2020koi,Grozdanov:2022npo,Grozdanov:2023txs}.
We will therefore expand our quasinormal frequencies for large values of $R_h$, $R_h\gg 1$, differently from the previous sections.
Defining 
\begin{equation}
\begin{aligned}
m&=\ell(\ell+1)-2\ \ \text{with}\ \ell\ge 2,
\end{aligned}
\end{equation}
the equation describing the scalar sector of gravitational perturbations in $\mathrm{AdS}_4$ can be written as (see \cite[eq.~(3.1)]{kodama2003master} for the definition of the master variable $\Phi$)
\begin{equation}
\label{eq:RW_AdS_Robin}
\left(\partial^2_r + \frac{f'(r)}{f(r)} \partial_r +\frac{\omega^2-V_S(r)}{f(r)^2}\right)\Phi(r)=0,
\end{equation}
where 
\begin{equation}
\begin{aligned}
f(r)&=1-\frac{2M}{r}+r^2,\\
V_S(r)&=\frac{f(r)}{\left(mr+6M\right)^2}\left[m^3+\left(2+\frac{6M}{r}\right)m^2+\frac{36M^2}{r^2}\left(m+2r^2+\frac{2M}{r}\right)\right].
\end{aligned}
\end{equation}
This equation has five regular singularities, located at $r=0,R_h,R_{\pm},R_5$, where
\begin{equation}
R_{\pm}=\frac{-R_h\pm i\sqrt{4+3R_h^2}}{2},\quad R_5=-\frac{3R_h\left(1+R_h^2\right)}{m}.
\end{equation}
The new singularity $R_5$, coming from the potential $V_S(r)$, is in the unphysical region $r<0$.
Similarly to the previous cases, we introduce the change of variables
\eq{z\brc{r}=\frac{R_h}{r}}
and the new wave function
\eq{\psi\brc{z}= r^{-1} \rme^{i\omega r_*} \Phi\brc{r}.}
The master equation \eqref{eq:RW_AdS_Robin}  then becomes 
\eqlb{eq:Robin_Eq_psi}{\psi''\brc{z} +\frac{\mathfrak{f}'\brc{z} -2 z^{-1} \mathfrak{f}\brc{z} + 2 i\,\omega R_h^{-1}}
{\mathfrak{f}\brc{z}} \, \psi'\brc{z} - \brc{\frac{\mathfrak{f}'\brc{z} -2 z^{-1} \mathfrak{f}\brc{z} + 2 i\,\omega R_h^{-1}}
{z\,\mathfrak{f}\brc{z}} + \frac{\mathfrak{V}\brc{z}} {\mathfrak{f}\brc{z}^2}} \psi\brc{z}=0,}
where 
\begin{equation}
\begin{aligned}
\mathfrak{f}\brc{z}&=\brc{1-z}\brc{1+z+z^2 + \frac{z^2}{R_h^2}},\\
\mathfrak{V}\brc{z}&=\frac{\mathfrak{f}\brc{z}}{\left(m R_h+ 6M z\right)^2}\left[m^3+\left(2+\frac{6M z}{R_h}\right)m^2+\frac{36M^2 z^2}{R_h^2}\left(m +\frac{2M z}{R_h}+ \frac{2 R_h^2}{z^2}\right)\right],
\end{aligned}
\end{equation}
and $M$ is related to $R_h$ via
\eq{2M=R_h\brc{1+R_h^2}.}
The boundary conditions in terms of the $\psi$ function are given by
\begin{equation}
\label{eq:Robin_BC}
\begin{aligned}
&\psi(z)\sim 1\ \ \text{for}\ z \sim 1,\\
&\biggl\{\frac{\mathrm{d}}{\mathrm{d}z}\left(\frac{\psi(z)}z\right)+\left[\frac{3 (1+R_h^2)}{m}+ \frac{i\omega}{R_h}\right]\frac{\psi(z)}{z}\biggr\}\bigg|_{z=0}=0.
\end{aligned}
\end{equation} 
The five regular singularities of the equation \eqref{eq:RW_AdS_Robin} have three different scalings with $R_h\to \infty$. The singularity at $r=0$ doesn't scale, the singularities $R_{\pm}$ and $R_h$ scale linearly, and $R_5$ scales as $R_h^3$. Hence, we will divide the space into three different regions and apply the perturbative method described in Section \ref{sec:pert_method}. 

\begin{figure}
\begin{center}
\includegraphics[width=0.8\linewidth]{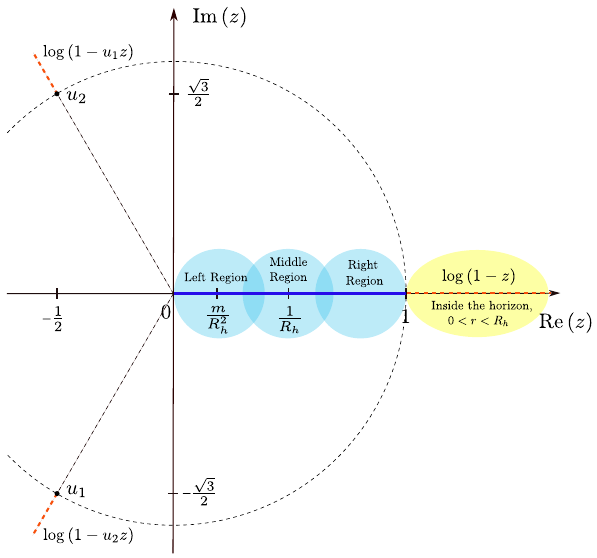}
\caption{Complex $z$ plane for scalar sector of gravitational perturbations in $\mathrm{SAdS}_4$.}
\label{fig:Robin_z}
\end{center}
\end{figure}

The three local variables are $x=R_h^3/(mr)+1/3$ for the left region (near the AdS boundary), $y=R_h^2/r$ for the middle region, and $z=R_h/r$ for the right one (near the BH horizon).\footnote{ We choose to add an intermediate region with local variable $y$ to increase the efficiency of the computation. According to our estimations \eqref{eq:Robin_left_fK}, without the middle region, one would need to compute at least $48$ orders in the expansion of the wave function in the left region \eqref{eq:Robin_psiL} to get the frequency expansion up to $\omega_5$ (assuming we do not increase the number of corrections computed in the right region). Adding the middle region allows us to get the same result by computing $\psi^L\brc{x}$ up to order $15$.}
Here the regions are labeled left and right as they appear on the complex $z$ plane (see Figure \ref{fig:Robin_z}).
From the point of view of the complex $z$ plane, the left and middle regions represent two zoomings close to the origin, with different scalings. Considering the normal form of the differential equation \eqref{eq:Robin_Eq_psi}, 
\begin{equation}
\psi''(z)+V_z(z)\psi(z)=0,
\end{equation}
the potential $V_z(z)$ has the following expansion in $1/R_h$
\begin{equation}
V_z(z)=\frac{z^6+16 z^3-8}{4 z^2 \left(z^3-1\right)^2}+\mathcal{O}\left(\frac{1}{R_h^2}\right).
\end{equation}
The two rescalings $x\sim\frac{R_h^2\,z}{m}$ and $y=R_h\,z$ are such that, in both variables, the differential equation in normal form has a potential, $V_x(x)$ and $V_y(y)$, respectively, with non-vanishing leading order in $1/R_h$,
\begin{equation}
\begin{aligned}
V_x(x)=&-\frac{2}{x^2}+\mathcal{O}\left(\frac{1}{R_h^2}\right),\\
V_y(y)=&-\frac{2}{y^2}+\mathcal{O}\left(\frac{1}{R_h^2}\right).
\end{aligned}
\end{equation}
Out of the three, the right region is the one in which it is more challenging to expand the solution of the differential equation. In particular, the solution involves multiple polylogarithms in several variables, which we analyze in Appendix \ref{appendixD}.

Since we will work with $R_h\gg 1$, the small parameter is $\alpha=1/R_h$, and the frequency expansion can be written as
\begin{equation}
\label{eq:Robin_omega_exp}
\omega=\sum_{k\ge 0}\omega_k\alpha^k.
\end{equation}
The intersections of the three regions and the boundary points $r=R_h, \infty$ determine three intervals in which the wave function should be continuous:
\begin{equation}
x\in \bsq{\frac13, \frac13 + \frac{1}{\alpha\,m}},\quad
y\in \bsq{1,\alpha^{-3/4}},\quad
z\in \bsq{\alpha^{1/4},1}.
\end{equation}
From the point of view of $x$ and $y$, the first two intervals have infinite lengths, their left endpoints are at finite values and their right endpoints are chosen to meet the next region (and so they become infinite because of the different scalings of the local variables in powers of $R_h$).
Finally, we will derive the low-lying QNM frequencies by requiring that the wave function and its first derivative are continuous at the intersection points $y=1$ and $z=\alpha^{1/4}$. As we explain later, the second intersection point $z=\alpha^{1/4}$ is chosen to avoid the reshuffling of terms in the wave function expansion \eqref{eq:Robin_psiR}. 

\subsection{Left Region}

The left region represents the region close to the $\mathrm{AdS}$ boundary, where we impose the Robin boundary condition.
The local variable in this region is
\begin{equation}
x=\frac{R_h^3}{m\,r}+\frac{1}{3}=\frac{\alpha^{-3}}{m\,r}+\frac{1}{3},
\end{equation}
and the $\mathrm{AdS}$ boundary is at $x=1/3$.
The master equation in the left region is obtained by applying the coordinate transformation $z=\alpha^2 m\brc{x-1/3}$ to \eqref{eq:Robin_Eq_psi} and substituting $\psi\brc{z}$ with $\psi^L\brc{x}$. In the leading order in $\alpha$, we get
\eq{\partial_x^2 \,\psi^L\brc{x} +\frac{6}{1-3\,x} \, \partial_x \,\psi^L\brc{x} -\frac{2\brc{1-6\,x}}{x^2 \brc{1-3\,x}^2}\,\psi^L\brc{x} + \mathcal{O}\brc{\alpha} =0.}
The two leading order solutions are
\eq{f_0^L \brc{x}=1-\frac{1}{3\,x},\quad
g_0^L \brc{x}= x^2 \brc{x-\frac{1}{3}}.}
Since $f_0^L$ satisfies the Robin boundary condition
\begin{equation}
\Biggl\{\frac{\mathrm{d}}{\mathrm{d}x}\left(\frac{\psi^L(x)}{x-\frac{1}{3}}\right)+\left[3(1+\alpha^2)+ i\,\alpha^3\, m\,\omega\right]\frac{\psi^L(x)}{x-\frac{1}{3}}\Biggr\}\bigg|_{x=\frac{1}{3}}=0,
\end{equation}
the following perturbative expansion for the wave function in the left region can be written:
\eqlb{eq:Robin_psiL}{\psi^L\brc{x} = f^L_0\brc{x}+\sum_{K\geq 1} f_K^L\brc{x} \alpha^K .}
We do not use \eqref{eq:psi_exp} to compute $f_K^L\brc{x}$ as they are simple Laurent polynomials in $x$. The form of these polynomials depends on whether $K$ is even or odd.  The following general result holds for the first $30$ computed orders:
\begin{equation}
\label{eq:Robin_left_fK}
\begin{aligned}
f_{2K}^L\brc{x} &= \brc{x-\frac13} \sum_{s=-K-1}^{K-1-\frac{4}{3} \sin\brc{K \frac{\pi}{3}}^2} \mathfrak{a}_{2K,s} \, x^s,
\\f_{2K-1}^L\brc{x} &= \brc{x-\frac13} \sum_{s=0}^{K-3+\frac{4}{3} \sin\brc{K \frac{\pi}{3}}^2} \mathfrak{a}_{2K-1,s} \, x^s,
\end{aligned}
\end{equation}
where the coefficients $\mathfrak{a}_{K,s}$ depend on the parameters $m$ and $\omega_i$. 
For example, we have for $K=1,2,3,4$:
\begin{equation}
\begin{aligned}
f_1^L\brc{x}&=0, & f_3^L\brc{x}&=-i m\, \omega_0 \left(x-\frac{1}{3}\right) ,\\
f_2^L\brc{x}&=\brc{x-\frac{1}{3}} \frac{1}{3\, x^2},\quad &
f_4^L\brc{x}&=\left(x-\frac{1}{3}\right) \left(\frac{1}{9\,x^3}-\frac{1}{3\,x^2} -i m\,\omega_1 \right).
\end{aligned}
\end{equation}
In each order in $\alpha$, the contribution of $g_0^L$  is fixed by the Robin boundary condition. The contribution of $f_0^L$ is arbitrary and can be absorbed into a normalization of the wave function $\psi^L\brc{x}$. We choose the normalization so that $f_0^L$ is only present in the leading order.

\subsection{Middle Region}

To match the wave function expansions in the left and right regions, we introduce an intermediate region with the local variable
\begin{equation}
y=\frac{R_h^2}{r}=\frac{\alpha^{-2}}{r}.
\end{equation}
The master equation in the middle region is obtained by applying the coordinate transformation $z=\alpha\,y$ to \eqref{eq:Robin_Eq_psi} and substituting $\psi\brc{z}$ with $\psi^M\brc{y}$. In the leading order in $\alpha$, we get
\eq{\partial_y^2 \,\psi^M\brc{y} -\frac{2}{y} \, \partial_y \,\psi^M\brc{y} + \mathcal{O}\brc{\alpha} =0.}
The two leading order solutions are
\eq{f_0^M \brc{y}=1,\quad
g_0^M \brc{y}=y^3.}
Strictly speaking, there is no boundary condition in the middle region. However, there is a way to use the expansion of the wave function in this region and apply the boundary condition near the horizon $y\sim \alpha^{-1}$. This requires a resummation of infinitely many terms, and the results agree with the ones obtained using three regions instead of just two. Here we focus on the procedure with three regions as it allows us to get more orders in the QNM frequency expansion. To justify our choice of functions $f_0^M$ and $g_0^M$, we can either use the gluing procedure or look at the behavior near the horizon. In the first couple of orders in $\alpha$, there is no resummation of terms in the wave function $\psi^M\brc{y}$ when we take $y\sim \alpha^{-1}$. Since near the horizon $g_0^M \brc{y}\sim \alpha^{-3}$, it can only appear in orders $\alpha^3$ and higher. This leads to the following perturbative expansion of the wave function:
\begin{equation}
\psi_{M}(y)=f_0^M(y)+\sum_{K\ge 1}f_{K}^M(y)\alpha^K.
\end{equation}
Similarly to the left region, the corrections $f_K^M(y)$ are Laurent polynomials of the form
\eq{f_{K}^M\brc{y} = \sum_{s=-K}^{K-\frac{4}{3} \sin\brc{K \frac{\pi}{3}}^2} \mathfrak{b}_{K,s} \, y^s,}
where coefficients $\mathfrak{b}_{K,s}$ also depend on the parameters $m$ and $\omega_i$. Starting from order $\alpha^3$, the gluing procedure fixes the contribution of $g_0^M$, so we keep the corresponding integration constants  $c_{K}^M$ in the expressions for $f_{K}^M$, $K\geq 3$. Out of the $27$ computed orders, we present the first $4$:
\begin{equation}
\begin{aligned}
f_{1}^M(y)&=-\frac{m}{3y},\\
f_{2}^M(y)&=\frac{m^2}{9\,y^2}-i\,\omega _0\,y,\\
f_{3}^M(y)&=-\frac{m^3}{27\,y^3}+\frac{m}{3\,y}-i\,\omega_1\,y + c_{3}^M y^3,\\
f_{4}^M(y)&=\frac{m^4}{81\, y^4}-\frac{2\, m^2}{9\, y^2}-i\, \omega_2\, y+ \frac{2\,m}{3}\, c_{3}^M y^2+c_{4}^M\, y^3.
\end{aligned}
\end{equation}

\subsection{Right Region}

The local variable in the right region is $z$, and the event horizon is at $z=1$. The leading order in $\alpha$ of \eqref{eq:Robin_Eq_psi} is
\eq{\partial_z^2 \,\psi\brc{z} +\frac{\left(z^3+2\right)}{z \left(z^3-1\right)}\, \partial_z \,\psi\brc{z} + \mathcal{O}\brc{\alpha} =0.}
The two leading order solutions are
\eq{f_0^R \brc{z}=1,\quad
g_0^R \brc{z}=\log\left(1-z^3\right).}
The Wronskian between these solutions is
\eq{W_0\brc{z} = \frac{3\,z^2}{z^3-1}.}

According to the boundary conditions \eqref{eq:Robin_BC}, the wave function in the right region is regular at $z=1$. The corresponding perturbative expansion of the wave function is then
\begin{equation}
\label{eq:Robin_psiR}
\psi^R(z)=f_0^R\brc{z}+ \sum_{K\ge 1}f_K^R\brc{z}\alpha^K.
\end{equation}
The corrections $f_K^R\brc{z}$ are computed with the help of \eqref{eq:psi_exp}, where the constants $b_K$ are fixed by the regularity condition at $z=1$. The integrals in \eqref{eq:psi_exp} can be described in terms of the multiple polylogarithms in several variables:
\eqlb{eq:Robin_Li_def}{\text{Li}_{s_1,\dots,s_k}\brc{z_1, \dots, z_k}=\sum_{n_1>n_2>\dots> n_k\geq 1}^{\infty} 
\frac{z_1^{n_1}\dots z_k^{n_k}}{n_1^{s_1}\dots n_{k}^{s_k}}.}
For $s_1\geq 2$, these functions satisfy 
\eq{z_1 \,\partial_{z_1} \text{Li}_{s_1,\dots,s_k}\brc{z_1, \dots, z_k}=\text{Li}_{s_1-1,\dots,s_k}\brc{z_1, \dots, z_k},}
and for $s_1=1$, $k\geq 2$,
\eq{\brc{1-z_1} \partial_{z_1} \text{Li}_{1,s_2,\dots,s_k}\brc{z_1, \dots, z_k}=\text{Li}_{s_2,\dots,s_k}\brc{z_1 z_2,z_3, \dots, z_k}.}
The weight and level of $\text{Li}_{s_1,\dots,s_k}\brc{z_1, \dots, z_k}$ are $s_1+\dots+s_k$ and $k$. When taking the integrals in \eqref{eq:psi_exp} with the input from this section, we will only encounter multiple polylogarithms with $s_1=s_2= \dots = s_k=1$ (see Appendix \ref{appendixD} for more details). In this case, the weight and level are the same. Moreover, all arguments $z_i$ with $i\geq 2$ are constants and can take one of the three possible values: $1$, $u_1$, and $u_2$. These constants are the third roots of unity
\eq{\label{rootsofunity}u_1= -\frac12 - \frac{i\sqrt{3}}{2}, \quad u_2= -\frac12 + \frac{i\sqrt{3}}{2}}
that arise in the following decomposition of $g^R_0\brc{z}$:
\eq{g^R_0\brc{z}= \log\brc{1-z} + \log\brc{1- u_1 z} +\log\brc{1-u_2 z}.}
Similarly to the previous cases with multiple polylogarithms, the corrections $f_K^R\brc{z}$ at order $\alpha^K$ are described in terms of functions $\text{Li}_{s_1,\dots,s_k}\brc{z_1, \dots, z_k}$ of weight $K$ and lower. This allows us to construct a linear basis of functions, in which $f_K^R\brc{z}$ can be expanded:
\begin{equation}
\begin{aligned}
&\frac{\sum_{m=-k_1}^{l_1} \zeta^R_m\, z^m}{\brc{1-u_1 z}^{i_1}\brc{1-u_2 z}^{j_1}} \, \log\brc{1-z}^{p_1} \log\brc{1-u_1 z}^{p_2} \log\brc{1-u_2 z}^{p_3},\\
&\frac{\sum_{m=-k_2}^{l_2} \xi^R_m\, z^m}{\brc{1-u_1 z}^{i_2} \brc{1-u_2 z}^{j_2}} \, \log\brc{1-z}^{p_4} \log\brc{1-u_1 z}^{p_5} \log\brc{1-u_2 z}^{p_6}
\text{Li}_{\{1\}_k}\brc{z_1 ,z_2,\dots,z_k},
\end{aligned}
\end{equation}
where $i_{1,2}$, $j_{1,2}$, $k_{1,2}$, $l_{1,2}$, $p_{j}$ are non-negative integers, and $0\leq p_1+p_2+p_3 \leq K$, $0\leq p_4+p_5+p_6 + k\leq K$.
Since the first argument in $\text{Li}_{\{1\}_k}\brc{z_1 ,z_2,\dots,z_k}$ can take one of the three possible forms
\eq{z_1 = z, \quad z_1= u_1 z,\quad \text{or} \quad z_1= u_2 z,}
we have $3^k$ functions that can enter the basis at level $k\geq 2$. However, this number is reduced due to the identities that involve multiplication by ordinary logarithm functions $\log\brc{1-z}$, $\log\brc{1-  u_1 z}$, and  $\log\brc{1- u_2 z}$ (see Appendix \ref{appendixD}). These identities allow us to use only two forms of the first argument $z_1 = u_1 z$ and $z_1 = u_2 z$. The reduced number of multiple polylogarithms that enter the basis is $8\times 3^{k-3}$ for $k\geq 3$, and just $3$ for $k=2$:
\eq{\text{Li}_{1,1}\brc{u_1 z, u_1},\quad \text{Li}_{1,1}\brc{u_1 z,u_2}, \quad \text{Li}_{1,1}\brc{u_2 z ,u_1}.}
Using {\tt Mathematica}, we compute $7$ corrections $f_K^R\brc{z}$; the first two are
\begin{equation*}
\begin{aligned}
f_1^R(z)=&\frac{\omega_0}{\sqrt{3}} \brc{ u_1 \log\brc{1-u_1 z}- u_2 \log\brc{1-u_2 z}},\\
f_2^R(z)=&-\frac{m}{3\, z}-\frac{i \,\omega_0^2}{3 \sqrt{3}} \bsq{\, \text{Li}_{1,1}\left(u_1 z,u_1\right)+u_1 \, \text{Li}_{1,1}\left(u_2 z,u_1 \right)-u_2 \,  \text{Li}_{1,1} \left(u_1 z,u_2\right) \,} + \\
&+\frac{i \,\omega_0^2}{3 \sqrt{3}} \bsq{ \log\brc{1-u_1 z}^2 - \log\brc{1-u_2 z}^2 - u_1  \log\brc{1-u_1 z}  \log\brc{1-u_2 z} } -
\end{aligned}
\end{equation*}
\begin{equation*}
\begin{aligned}
&  -\frac{i \,\omega_0^2}{3 \sqrt{3}} \log\brc{1-z}\bsq{\, u_2 \log\brc{1-u_1 z}-u_1 \log\brc{1-u_2 z} \, } + \frac{\omega_1 - i\,\omega_0^2}{\sqrt{3}}\, \log\brc{1-u_2 z} -\\
&- \frac{u_1 \,\omega_1 - i\,u_2 \, \omega_0^2}{\sqrt{3}} \, \log\brc{1-z} + b^R_2\, g^R_0\brc{z},
\end{aligned}
\end{equation*}
where
\eq{b^R_2= \frac{u_1\,\omega_1}{\sqrt{3}}+ \frac{i \,\omega_0^2}{3 \sqrt{3}} \bsq{\,u_2 \log\brc{1-u_1}-u_1 \log\brc{1-u_2}-3\, u_2\,}.}
We estimate the following behavior of $f^R_K\brc{z}$ as $z\rightarrow 0$ based on the obtained results:
\begin{equation}
\label{eq:RobinR_small_z}
K\geq 1:\quad
f^R_{2K-1} \brc{z}\sim z^{2-K},\quad 
f^R_{2K} \brc{z}\sim z^{-K}.
\end{equation}
Thus, to avoid the reshuffling of terms, we choose the gluing point between the middle and the right region to be $z=\alpha^{1/4}$. 

\subsection{Results for QNM frequencies}\label{resultsAdSbig}

We need two continuity conditions to determine the QNM frequencies, at $z=\alpha^{1/4}$ and $z=\alpha$:
\begin{equation}\small
\label{eq:Robin_cont}
\begin{aligned}
&\psi^{M}\brc{\alpha^{-3/4}}=C^{M}_R\brc{\alpha} \psi^{R}\brc{\alpha^{1/4}}, & 
&\partial_z\psi^{M}\brc{z/\alpha }\bigg|_{z=\alpha^{1/4}} = C^{M}_R\brc{\alpha}\partial_z\psi^{R}\brc{z} \bigg|_{z=\alpha^{1/4}},\\
&\psi^{L}\brc{1/3+\brc{\alpha \, m}^{-1}}=C^{L}_M\brc{\alpha} \psi^{M}\brc{1}, & 
&\partial_z\psi^{L}\brc{1/3+z \brc{\alpha^2 m}^{-1} }\bigg|_{z=\alpha} = C^{L}_M\brc{\alpha}\partial_z\psi^{M}\brc{z/\alpha} \bigg|_{z=\alpha}.
\end{aligned}
\end{equation}
The first condition in \eqref{eq:Robin_cont} is used to fix the integration constants $c_{K}^M$, and the second one gives the coefficients $\omega_k$ in the QNM frequency expansion \eqref{eq:Robin_omega_exp}.
The first seven computed orders of the wave function expansion in the right region allow us to determine $\omega_k$ up to $k=6$:  
\begin{equation}\label{QNMhydro}
\begin{aligned}
\omega_0=&\sqrt{\frac{m+2}{2}},\quad  \omega_1=-\frac{i m}{6},\\
\omega_2=&\frac{\sqrt{2} \,m}{36\, \sqrt{m+2}}+\frac{m\, \sqrt{m+2}}{108\, \sqrt{2}} \bsq{ 15+\sqrt{3}\,\pi-9\, \log\brc{3}},\\
\omega_3=&-\frac{m \brc{m+2}}{18 \sqrt{3}} \bsq{ \text{Li}_{1,1}\left(u_1,u_1\right)+u_1 \text{Li}_{1,1}\left(u_2,u_1\right)-u_2 \text{Li}_{1,1}\left(u_1,u_2\right)}+\\
&+\frac{m \brc{m+2}}{1296 \sqrt{3}} \bsq{\pi^2-6 i \,\pi \log\brc{3} +9\brc{u_2-3\,u_1} \log\brc{3}^2}+\\
&+\frac{i m \brc{m+3}}{162} \bsq{9+\sqrt{3}\, \pi - 9 \log\brc{3}},\\
\omega_4=&-\frac{i\,m \brc{m+2}^{3/2}}{54 \sqrt{6}} \biggl[ \text{Li}_{\bfi{1}_3}\brc{u_1,u_1,u_1}-u_1 \,\text{Li}_{\bfi{1}_3}\brc{u_1,u_1,1} -u_1 \, \text{Li}_{\bfi{1}_3}\brc{u_1,1,u_2} - \biggr.\\
&\biggl. - 2\, u_2\, \text{Li}_{\bfi{1}_3}\brc{u_1,u_2,1} 
 - \brc{\mathbf{u_1 \leftrightarrow u_2}}\biggr] + \dots\, ,\\
\omega_5=& \frac{m \brc{m+2}^2}{162 \sqrt{3}} \biggl[ \text{Li}_{\bfi{1}_4}\brc{u_1,u_1,u_1,1} + u_2 \,\text{Li}_{\bfi{1}_4}\brc{u_1,1,u_1,u_1} -2 \, \text{Li}_{\bfi{1}_4}\brc{u_1,1,1,u_2}-\biggr.\\
&-u_1\, \text{Li}_{\bfi{1}_4}\brc{u_1,1,u_2,1}- 2\,u_2\, \text{Li}_{\bfi{1}_4}\brc{u_1,u_2,1,1} - u_1\, \text{Li}_{\bfi{1}_4}\brc{u_1,u_2,u_2,1} -\\
&\biggl.- \brc{\mathbf{u_1 \leftrightarrow u_2}}\biggr] + \frac{m \brc{m+2}^2}{486 \sqrt{3}} \biggl[ 3\, \text{Li}_{\bfi{1}_4}\brc{u_1,1,u_1,1} + 6\, u_1 \, \text{Li}_{\bfi{1}_4}\brc{u_1,1,1,u_1} - \biggr.\\
&\biggl.- 2\, u_2 \, \text{Li}_{\bfi{1}_4}\brc{u_2,u_2,u_1,1}  \biggr] +\dots\, ,
\end{aligned}
\end{equation}
where we shortened the results for $\omega_4$ and $\omega_5$ for readers convenience. The full results, including the result for $\omega_6$,  can be found in the attached {\tt Mathematica} files.
Notice that, as compared to the QNM frequencies computed in Sec.~\ref{section3} and Sec.~\ref{section4}, here the frequencies involve different irrational numbers, for instance, $\log 3$, $\sqrt{3}$, as well as colored multiple zeta values of level 3.

Upon taking the  scaling limit 
\eq{R_h\rightarrow \infty,\quad \ell \rightarrow \infty, \quad \frac{2\,\ell}{3\,R_h} \rightarrow \mathfrak{q},}
where $\mathfrak{q}$ stays constant, we reproduce the results for the QNM frequencies of the M2-brane in the $\mathrm{AdS}_4$ background (see Table IV in \cite{natsuume2008causal}) which are directly linked to hydrodynamics \cite{policastro2002ads,policastro2003ads,son2002minkowski,Bhattacharyya:2007vjd}.
Also, the following rescaling of the frequency is needed: 
\begin{equation} 
\mathfrak{w}=\frac{2\,\omega}{3\, R_h}.
\end{equation}
Applying this limit to \eqref{QNMhydro}, we obtain an expansion of $\mathfrak{w}$ in $\mathfrak{q}$:
\eq{\mathfrak{w}=\sum_{k\geq 1} \mathfrak{w}_k \, \mathfrak{q}^k,}
where $\mathfrak{w}_1$, $\mathfrak{w}_2$, and $\mathfrak{w}_3$ agree with the results from  \cite{natsuume2008causal}, and the new results are
\be
\ba
\mathfrak{w}_4=&-\frac{\sqrt{3}}{16} \bsq{ \text{Li}_{1,1}\left(u_1,u_1\right)+u_1 \text{Li}_{1,1}\left(u_2,u_1\right)-u_2 \text{Li}_{1,1}\left(u_1,u_2\right)}+\frac{72\,i\sqrt{3} +24\,i \,\pi +\pi^2}{384 \sqrt{3}} - \\
&- \frac{12\,i\sqrt{3} +i\, \pi} {64 \sqrt{3}}\, \log\brc{3} +\frac{\sqrt{3}}{128 }\brc{u_2-3\,u_1} \log\brc{3}^2,
\\
\mathfrak{w}_5=&-\frac{i \sqrt{3}}{32 \sqrt{2}} \left[ \text{Li}_{\bfi{1}_3}\brc{u_1,u_1,u_1}-u_1 \,\text{Li}_{\bfi{1}_3}\brc{u_1,u_1,1} -u_1 \, \text{Li}_{\bfi{1}_3}\brc{u_1,1,u_2} - \right.\\
& \left. - 2\, u_2\, \text{Li}_{\bfi{1}_3}\brc{u_1,u_2,1} 
 - \brc{\mathbf{u_1 \leftrightarrow u_2}}\right] + \dots\, , \\
\mathfrak{w}_6=&\, \frac{\sqrt{3}} {64} \biggl[ \text{Li}_{\bfi{1}_4}\brc{u_1,u_1,u_1,1} + u_2 \,\text{Li}_{\bfi{1}_4}\brc{u_1,1,u_1,u_1} -2 \, \text{Li}_{\bfi{1}_4}\brc{u_1,1,1,u_2}-\biggr.\\
&-u_1\, \text{Li}_{\bfi{1}_4}\brc{u_1,1,u_2,1}- 2\,u_2\, \text{Li}_{\bfi{1}_4}\brc{u_1,u_2,1,1} - u_1\, \text{Li}_{\bfi{1}_4}\brc{u_1,u_2,u_2,1} -\\
&\biggl.- \brc{\mathbf{u_1 \leftrightarrow u_2}}\biggr] + \frac{1}{64 \sqrt{3}} \biggl[ 3\, \text{Li}_{\bfi{1}_4}\brc{u_1,1,u_1,1} + 6\, u_1 \, \text{Li}_{\bfi{1}_4}\brc{u_1,1,1,u_1} - \biggr.\\
&\biggl.- 2\, u_2 \, \text{Li}_{\bfi{1}_4}\brc{u_2,u_2,u_1,1}  \biggr] +\dots\, ,\\
\end{aligned}
\end{equation}
where we shortened the results for $\mathfrak{w}_5$ and $\mathfrak{w}_6$ for readers convenience. The full results, including the result for $\mathfrak{w}_7$, can be found in the attached {\tt Mathematica} files. The numerical values of these coefficients are  
\begin{equation}
\begin{aligned}
\mathfrak{w}_1 & =\frac{1}{\sqrt{2}},\\
\mathfrak{w}_2 & =-\frac{i}{4}~,\\
\mathfrak{w}_3 & = 0.155473446153645...,\\
\mathfrak{w}_4 & = 0.067690388847266...\,\cdot i,\\
\mathfrak{w}_5 &=- 0.010733416957692...,\\
\mathfrak{w}_6 &= 0.013959543659902...\,\cdot i,\\
\mathfrak{w}_7 &= - 0.016615814626711...\, .
\end{aligned}
\end{equation}
These alternate between real and imaginary parts, precisely as predicted in \cite{Grozdanov:2019uhi,Grozdanov:2019kge}.\footnote{We would like to thank S.~Grozdanov for useful discussions on this point and for providing us with approximate numerical values against which we could check our results.}

\section{Conclusions}

This paper focuses on analytical aspects of spectral problems associated with perturbation theory for four-dimensional (A)dS black holes. We explore these problems using two analytic strategies: one based on the NS functions and one based on a recursive structure involving multiple polylogarithms. 
Thanks to these tools, we can compute the quasinormal mode frequencies and their eigenfunctions analytically in various regimes. For instance, we can obtain the series expansion at large  $R_h$, small $R_h$, or large spin $\ell$ ($R_h$ being the BH horizon). 

We use the approach based on the NS functions in the context of four-dimensional dS Schwarzschild black holes. In this setup, the NS functions allow us to compute the large $\ell$ expansion of QNMs systematically. We find that, up to non-perturbative effects in $\ell$, the QNMs are (negative) imaginary numbers that are even functions of $R_h$.
To include non-perturbative effects in the spin, switching to the polylog approach is convenient. Once non-perturbative effects are included, QNMs are no longer even in $R_h$. But we still find a branch of purely imaginary modes, thereby providing analytical confirmation of the results obtained through numerical studies in \cite{KZ'22,Brady:1996za,Jansen:2017oag, Cardoso:2017soq}. Exploring the interplay between  the NS and polylog approaches would be interesting. In particular, the appearance of multiple polylogarithms and multiple zeta values may be related to the behavior of the NS functions close to their singular points, see e.g.~\cite{Alekseev:2019gkl,Gorsky:2017ndg,Beccaria:2016wop}.

We extend the polylog method to study conformally coupled scalar, electromagnetic, and vector-type gravitational
perturbations in asymptotically AdS$_4$ Schwarzschild black holes. The NS functions are less effective for these perturbations because the point at spatial infinity is not a singular point of the equation.\footnote{If we consider massive scalar perturbation instead, the underlying equation has five regular singular points and spatial infinity is mapped to one of them. In this case, one can use the NS functions for an $SU(2)\times SU(2)$ linear quiver \cite{AlQi}. In addition, for generic mass, the leading order solution does not reduce to a rational function; hence the polylog approach presented can not be applied straightforwardly.}  Hence, we switch to the polylog method for Dirichlet and Robin boundary conditions. As an application, we use this technique to study the low-lying modes of the scalar sector of gravitational perturbations and compute  several orders in the $1/R_h$ expansion. Even in the hydrodynamic expansion, this allowed us to go beyond the results presently available in the literature.
From the point of view of holography, the polylog method presents finite spin predictions for the dual 3d CFT. It would be interesting to explore this further in higher spacetime dimensions and make contact with past and recent developments in the study of holographic CFTs \cite{son2002minkowski,Nunez:2003eq, witten:2001ua, Gubser:2002vv,hartman:2006dy,Li:2020dqm,Dodelson:2022eiz,Dodelson:2022yvn,Esper:2023jeq,Karlsson:2021duj,Kulaxizi:2018dxo,Karlsson:2019qfi,Kulaxizi:2019tkd,Karlsson:2019dbd,Karlsson:2020ghx}.

The technical result we obtained about the perturbation theory of second-order linear differential equations with Fuchsian (or irregular\footnote{A detailed analysis of the case with irregular singularities, which is relevant for asymptotically flat black holes, will appear in \cite{GlPa}.}) singularities points to the existence of a recursive structure for their solution involving multiple polylogarithms.
It raises the question of whether there exists a deeper algebraic structure beyond this that could improve  the algorithm. This should allow us to have a higher level of analytic control over the problem at hand.
For instance, it would be interesting to quantify the precise analytic properties of QNM frequencies as functions of the BH radius and/or other relevant parameters to understand their physical meaning better. For example, this would allow to detect phase transitions and/or (in)stabilities. 
These considerations become especially interesting when considering rotating and/or charged black holes. Indeed, these exhibit a richer structure with intriguing (in)stability features. It would be interesting to revisit these problems within the approaches presented in this paper.

Let us remark that the polylog method we developed shares 
similarities with the techniques used to compute Feynman integrals in Quantum Chromo-Dynamics, see e.g.~\cite{Duhr:2019tlz,Panzer:2014caa,Frellesvig:2018lmm,c1,c2,c3} and references therein.\footnote{
During the writing of this paper, we were informed  that Saso Grozdanov is also exploring similar ideas in the context of AdS$_5$ black holes (particularly the hydrodynamic limit)\cite{informal}.}
Similar techniques also recently appeared in studying higher curvature corrections to the effective low-energy gravitational theory arising from string scattering diagrams \cite{Alday:2022xwz,Alday:2023jdk,Alday:2023mvu,Bern:2019nnu,Parra-Martinez:2020dzs,DiVecchia:2021bdo,Herrmann:2021tct}. 
Although these results are directly related to hyperbolic trajectories, one can use 
the data extracted from the amplitudes to determine the parameters of the effective one-body potential of \cite{Buonanno:1998gg} to be used to describe the gravitational bound states.
From the computational viewpoint, the resulting polylog expansion in such approximation is naturally obtained by computing the relevant Feynman multiloop integrals in the proper kinematic regime.
On the contrary, in the QNMs regime, the appearance of multiple polylogarithms  does not seem to have a direct interpretation in terms of Feynman multiloop integrals.   Moreover, different types of special functions arise for other gravitational backgrounds and/or other perturbations. For example, when considering asymptotically flat black holes, there is the appearance of multiple polyexponential functions as well \cite{GlPa}. It would be interesting to understand this better.

Finally, one of the most challenging and exciting questions would be to go beyond linear perturbation theory. The NS and polylog methods allow for the computation of the eigenfunctions and the Green functions, which are essential inputs to go beyond the linear theory.

\appendix
\section{NS functions}\label{appendixA}

This appendix reports the notations and conventions used in Sec.~\ref{section3.1}, where the gauge theory approach is applied to the Heun connection problem. The relevant theory is $\mathcal{N}=2$ $SU(2)$ gauge theory with $N_f=4$ fundamental hypermultiplets. 

If $Y$ is a Young diagram, we denote with $(Y_1\ge Y_2\ge\dots)$ the heights of its columns and with $(Y'_1\ge Y'_2,\dots)$ the lengths of its rows. For every Young diagram $Y$ and for every box $s=(i,j)$, we denote the arm length and the leg length of $s$ with respect to the diagram $Y$ as
\begin{equation}
A_Y(i, j) = Y_j -  i, \quad L_Y(i, j) =Y'_i - j.
\end{equation}
Note that we do not require $s$ to be in $Y$: if this is the case, the arm length and the leg length are non-negative quantities, but this is not true in general.

We now introduce the main contributions coming into play for the definition of the instanton partition function of $\mathcal{N}=2$ $SU(2)$  gauge theory with fundamental matter. Let us denote with $\vec{Y}=\left( Y_1, Y_2 \right)$ a pair of Young diagrams and with $| \vec{Y} | = | Y_1 | + | Y_2 |$ the total number of boxes. We denote with $\vec{a}=(a_1,a_2)$ the v.e.v. of the scalar in the vector multiplet and with $\epsilon_1,\epsilon_2$ the parameters characterizing the $\Omega$-background. We define the hypermultiplet and vector contribution as
\cite{Flume:2002az,Bruzzo:2002xf}
\begin{equation}\small
\begin{aligned}
    &z_{\text{hyp}} \left( \vec{a}, \vec{Y}, m \right) = \prod_{k= 1,2} \prod_{(i,j) \in Y_k} \left[ a_k + m + \epsilon_1 \left( i - \frac{1}{2} \right) + \epsilon_2 \left( j - \frac{1}{2} \right) \right] \,, \\
    &z_{\text{vec}} \left( \vec{a}, \vec{Y} \right) = \prod_{i,j=1}^2\prod_{s\in Y_i}\frac{1}{a_i-a_j-\epsilon_1L_{Y_j}(s)+\epsilon_2(A_{Y_i}(s)+1)}\prod_{t\in Y_j}\frac{1}{-a_j+a_i+\epsilon_1(L_{Y_i}(t)+1)-\epsilon_2A_{Y_j}(s)}\,.
\end{aligned}
\end{equation}
We will always take $\epsilon_1=1$ and $\vec{a}=(a,-a)$.
Let us denote with $m_1,m_2,m_3,m_4$ the masses of the four hypermultiplets and let us introduce the gauge parameters $a_0,a_t,a_1,a_{\infty}$ satisfying
\begin{equation}\label{gaugemasses}
\begin{aligned}
m_1&=-a_t-a_0,\\
m_2&=-a_t+a_0,\\
m_3&=a_{\infty}+a_1,\\
m_4&=-a_{\infty}+a_1.
\end{aligned}
\end{equation}
Moreover, we denote with $t$ the instanton counting parameter $t=e^{2\pi i\tau}$, where $\tau$ is related to the gauge coupling by 
\begin{equation}
    \tau=\frac{\theta}{2\pi}+i\frac{4\pi}{g_{\rm YM}^2}.
\end{equation}
In the {\tt Mathematica} programs available at {\url{https://github.com/GlebAminov/BH_PolyLog}}, we also use the redefined masses $M_i$, which are related to $m_i$ via
\be m_i= M_i+\frac{t}{2\brc{1-t}}\sum_{j=0}^4 M_j .\ee
The instanton part of the NS free energy is then given as a power series in $t$ by
\begin{equation}
F(t)=\lim_{\epsilon_2\to 0}\epsilon_2\log\Biggl[(1-t)^{-2\epsilon_2^{-1}\left(\frac{1}{2}+a_1\right)\left(\frac{1}{2}+a_t\right)}\sum_{\vec{Y}}t^{|\vec{Y}|}z_{\text{vec}} \left( \vec{a}, \vec{Y} \right)\prod_{i=1}^4z_{\text{hyp}} \left( \vec{a}, \vec{Y}, m_i \right)\Biggr].
\end{equation}
In the text, we will also refer to the \emph{full NS free energy}, which contains not only the instanton part but also the classical and one-loop contributions. This is explicitly given by
\begin{equation}
\begin{aligned}
F_{\mathrm{full}}(t)=F(t)-a^2\log(t)-\sum_{i=1}^4\psi^{(-2)}\left(\frac{1}{2}-a-m_i\right)-\sum_{i=1}^4\psi^{(-2)}\left(\frac{1}{2}+a-m_i\right)+\\
+\psi^{(-2)}\left(1+2a\right)+\psi^{(-2)}\left(1-2a\right),
\end{aligned}
\end{equation}
where
\begin{equation}
\psi^{(-2)}(z)=\int_0^z\mathrm{d}t\log\left[\Gamma(t)\right].
\end{equation}

The gauge parameter $a$ is expressed in a series expansion in the instanton counting parameter $t$, obtained by inverting the \emph{Matone relation} \cite{Matone:1995rx,Flume:2004rp}
\begin{equation}
u^{(0)} =-\frac{1}{4} - a^2 + a_t^2 + a_0^2 + t \partial_t F(t),
\end{equation}
where the parameter $u^{(0)}$ is the complex moduli parametrizing the corresponding SW curve.
Explicitly, the expansion reads as follows
\begin{equation} 
a=\pm\Biggl\{\sqrt{-\frac{1}{4}-u^{(0)}+a_t^2+a_0^2}+\frac{\bigl(\frac{1}{2}+u^{(0)}-a_t^2-a_0^2-a_1^2+a_{\infty}^2\Bigr)\Bigl(\frac{1}{2}+u^{(0)}-2a_t^2\Bigr)}{2(1+2u^{(0)}-2a_t^2-2a_0^2)\sqrt{-\frac{1}{4}-u^{(0)}+a_t^2+a_0^2}}t+\mathcal{O}(t^2)\Biggr\}.
\end{equation} 

\section{Useful facts about multiple polylogarithms in a single variable}\label{appendixB}
There are many identities between polylogarithms and multiple polylogarithms.
Below is the list of identities that are relevant in our case.
First, for multiple polylogarithms of the form $\text{Li}_{1,s_2,\dots,s_k}\brc{z}$, we have:
\eqlb{eq:Li_Id1n}{\text{Li}_{\bfi{1}_n}\brc{z}=\frac{\brc{-1}^{n}}{n!} \log\brc{1-z}^n.}
Taking derivatives and using \eqref{eq:propertypolylog} and \eqref{eq:propertypolylog2}, it is easy to show by induction that
\eq{\label{eq:B2}n\geq1:\quad \sum_{k=1}^{n-1} \text{Li}_{k,n-k+1} \brc{z}+2\,\text{Li}_{n,1}\brc{z} +\log\brc{1-z} \text{Li}_{n}\brc{z}=0,}
\eqn{\left\{\at{m\geq1,}{n\geq1}\right.:\quad \sum_{k=1}^{m-1} \text{Li}_{k,m-k+1,n} \brc{z}+\sum_{k=1}^{n-1} \text{Li}_{m,k,n-k+1} \brc{z} + \text{Li}_{m,1,n}\brc{z} + 2\,\text{Li}_{m,n,1}\brc{z}+}
\eq{+\log\brc{1-z} \text{Li}_{m,n}\brc{z}=0.}
Generalizing the last two identities to an arbitrary level, one gets the following identity, which we use to express $\text{Li}_{1,s_1,\dots,s_n}\brc{z}$ in terms of multiple polylogarithms $\text{Li}_{r_1,\dots,r_{n+1}}\brc{z}$ with $r_1\geq 2$:
\eqn{\sum_{i=1}^{n} \sum_{k=1}^{s_i-1} \text{Li}_{s_1,\dots,s_{i-1},k,s'_{i},s_{i+1},\dots,s_n}\brc{z} +
\sum_{i=1}^{n-1} \text{Li}_{s_1,\dots,s_{i},1,s_{i+1},\dots,s_n}\brc{z}+ 2\,\text{Li}_{s_1,\dots,s_n,1}\brc{z} +}
\eqlb{eq:Li_Id1}{+\log\brc{1-z} \text{Li}_{s_1,\dots,s_n}\brc{z}=0,}
where in the first double sum, we insert index $k$ in the position of $s_i$ and then move $s_i$ to the next position while modifying it as
\eq{s'_i=s_i -k+1.}
Up to weight 4, all multiple polylogarithms in a single variable can be expressed as ordinary polylogarithms by combining the above identities and the following ones \cite{Lewin'81,Gangl'16}:
\begin{equation}
\label{eq:Id_MLtoL1}
\text{Li}_{2,1}\brc{z}+ \text{Li}_3\brc{1-z}-\log\brc{1-z} \text{Li}_2\brc{1-z} - \frac12 \log\brc{z} \log\brc{1-z}^2 -\zeta\brc{3}=0,
\end{equation}
\begin{equation}
\begin{aligned} \text{Li}_{3,1}\brc{z}-\text{Li}_4\brc{z}+ \text{Li}_4\brc{1-z}- \text{Li}_4\brc{\frac{z}{z-1}}
+\log\brc{1-z} \text{Li}_3\brc{z} =\frac{1}{24} \log\brc{1-z}^4 \\
- \frac16 \log\brc{z} \log\brc{1-z}^3 + \frac{\pi^2}{12} \log\brc{1-z}^2 +
\zeta\brc{3} \log\brc{1-z} + \frac{\pi^4}{90},
\end{aligned}
\end{equation}
\begin{equation}
\begin{aligned}
\text{Li}_{2,1,1}\brc{z}+ \text{Li}_4\brc{1-z}- \log\brc{1-z} \text{Li}_3\brc{1-z}
+ \frac12 \log\brc{1-z}^2 \text{Li}_2\brc{1-z} = \\
= \frac{\pi^4}{90} - \frac16 \log\brc{z} \log\brc{1-z}^3 ,
\end{aligned}
\end{equation}
\begin{equation}
\label{eq:Id_MLtoL2}
4\, \text{Li}_{3,1}\brc{z} +2\, \text{Li}_{2,2}\brc{z} - \text{Li}_{2}\brc{z}^2 =0.
\end{equation}
There are identities for weight higher than $4$, but not enough to express all multiple polylogarithms as ordinary polylogarithms.
For example, we have for weight $5$:
\begin{equation}
\begin{aligned} \text{Li}_{2,1,1,1}\brc{z}+ \text{Li}_{5}\brc{1-z} - 
\log\brc{1-z} \text{Li}_{4}\brc{1-z}+ \frac12\,\log\brc{1-z}^2 \text{Li}_{3}\brc{1-z}= \\
\frac16\, \log\brc{1-z}^3 \text{Li}_{2}\brc{1-z} +\frac1{24}\, \log\brc{z} \log\brc{1-z}^4 +\zeta\brc{5}.
\end{aligned}
\end{equation}
The latter can be checked by taking a derivative and using identity \ref{eq:Li_Id1n}.
Throughout the paper, we choose not to use the powers of polylogarithms in any basis, which reduces the number of relevant identities. 

 Multiple zeta values (MZVs) and Euler sums arise when evaluating the quasinormal mode frequencies:
\eqlb{MZV_def}{\text{Li}_{s_1,\dots,s_k}\brc{1}\equiv \zeta\brc{s_1,\dots,s_k},\quad
\text{Li}_{s_1,\dots,s_k}\brc{-1}\equiv \zeta\brc{-s_1,s_2,\dots,s_k}.}
Some of these values can be computed using the known relations \cite{Euler1911,Nielsen2005,Hoffman1992,BBG1995} of the form:
\eqlb{eq:MZV_Id_1}{a,b>1:\quad \zeta\brc{a,b}+\zeta\brc{b,a}= \zeta\brc{a}\zeta\brc{b}-\zeta\brc{a+b},}
\eq{\zeta\brc{-2\,n,1}=\frac{1}{2} \zeta\brc{2\,n+1} -\frac{2\,n-1}{2} \,\eta\brc{2\,n+1}
+\sum_{k=1}^{n-1} \eta\brc{2\,k} \zeta\brc{2\,n+1-2\,k} ,}
where
\eq{\eta\brc{x}=\brc{1-2^{1-x}} \zeta\brc{x}.}
In particular, the following  MZVs and Euler sums of  weight $5$ can be written in terms of Riemann $\zeta$-functions \cite{BBG1995}:
\eq{\zeta\brc{2,3}= \frac{9}{2} \,\zeta\brc{5}-\frac{\pi ^2 }{3}\,\zeta\brc{3},\quad
\zeta\brc{3,2}= \frac{\pi ^2 }{2}\,\zeta\brc{3}-\frac{11}{2} \,\zeta\brc{5},\quad
\zeta\brc{4,1}=2 \,\zeta\brc{5}-\frac{\pi ^2 }{6}\,\zeta\brc{3},}
\eqlb{eq:MZV_Id_2}{\zeta\brc{-2,3}= \frac{51}{32} \,\zeta\brc{5}-\frac{\pi ^2 }{8}\,\zeta\brc{3},
\quad \zeta\brc{-3,2}= \frac{41}{32} \,\zeta\brc{5}-\frac{5\,\pi ^2 }{48}\,\zeta\brc{3},\quad
\zeta\brc{-4,1}=\frac{\pi ^2 }{12}\,\zeta\brc{3}-\frac{29}{32} \,\zeta\brc{5}.}

Lastly, we need expansions of multiple polylogarithms around $z=1$. Such an expansion for the polylogarithm $\text{Li}_n\brc{z}$ with $n\geq 1$ is given by \cite{Wood1992,Gradshteyn2014}
\eq{\text{Li}_n\brc{\rme^\mu}=\frac{\mu^{n-1}}{\brc{n-1}!}\bsq{H_{n-1}-\log\brc{-\mu}} +
\sum_{\at{k=0}{k\neq n-1}}^{\infty}\zeta\brc{n-k} \frac{\mu^k}{k!} ,}
where $H_n$ is the $n$-th harmonic number and $\abs{\mu}<2\pi$. To derive the same for $\text{Li}_{1,n}\brc{z}$, we integrate both sides of the following equation:
\eq{\frac{\rmd}{\rmd \mu}\, \text{Li}_{1,n}\brc{\rme^\mu} = \frac{\rme^\mu}{1-\rme^\mu}\, \text{Li}_n\brc{\rme^\mu},}
where
\eq{\frac{\rme^\mu}{1-\rme^\mu}=-\frac12 -\frac1{\mu}-\sum_{j=1}^{\infty} B_{2j}\, \frac{\mu^{2j-1}}{\brc{2 j}!}.}
Up to a constant of integration $c_{1,n}$ we get:
\eqn{n \geq2:\quad \text{Li}_{1,n}\brc{\rme^\mu}=c_{1,n}-\zeta\brc{n} \log\brc{-\mu} -\frac{1}{2}\text{Li}_{n+1}\brc{\rme^\mu} -
\sum_{\at{k=1}{k\neq n-1}}^{\infty}\zeta\brc{n-k} \frac{\mu^k}{k!\, k} }
\eqlb{eq:exp_Li1n}{-\frac{1}{\brc{n-1}!}\sum_{j=0}^{\infty} \frac{B_{2j}}{\brc{2 j}!}\, \frac{\mu^{2j+n-1}}{2j+n-1}
\bsq{H_{n-1}+\frac{1}{2j+n-1}-\log\brc{-\mu}}}
\eqn{-\sum_{j=1}^{\infty}\sum_{\at{k=2j}{k\neq 2j+ n-1}}^{\infty} \frac{B_{2j}}{\brc{2 j}!}\,
\frac{\zeta\brc{2j+n-k}}{\brc{k-2j}!}\,\frac{\mu^k}{k}.}
Using \eqref{MZV_def} and the above polylogarithm identities, one obtains the first few coefficients $c_{1,n}$. For example, from \eqref{eq:B2} and (\ref{eq:Id_MLtoL1})--(\ref{eq:Id_MLtoL2}), we get
\eq{c_{1,2} = -\frac{3}{2}\, \zeta\brc{3},\quad c_{1,3} = -\frac{\pi^4}{120}.}
Now, we can get the expansion for $\text{Li}_{m,n}\brc{\rme^\mu}$ by consecutively integrating (\ref{eq:exp_Li1n}):
\begin{equation*}\small
\begin{aligned}
m \geq 1,\,n \geq2:\quad \text{Li}_{m,n}\brc{\rme^\mu}=\sum_{k=0}^{m-1} c_{m-k,n}\,\frac{\mu^k}{k!}+
\zeta\brc{n} \frac{\mu^{m-1}}{\brc{m-1}!}\bsq{H_{m-1}-\log\brc{-\mu}}-\frac12\, \text{Li}_{m+n}\brc{\rme^\mu}\\
-\sum_{\at{k=1}{k\neq n-1}}^{\infty}\zeta\brc{n-k} \frac{\mu^{k+m-1}}{k\, \brc{k+m-1}!}
-\sum_{j=1}^{\infty}\sum_{\at{k=2j}{k\neq 2j+ n-1}}^{\infty} \frac{B_{2j}}{\brc{2 j}!}\,
\frac{\zeta\brc{2j+n-k}}{\brc{k-2j}!}\,\frac{\brc{k-1}!}{\brc{k+m-1}!}\,\mu^{k+m-1}\\
-\frac{\mu^{n+m}}{\brc{n-1}!}\sum_{j=0}^{\infty} \frac{B_{2j}}{\brc{2 j}!}\, \frac{\brc{2j+n-2}!\, \mu^{2j-2}}{\brc{2j+n+m-2}!}
\bsq{H_{2j+n+m-2}+H_{n-1}-H_{2j+n-2}-\log\brc{-\mu}}.
\end{aligned}
\end{equation*}
Again, the integration constants $c_{m,n}$ can be computed with the help of the known identities:
\eqn{c_{2,2}=\frac{\pi^4}{72},\quad c_{1,4}=\frac{\pi^2}{6}\,\zeta\brc{3}-\frac52\, \zeta\brc{5},
\quad c_{2,3}=-\frac{\pi^2}{3}\,\zeta\brc{3}+5\, \zeta\brc{5},
\quad c_{3,2}=\frac{\pi^2}{2}\,\zeta\brc{3}-5\, \zeta\brc{5}.}
In the same way, one can derive the expansion for $\text{Li}_{m,1}$ by consecutively integrating $\text{Li}_{1,1}$:
\begin{equation} \small
\begin{aligned} 
m \geq 1:\quad \text{Li}_{m,1}\brc{\rme^\mu}=&\sum_{k=0}^{m-2} \zeta\brc{m-k,1} \,\frac{\mu^k}{k!} -
\sum_{k=1}^{\infty} \zeta\brc{1-k} \frac{\mu^{k+m-1}}{\brc{k+m-1}!} \bsq{\log\brc{-\mu} +H_k-H_{k+m-1}}\\
&+\frac{\mu^{m-1}}{\brc{m-1}!} \bsq{\frac12\,\log\brc{-\mu}^2 -H_{m-1} \log\brc{-\mu}  +H_{m-1,2} +
\sum_{k=1}^{m-1} \frac{H_{k-1}}{k}}\\
&+\frac12 \sum_{j=1}^{\infty} \sum_{k=j+1}^{\infty} \zeta\brc{1-j} \zeta\brc{j-k+1} \frac{k!}{\brc{k+m-1}!}\,
\frac{\mu^{k+m-1}}{j!\brc{k-j}!},
\end{aligned}
\end{equation}
where $H_{m,2}$ is the generalized harmonic number of the form
\eq{H_{m,2}=\sum_{k=1}^{m} \frac{1}{k^2}.}

\section{Solving integral recurrence relations}
\label{app:recurrence_rel}

In sections \ref{sec:dS_Left} and \ref{sec:AdS_Left}, we claimed that the wave functions $\psi^{L}\brc{z}$ at order $t^K$ (or, equivalently, $R_h^K$) are described in terms of multiple polylogarithms of weight $K$ and lower. In this section, we will prove this claim, but first, let us clarify the terminology.
The notion of weight is related to the power of a logarithm function, as seen in the following identity:
\eq{\text{Li}_{\bfi{1}_n}\brc{z}=\frac{\brc{-1}^{n}}{n!} \log\brc{1-z}^n.}
Thus, we will ascribe weight to the ordinary logarithm functions as follows. For any product of two logarithms
\eq{m,n\geq0:\quad \log\brc{z}^m \log\brc{z-1}^n,}
the weight equals $n+m\geq0$. For the product of a logarithm and a multiple polylogarithm
\eq{k\geq1,\,n\geq0:\quad\log\brc{z-1}^n \text{Li}_{s_1,\dots,s_k}\brc{1-z},}
the weight is $n+s_1+\dots+s_k > 0$. Here we do not consider the other possible product $\log\brc{z}^m \text{Li}_{s_1,\dots,s_k}\brc{1-z}$ because, due to the identities of the form (\ref{eq:Li_Id1}), this product can always be rewritten as a linear combination of multiple polylogarithms. Some simple examples are:
\eq{\log\brc{z} \,\text{Li}_2 \brc{1-z} = - \text{Li}_{1,2} \brc{1-z}-2\, \text{Li}_{2,1} \brc{1-z},}
\eq{\frac12 \,\log\brc{z}^2 \,\text{Li}_2 \brc{1-z} = \text{Li}_{1,1,2} \brc{1-z}+2\,\text{Li}_{1,2,1} \brc{1-z}
+3 \,\text{Li}_{2,1,1} \brc{1-z}.}
In general, multiple polylogarithm functions can not be rewritten as powers of ordinary logarithm functions. We will use both logarithms and multiple polylogarithms of a certain weight to build a linear basis in which the wave function can be expanded at a certain order in $t$. In what follows, all powers of logarithms are non-negative integers.

We are going to prove our claim by induction. In the first order in $t$, the integrands in the recurrence relations are just rational functions of the form
\eq{\frac{\sum_{m=0}^{r_0} \alpha_m\, z^m}{z^{i_0} \brc{z-1}^{j_0}}}
with non-negative integers $r_0,i_0,j_0$ that depend on the quantum numbers of the scalar, electromagnetic, or gravitational perturbations. These rational functions can be broken up into a sum of monomials in $z$ and poles at $z=0,1$ with the help of the identities
\eq{n,m \geq 0:\quad \frac{z^n}{\brc{z-1}^m} = \sum_{k=0}^{n} \brc{\at{n}{k}} \brc{z-1}^{k-m},}
\eqlb{eq:poles_Id}{\frac{1}{z^n \brc{1-z}^m}= \sum_{k=1}^{n} \brc{\at{n+m-k-1}{m-1}} \frac{1}{z^k} +
\sum_{j=1}^{m}\brc{\at{n+m-j-1}{n-1}}\frac{1}{\brc{1-z}^{j}},}
where in the last identity $n,m \geq 1$. Thus, the wave function $\psi^{L}\brc{z}$ ar order $t$  is described in terms of rational functions and logarithms of weight $1$: $\log\brc{z}$ and $\log\brc{z-1}$. Next, we assume that the integrands in the recurrence relations at order $t^{K+1}$ are linear combinations of functions with maximum weight $K$ :
\eqn{\frac{\sum_{m=0}^{r_1} \alpha_m\, z^m}{z^{i_1} \brc{z-1}^{j_1}} \, \log\brc{z-1}^{p_1} \log\brc{z}^{p_2},}
\eq{\frac{\sum_{m=0}^{r_2} \beta_m\, z^m}{z^{i_2} \brc{z-1}^{j_2}} \, \log\brc{z-1}^{p_3} \text{Li}_{s_1,\dots,s_k}\brc{1-z}.}
After breaking up rational functions with the help of (\ref{eq:poles_Id}), we will consider all possible integrals case by case and show that the maximum weight after the integration is $K+1$. Splitting this last part of the proof into three steps is helpful. In each step, we will deal with the following integrals:
\begin{enumerate}
\item Integrals that increase the maximum weight by one.
\item Integrals that do not increase the maximum weight and involve only one logarithm or multiple polylogarithm:
$\log\brc{z}^m$, $\log\brc{z-1}^n$, or $\text{Li}_{s_1,\dots,s_k}\brc{1-z}$.
\item Integrals that do not increase the maximum weight and involve the following products of logarithms:
$\log\brc{z}^m \log\brc{z-1}^n$ and $\log\brc{z-1}^n \text{Li}_{s_1,\dots,s_k}\brc{1-z}$.
\end{enumerate}

{\it Step 1}. Four types of integrals increase the maximum weight. In each case the integrand has a factor of $z^{-1}$ or $\brc{z-1}^{-1}$. For the product of two logarithms of weight $n+m$, $n,m\geq 0$  we have:
\eq{\int \frac{\log\brc{z}^m \log\brc{z-1}^n}{z}\, \rmd z =
\brc{-1}^{m+n+1}\, m!\,n!\,\sum_{j=0}^{n} \frac{\brc{-1}^j}{j!} \log\brc{z-1}^j
\text{Li}_{n-j+1,\bfi{1}_{m}}\brc{1-z},}
\eq{\int \frac{\log\brc{z}^m \log\brc{z-1}^n}{z-1}\, \rmd z =
\brc{-1}^{m+n}\, m!\,n!\,\sum_{j=0}^{n} \frac{\brc{-1}^j}{j!} \log\brc{z-1}^j
\text{Li}_{n-j+2,\bfi{1}_{m-1}}\brc{1-z},}
where in the last integral $m\geq 1$ and $\text{Li}_{n,\bfi{1}_{0}}\equiv \text{Li}_n$. The resulting weight after the integration is $1+n+m$. In the more general case of integrals involving multiple polylogarithms, we have ($n\geq0$):
\eq{\int \frac{\log\brc{z-1}^n}{z}\, \text{Li}_{s_1,\dots,s_k} \brc{1-z}\,\rmd z =
\brc{-1}^{n+1}\,n!\,\sum_{j=0}^{n} \frac{\brc{-1}^j}{j!} \log\brc{z-1}^j
\text{Li}_{n-j+1,s_1,\dots,s_k}\brc{1-z},}
\eq{\int \frac{\log\brc{z-1}^n}{z-1}\, \text{Li}_{s_1,\dots,s_k} \brc{1-z}\,\rmd z =
\brc{-1}^{n}\,n!\,\sum_{j=0}^{n} \frac{\brc{-1}^j}{j!} \log\brc{z-1}^j
\text{Li}_{s_1+n-j+1,s_2,\dots,s_k}\brc{1-z}.}
Again, after the integration, the weight was increased by $1$ from $n+s_1+\dots +s_k$ to $1+n+s_1+\dots +s_k$.
The above identities were obtained by repeated integrations by parts.

{\it Step 2}. The integrands in this step are products of one logarithm or multiple polylogarithm with $z^n$ or $\brc{z-1}^n$, with $n\neq -1$.
Moreover, it is enough to consider only the negative powers of $\brc{z-1}$ since all positive powers can be reduced to monomials in $z$.
We start with integrals involving the $\log\brc{z}$ function:
\eqlb{eq:intLog_zn}{n\neq-1:\quad \int z^n \log\brc{z}^m  \rmd z = \brc{-1}^m\, m!\, \frac{z^{n+1}}{\brc{n+1}^{m+1}}
\sum_{j=0}^{m} \frac{\brc{-1}^j}{j!} \brc{n+1}^j \log\brc{z}^j,}
\eqn{n\geq 2:\quad \int \frac{\log\brc{z}}{\brc{z-1}^n}\,  \rmd z =
\frac{1}{1-n}\brc{\brc{-1}^n+\frac1{\brc{z-1}^{n-1}}} \log\brc{z} - }
\eqlb{eq:intLog_z1n}{-\frac{\brc{-1}^n}{1-n}\,\log\brc{z-1} + \frac{\brc{-1}^n}{1-n}\,\sum_{j=1}^{n-2} \frac{1}{j\brc{z-1}^j},}
\eqn{m\geq 2:\quad \int \frac{\log\brc{z}^m}{\brc{z-1}^2}\,  \rmd z =\frac{z}{1-z}\,\log\brc{z}^m -
\brc{-1}^m\, m!\, \text{Li}_{2,\bfi{1}_{m-2}}\brc{1-z},}
\eqn{n,m\geq 2:\quad \int \frac{\log\brc{z}^m}{\brc{z-1}^n}\,  \rmd z =
\frac{1}{1-n}\brc{\brc{-1}^n+\frac1{\brc{z-1}^{n-1}}} \log\brc{z}^m +}
\eqlb{eq:intLogz_mn}{+\brc{-1}^{m+n} \frac{m!}{1-n}\, \text{Li}_{2,\bfi{1}_{m-2}}\brc{1-z} + \frac{m}{1-n}
\sum_{k=2}^{n-1}\brc{-1}^{k+n} \int \frac{\log\brc{z}^{m-1}}{\brc{z-1}^k}\,  \rmd z,}
where the last equation allows us to take the corresponding integral recursively. In principle, the integrals with the other logarithm $\log\brc{z-1}$ can be obtained from (\ref{eq:intLog_z1n})--(\ref{eq:intLogz_mn}) by shifting the variable $z\rightarrow 1-z$. This, however, would change the argument of multiple polylogarithms from $\brc{1-z}$ to $z$. Since we want our multiple polylogarithms to converge in the disk $\abs{1-z}<1$ (or $\abs{1-z}\leq1$ when $s_1\geq 2$), we rewrite (\ref{eq:intLog_z1n})--(\ref{eq:intLogz_mn}) using the function $\log\brc{z-1}$:
\eqn{n\geq 2:\quad \int \frac{\log\brc{z-1}}{z^n}\,  \rmd z =
\frac{z^{1-n}-1}{1-n} \log\brc{z-1} +\frac{\log\brc{z}}{1-n} +
 \frac{1}{n-1}\,\sum_{j=1}^{n-2} \frac{1}{j\,z^j},}
\eqn{m\geq 0:\quad \int \frac{\log\brc{z-1}^m}{z^2}\,  \rmd z =\frac{z-1}{z}\,\log\brc{z-1}^m -}
\eqn{-\brc{-1}^m m! \sum_{j=0}^{m-1}\frac{\brc{-1}^j}{j!}\, \log\brc{z-1}^j \text{Li}_{m-j}\brc{1-z},}
\eqn{\at{n\geq 2,}{m\geq 0}:\quad \int \frac{\log\brc{z-1}^m}{z^n}\,  \rmd z =
\frac{z^{1-n}-1}{1-n} \log\brc{z-1}^m +}
\eqn{+\brc{-1}^m \, \frac{m!}{1-n} \sum_{j=0}^{m-1}\frac{\brc{-1}^j}{j!}\, \log\brc{z-1}^j \text{Li}_{m-j}\brc{1-z}+}
\eq{ + \frac{m}{1-n} \sum_{k=2}^{n-1} \int \frac{\log\brc{z-1}^{m-1}}{z^k}\,  \rmd z.}
In all the integrals taken so far in step $2$, we can explicitly see that the maximum weights before and after integration are the same.

For the integrals that involve multiple polylogarithms $\text{Li}_{s_1,\dots,s_k} \brc{1-z}$, we consider the two cases $s_1=1$ and $s_1\geq 2$ separately. First, we look at the integrals with non-negative powers of $z$:
\eqn{\left\{\at{k\geq 2,}{n\geq 0}\right.:\quad \int z^n \, \text{Li}_{1,s_2,\dots,s_k} \brc{1-z} \rmd z =
\frac{z^{n+1}}{n+1}\, \text{Li}_{1,s_2,\dots,s_k} \brc{1-z}+}
\eqlb{eq:Li1_st2}{+\frac{1}{n+1}\,\int z^n \,\text{Li}_{s_2,\dots,s_k} \brc{1-z} \rmd z,}
\eqn{\left\{\at{s_1\geq 2,}{n\geq 0}\right.:\quad \int z^n\, \text{Li}_{s_1,\dots,s_k} \brc{1-z} \rmd z =
\frac{z^{n+1}-1}{n+1}\, \text{Li}_{s_1,\dots,s_k} \brc{1-z}-}
\eq{-\frac{1}{n+1}\,\sum_{m=0}^{n}\int z^m \, \text{Li}_{s_1-1,s_2,\dots,s_k} \brc{1-z} \rmd z.}
The above integrals can be taken recursively until one gets integrals of the form (\ref{eq:intLog_zn}) and the maximum weight of the final result is equal to $K$. Similarly, we have for the integrals with negative powers of $z$ (except for $1/z$):
\eqn{\left\{\at{k\geq 2,}{n\geq 2}\right.:\quad \int \frac1{z^n}\, \text{Li}_{1,s_2,\dots,s_k} \brc{1-z} \rmd z =
\frac{z^{1-n}}{1-n}\, \text{Li}_{1,s_2,\dots,s_k} \brc{1-z}+}
\eq{+\frac{1}{1-n}\,\int \frac1{z^n}\, \text{Li}_{s_2,\dots,s_k} \brc{1-z} \rmd z,}
\eqn{\left\{\at{s_1\geq 2,}{n\geq 2}\right.:\quad \int \frac1{z^n}\, \text{Li}_{s_1,\dots,s_k} \brc{1-z} \rmd z =
\frac{z^{1-n}-1}{1-n}\, \text{Li}_{s_1,\dots,s_k} \brc{1-z}+\frac{\text{Li}_{1,s_1-1,s_2,\dots,s_k} \brc{1-z}}{n-1} +}
\eq{+\frac{1}{1-n}\,\sum_{m=2}^{n-1}\int \frac1{z^m} \,\text{Li}_{s_1-1,s_2,\dots,s_k} \brc{1-z} \rmd z.}
Finally, we consider the integrals with multiple polylogarithms divided by $\brc{z-1}^n$, $n\geq 2$:
\eqn{\left\{\at{k\geq 2,}{n\geq 2}\right.:\quad \int \frac1{\brc{z-1}^n}\, \text{Li}_{1,s_2,\dots,s_k} \brc{1-z} \rmd z =
\frac{\brc{-1}^n+\brc{z-1}^{1-n}}{1-n}\, \text{Li}_{1,s_2,\dots,s_k} \brc{1-z}}
\eq{+\frac{\brc{-1}^n}{1-n}\, \text{Li}_{s_2+1,\dots,s_k} \brc{1-z}+ \frac{\brc{-1}^n}{n-1}\,\sum_{m=2}^{n-1}\int \frac{\brc{-1}^m}{\brc{z-1}^m} \, \text{Li}_{s_2,\dots,s_k} \brc{1-z} \rmd z,}
\eqn{\left\{\at{s_1\geq 2,}{n\geq 2}\right.:\quad \int \frac1{\brc{z-1}^n}\, \text{Li}_{s_1,\dots,s_k} \brc{1-z} \rmd z =
\frac{\brc{z-1}^{1-n}}{1-n}\, \text{Li}_{s_1,\dots,s_k} \brc{1-z}}
\eqlb{eq:Li2_zm1_st2}{+\frac{1}{n-1}\,\int \frac1{\brc{z-1}^n}\, \text{Li}_{s_1-1,s_2,\dots,s_k} \brc{1-z} \rmd z.}

{\it Step 3}. Here we have to deal with four types of integrals:
\eqlb{eq:2Log_z}{\left\{\at{k\neq -1}{n,m\geq1}\right.:\quad
\int z^k \log\brc{z}^m \log\brc{z-1}^n  \rmd z,}
\eqlb{eq:2Log_zm1}{\left\{\at{k\geq 2}{n,m\geq1}\right.:\quad
\int \frac{\log\brc{z}^m \log\brc{z-1}^n}{\brc{z-1}^k} \, \rmd z,}
\eqlb{eq:LogLi_z}{\left\{\at{m\neq -1}{n\geq1}\right.:\quad
\int z^m \log\brc{z-1}^n \text{Li}_{s_1,\dots,s_k} \brc{1-z} \rmd z,}
\eqlb{eq:LogLi_zm1}{\left\{\at{m\geq 2}{n\geq1}\right.:\quad
\int \frac{\log\brc{z-1}^n}{\brc{z-1}^m}\, \text{Li}_{s_1,\dots,s_k} \brc{1-z} \rmd z.}
In each case, we can use integration by parts to reduce the weight of one of the logarithms by $1$. Applying integration by parts recursively allows us to reduce all integrals of the form (\ref{eq:2Log_z})--(\ref{eq:LogLi_zm1}) to one of the integrals from step $2$ or $1$. For example, in the case of (\ref{eq:2Log_z}), we have
\eqn{\int z^k \log\brc{z}^m \log\brc{z-1}^n  \rmd z = \brc{-1}^m\, m!\,
\frac{z^{k+1}}{\brc{k+1}^{m+1}} \sum_{j=0}^{m} \frac{\brc{-1}^j}{j!} \brc{k+1}^j \log\brc{z}^j \log\brc{z-1}^n }
\eqlb{eq:Int1_st3}{-\brc{-1}^m\,\frac{m!\, n}{\brc{k+1}^{m+1}}\, \sum_{j=0}^{m} \frac{\brc{-1}^j}{j!} \brc{k+1}^j
\int \frac{z^{k+1}}{z-1} \, \log\brc{z}^j \log\brc{z-1}^{n-1} \rmd z.}
We simplify the integral in the {\it rhs} of (\ref{eq:Int1_st3}) by breaking up the rational function $z^{k+1}/\brc{z-1}$ into a sum of monomials in $z$ and poles at $z=0,1$:
\eq{k\geq 0:\quad \frac{z^{k+1}}{z-1} = \frac{1}{z-1} +\sum_{j=0}^{k} z^j,}
\eq{k\leq -2:\quad \frac{z^{k+1}}{z-1} = \frac{1}{z-1}-\sum_{j=1}^{-k-1} z^{-j}.}
Almost all the resulting integrals are of the first type (\ref{eq:2Log_z}) with the power of $\log\brc{z-1}$ reduced by $1$. The remaining two integrals
\eqlb{eq:IntLL_3}{\int \frac{\log\brc{z}^j \log\brc{z-1}^{n-1}}{z}\, \rmd z\quad \text{and}\quad
\int \frac{\log\brc{z}^j \log\brc{z-1}^{n-1}}{z-1}\, \rmd z}
were already taken in step $1$, and the maximum weight is $n+m=K$. Recursively applying this procedure, one can either express (\ref{eq:2Log_z}) in terms of integrals like (\ref{eq:IntLL_3}) or reduce it to (\ref{eq:intLog_zn}). In the same way, (\ref{eq:2Log_zm1}) can be essentially reduced to (\ref{eq:intLog_zn}) with $z$ replaced by $\brc{z-1}$. Finally, the last two types of integrals, (\ref{eq:LogLi_z}) and (\ref{eq:LogLi_zm1}), are reducible to a combination of integrals from (\ref{eq:Li1_st2})--(\ref{eq:Li2_zm1_st2}). 

To summarize, we have shown by induction that the wave function $\psi^{L}\brc{z}$ at any order $t^K$ is a linear combination of certain functions of weight $K$ or lower. The only special functions needed are multiple polylogarithms with argument $\brc{1-z}$ (another possible argument would be $z$, but it would not be consistent with the boundary condition at the AdS boundary in the SAdS case). The same can be done for the wave function in the right region, $\psi^{R}\brc{z}$.

\section{Multiple polylogarithms for hydrodynamic QNMs}\label{appendixD}

For the computations of gravitational QNMs in the scalar sector (Sec.~\ref{section5}), we introduced an expansion of the solution using the multiple polylogarithms in several variables \eqref{eq:Robin_Li_def}.
An alternative definition could be given in terms of one-forms\footnote{For $\omega_{z_1}, \dots, \omega_{z_p}$ differential one-forms, with $\omega_{z_i}=f_{z_i}(t)\mathrm{d}t$ 
for some function $f_{z_i}$, we define inductively $\int_0^x\omega_{z_1} \dots \omega_{z_p}=\int_0^x f_{z_1}(t)\mathrm{d}t \int_0^t \omega_{z_2}\dots\omega_{z_p}$.} \cite{waldschmidt2002multiple}
\begin{equation}
\begin{aligned}
&\text{Li}_{s_1,\dots,s_k}\brc{z_1, \dots, z_k}=\int_0^1\omega_0^{s_1-1}\omega_{z_1}\omega_0^{s_2-1}\omega_{z_1z_2}\dots\omega_0^{s_k-1}\omega_{z_1\dots z_k},
\end{aligned}
\end{equation}
where
\begin{equation}
\omega_{z}=
\left\{
\begin{aligned}
&\frac{z\mathrm{d}t}{1-z t}, & z\neq 0,\\
&\frac{\mathrm{d}t}{t}, & z= 0.
\end{aligned}
\right.
\end{equation}
All the integrals in Section \ref{section5} do not include $\omega_0$ after the simplification, which means that $s_1=s_2=\dots = s_k=1$. We define the relevant multiple polylogarithms as
\eq{\label{1formproduct}\text{Li}_{\{1\}_k}(z_1 z,z_2,\dots,z_k)=\int_0^z\omega_{z_1}\omega_{z_1 z_2}\dots\omega_{z_1\dots z_k},}
where  $z_i\in\{1,u_1,u_2\}$ for every $i=1,\dots,k$
and $u_1$, $u_2$ are the third roots of unity \eqref{rootsofunity}.
We consider the following products of ordinary logarithm functions and multiple polylogarithms to describe the wave function:
\begin{equation}
\label{eq:Robin_Li_prod}
\begin{aligned}
&\log\brc{1-z}^{p_1} \log\brc{1-u_1\, z}^{p_2} \log\brc{1-u_2\, z}^{p_3}\\
&\log\brc{1-z}^{p_4} \log\brc{1-u_1 \, z}^{p_5} \log\brc{1-u_2 \,z}^{p_6}
\text{Li}_{\{1\}_k}\brc{z_1 z,z_2,\dots,z_k}.
\end{aligned}
\end{equation}
At order $\alpha^K$, only functions with maximum weight $K$ appear, so that $0\leq p_1+p_2+p_3 \leq K$ and $0\leq p_4+p_5+p_6 + k\leq K$. However, at a fixed weight, some identities make the functions listed in \eqref{eq:Robin_Li_prod} linearly dependent. 
For example, at level $k=2$, we have the following identities: 
\begin{equation}
\begin{aligned}
\text{Li}_{1,1}(z,u_1)&=\log(1-u_1\,z)\log(1-z)-\text{Li}_{1,1}(u_1\,z,u_2),\\
\text{Li}_{1,1}(z,u_2)&=\log(1-u_2\,z)\log(1-z)-\text{Li}_{1,1}(u_2\,z,u_1),\\
\text{Li}_{1,1}(u_2\,z,u_2)&=\log(1-u_1\,z)\log(1-u_2\,z)-\text{Li}_{1,1}(u_1\,z,u_1),
\end{aligned}
\end{equation}
including the ones that reduce to the single variable case: 
\begin{equation}
\begin{aligned}\label{oldappd6}
\text{Li}_{1,1}(z,1)&=\frac12 \, \log\brc{1-z}^2~.
\end{aligned}
\end{equation}
Thus, out of $9$ possible functions $\text{Li}_{1,1}(z_1 z,z_2)$ at level $k=2$, we only need $3$:
\begin{equation}
\text{Li}_{1,1}\left(u_1z,u_2\right),\quad \text{Li}_{1,1}\left(u_1z,u_1\right),\quad \text{Li}_{1,1}\left(u_2z,u_1\right).
\end{equation}
In the rest of the appendix, we will try to classify the identities arising at a given level $k$ and find what multiple polylogarithms are needed to form a linear basis in \eqref{eq:Robin_Li_prod}. 

According to \eqref{1formproduct}, there is a one-to-one correspondence between multiple polylogarithms and ordered multisets of one-forms $\{\omega_{z_1}, \omega_{z_1 z_2},\dots, \omega_{z_1\dots z_k}\}$. If two multiple polylogarithms are related by the permutation of the one-forms in the corresponding ordered multisets, then an identity exists between these two. However, this identity could be reducible in the sense that it can be split into smaller ones. To show this, we integrate by parts the right-hand side of \eqref{1formproduct}: 
\begin{equation}
\begin{aligned}
&\int_0^z\omega_{z_1}\omega_{z_1z_2}\dots\omega_{z_1\dots z_k}=\int_0^z\frac{\mathrm{d}}{\mathrm{d} t}\text{Li}_1(z_1 t)\mathrm{d}t\,\omega_{z_1z_2}\dots\omega_{z_1\dots z_k}=\\
&=\text{Li}_1(z_1 z)\int_0^z\omega_{z_1z_2}\dots\omega_{z_1\dots z_k}-\int_0^z\mathrm{d}t\,\text{Li}_1(z_1 t)\frac{z_1z_2}{1-z_1z_2 t}\omega_{z_1z_2z_3}\dots\omega_{z_1\dots z_k},
\end{aligned}
\end{equation}
where $\text{Li}_1\brc{z_1 t}$ is the ordinary logarithm function:
\eq{\text{Li}_1\brc{z_1 t} = - \log\brc{1-z_1 t}}
and
\eq{\text{Li}_1(z_1 z)\int_0^z\omega_{z_1z_2}\dots\omega_{z_1\dots z_k} = - \log\brc{1-z_1 z} \text{Li}_{\{1\}_{k-1}}(z_1 z_2 z,z_3,\dots,z_k).}
Continuing the integration by parts, we obtain
\begin{equation}
\begin{aligned}
&\int_0^z\omega_{z_1}\omega_{z_1z_2}\dots\omega_{z_1\dots z_k}\supset \int_0^z\mathrm{d}t\text{Li}_1(z_1 t)\text{Li}_1(z_1z_2 t)\dots\text{Li}_1(z_1\dots z_{k-1} t)\frac{z_1\dots z_k}{1-z_1\dots z_k t}=\\
&=\int_0^z\mathrm{d}t\text{Li}_1(y_1 t)\text{Li}_1(y_2 t)\dots\text{Li}_1(y_{k-1} t)\frac{y_k}{1-y_k t},
\end{aligned}
\end{equation}
where $y_j=z_1\cdots z_j$ for $j=1,\dots,k$. 
From this last integral, one can reconstruct by the reverse process any other multiple polylogarithm for which the representation in \eqref{1formproduct} involves the integrals of the same one-forms in a different order. 
In the intermediate steps of this procedure, there appear products of the form
\begin{equation}
\text{Li}_{\{1\}_{m_1}}(z^{(1)}_1\,z,\dots,z^{(1)}_{m_1})\cdot\,\dots\, \cdot\text{Li}_{\{1\}_{m_r}}(z^{(r)}_1\,z,\dots,z^{(r)}_{m_r}),\ \ \text{with}\ \ m_1+\dots+m_r=k.
\end{equation}
It is possible to rewrite these in terms of products in \eqref{eq:Robin_Li_prod} using \emph{shuffle relations} (see for example Eq. (5.4) in \cite{borwein2001special}). The result is an identity involving two multiple polylogarithms that are related by the permutation of the corresponding one-forms.

Let us describe with a concrete identity at level 4 how this works. We prove that
\begin{equation}\label{D13}\small
\text{Li}_{1,1,1,1}(z,u_1,1,u_2)=-2\text{Li}_{1,1,1,1}(u_1\,z,1,u_2,1)-\text{Li}_{1,1,1,1}(u_1\,z,u_2,u_1,u_2)-\text{Li}_{1,1,1}(u_1\,z,1,u_2)\log(1-z).
\end{equation}
By definition, the lhs is 
\begin{equation}
\text{Li}_{1,1,1,1}(z,u_1,1,u_2)=\int_0^z\omega_1\omega_{u_1}\omega_{u_1}\omega_1=-\text{Li}_{1,1,1}(u_1\,z,1,u_2)\log(1-z)-\int_0^z\text{Li}_1(t)\frac{u_1\,\mathrm{d}t}{1-u_1\,z}\omega_{u_1}\omega_1,
\end{equation}
where in the last equality we integrated by parts. Therefore, we reduce to proving that
\begin{equation}\label{eq:ex1}
\int_0^z\text{Li}_1(t)\frac{u_1\,\mathrm{d}t}{1-u_1\,z}\omega_{u_1}\omega_1=2\text{Li}_{1,1,1,1}(u_1\,z,1,u_2,1)+\text{Li}_{1,1,1,1}(u_1\,z,u_2,u_1,u_2).
\end{equation}
We have
\begin{equation}\label{eq:ex2}
\begin{aligned}
&\int_0^z\text{Li}_1(t)\frac{u_1\,\mathrm{d}t}{1-u_1\,z}\omega_{u_1}\omega_1=\int_0^z\frac{\mathrm{d}}{\mathrm{d}t}\text{Li}_{1,1}(u_1\,t,u_2)\omega_{u_1}\omega_1=\\
&\text{Li}_{1,1}(u_1\,z,u_2)\text{Li}_{1,1}(u_1\,z,u_2)-\int_0^z\text{Li}_{1,1}(u_1\,t,u_2)\frac{u_1\,\mathrm{d}t}{1-u_1\,t}\omega_1.
\end{aligned}
\end{equation}
Moreover,
\begin{equation}\label{eq:ex3}\small
\begin{aligned}
&\int_0^z\text{Li}_{1,1}(u_1\,t,u_2)\frac{u_1\,\mathrm{d}t}{1-u_1\,t}\omega_1=\int_0^z\frac{\mathrm{d}}{\mathrm{d}t}\text{Li}_{1,1,1}(u_1\,t,1,u_2)\omega_1=\\
&\text{Li}_{1,1,1}(u_1\,z,1,u_2)\text{Li}_1(z)-\int_0^z\text{Li}_{1,1,1}(u_1\,t,1,u_2)\frac{\mathrm{d}t}{1-t}=\text{Li}_{1,1,1}(u_1\,z,1,u_2)\text{Li}_1(z)-\text{Li}_{1,1,1,1}(z,u_1,1,u_2).
\end{aligned}
\end{equation}
Putting together \eqref{eq:ex1}-\eqref{eq:ex2}-\eqref{eq:ex3}, it remains to prove that
\begin{equation}\label{eq:ex4}
\begin{aligned}
&\text{Li}_{1,1}(u_1\,z,u_2)\text{Li}_{1,1}(u_1\,z,u_2)-\text{Li}_{1,1,1}(u_1\,z,1,u_2)\text{Li}_1(z)+\text{Li}_{1,1,1,1}(z,u_1,1,u_2)=\\
&2\text{Li}_{1,1,1,1}(u_1\,z,1,u_2,1)+\text{Li}_{1,1,1,1}(u_1\,z,u_2,u_1,u_2).
\end{aligned}
\end{equation}
Applying the shuffle relation to the first two terms in lhs, we have
\begin{equation}
\begin{aligned}
&\text{Li}_{1,1}(u_1\,z,u_2)\text{Li}_{1,1}(u_1\,z,u_2)=4\text{Li}_{1,1,1,1}(u_1\,z,1,u_2,1)+2\text{Li}_{1,1,1,1}(u_1\,z,u_2,u_1,u_2),\\
&\text{Li}_{1,1,1}(u_1\,z,1,u_2)\text{Li}_1(z)=2\text{Li}_{1,1,1,1}(u_1\,z,1,u_2,1)+\text{Li}_{1,1,1,1}(u_1\,z,u_2,u_1,u_2)+\text{Li}_{1,1,1,1}(z,u_1,1,u_2).
\end{aligned}
\end{equation}
Therefore, as we wanted, the lhs of \eqref{eq:ex4} becomes
\begin{equation}\small
\begin{aligned}
&4\text{Li}_{1,1,1,1}(u_1\,z,1,u_2,1)+2\text{Li}_{1,1,1,1}(u_1\,z,u_2,u_1,u_2)-[2\text{Li}_{1,1,1,1}(u_1\,z,1,u_2,1)+\text{Li}_{1,1,1,1}(u_1\,z,u_2,u_1,u_2)+\\
&+\text{Li}_{1,1,1,1}(z,u_1,1,u_2)]+\text{Li}_{1,1,1,1}(z,u_1,1,u_2)=2\text{Li}_{1,1,1,1}(u_1\,z,1,u_2,1)+\text{Li}_{1,1,1,1}(u_1\,z,u_2,u_1,u_2).
\end{aligned}
\end{equation}

Let us remark that with the previous procedure, one can find several identities at a fixed level involving the same multiple polylogarithm. 
To choose which elements to add to the basis, we followed the criterium that we omit the multiple polylogarithms with the first argument $z$.  This criterium comes from the regularity condition on the wave function at $z=1$. Moreover, when possible, we tried to include the same number of multiple polylogarithms with the first argument $u_1\,z$ and with the first argument $u_2\,z$  (for example, it is not possible at level $k=2$).

For completeness, let us write the elements of level 3 that we add to our basis:
\begin{equation}
\begin{aligned}
&\text{Li}_{1,1,1}\left(u_1\,z,1,u_2\right),\quad \text{Li}_{1,1,1}\left(u_1\,z,1,u_1\right),\quad \text{Li}_{1,1,1}\left(u_1\,z,u_2,1\right),\quad \text{Li}_{1,1,1}\left(u_1\,z,u_1,u_1\right),\\
&\text{Li}_{1,1,1}\left(u_1\,z,u_1,1\right),\quad \text{Li}_{1,1,1}\left(u_2\,z,1,u_1\right),\quad \text{Li}_{1,1,1}\left(u_2\,z,u_1,1\right),\quad \text{Li}_{1,1,1}\left(u_2\,z,u_2,u_2\right),
\end{aligned}
\end{equation}
and the nontrivial identities with the other functions of the same level (other identities are obtained by exchanging $u_1$ with $u_2$):
\begin{equation}\small
\begin{aligned}
&\text{Li}_{1,1,1}(u_2\,z,u_2,1)=\text{Li}_{1,1,1}(u_1\,z,1,u_1)+\text{Li}_{1,1}(u_1\,z,u_1)\log(1-u_1\,z)-\frac{\log(1-u_1\,z)^2\log(1-u_2\,z)}{2},\\
&\text{Li}_{1,1,1}(z,u_2,1)=\text{Li}_{1,1,1}(u_2\,z,1,u_1)+\text{Li}_{1,1}(u_2\,z,u_1)\log(1-u_2\,z)-\frac{\log(1-u_2\,z)^2\log(1-z)}{2},\\
&\text{Li}_{1,1,1}(z,u_1,1)=\text{Li}_{1,1,1}(u_1\,z,1,u_2)+\text{Li}_{1,1}(u_1\,z,u_2)\log(1-u_1\,z)-\frac{\log(1-u_1\,z)^2\log(1-z)}{2},\\
&\text{Li}_{1,1,1}(u_2\,z,u_2,u_1)=-2\text{Li}_{1,1,1}(u_1\,z,u_1,1)-\text{Li}_{1,1}(u_1\,z,u_1)\log(1-u_2\,z),\\
&\text{Li}_{1,1,1}(z,u_1,u_1)=\text{Li}_{1,1,1}(u_2\,z,u_2,u_2)-\text{Li}_{1,1}(u_1\,z,u_1)\log(1-z)+\text{Li}_{1,1}(u_1\,z,u_2)\log(1-u_2\,z),\\
&\text{Li}_{1,1,1}(z,u_2,u_1)=-2\text{Li}_{1,1,1}(u_2\,z,u_1,1)-\text{Li}_{1,1}(u_2\,z,u_1)\log(1-z),\\
&\text{Li}_{1,1,1}(z,u_1,u_2)=-2\text{Li}_{1,1,1}(u_1\,z,u_2,1)-\text{Li}_{1,1}(u_1\,z,u_2)\log(1-z).
\end{aligned}
\end{equation}
Computing all the identities up to level $k=7$, we arrive at the following conclusion.
The number of multiple polylogarithms needed to form a basis in \eqref{eq:Robin_Li_prod}  at level $k\ge 3$ is $8\times 3^{k-3}$. Even though this significantly reduces the number of functions used at a certain level $k$, we still need to compute the identities for all $3^k$ functions to go to the next level $k+1$.

\section{Connection Formula for \texorpdfstring{$\mathrm{SAdS}_4$}{} } \label{appendixconnection}

In this appendix, we write the connection formula relevant for the Schwarzschild anti-de Sitter black hole, and we compute from it the first order correction in $R_h$ of the quasinormal mode frequency $\omega_{0,1,1}$.

Using the notations introduced in Sec.~\ref{sec:gaugetheory} for the Heun solutions and the dictionary in \eqref{dictionaryAdS1}-\eqref{dictionaryAdS2} along with
\begin{equation}\small
\begin{aligned}
&t=\frac{R_h(R_+-R_-)}{R_-(R_+-R_h)},\quad
z_{\infty}=1-\frac{R_+}{R_-},\quad
a_0=s,\\
&a_1=\frac{i\omega R_-}{(R_--R_h)(R_--R_+)},\quad
a_t=\frac{i\omega R_h}{(R_h-R_+)(R_h-R_-)},\quad 
a_{\infty}=\frac{i\omega R_+}{(R_+-R_h)(R_+-R_-)},\\
&u=\frac{\left(R_--R_h\right)^2\left(R_h-R_+\right)\left[2\ell(\ell+1)+2R_+(R_++R_h)s(s-1)+R_-(R_+-2sR_++R_h)\right]+4R_-R_h^2\omega^2}{2R_+(R_--R_h^3)(R_+-R_h)},
\end{aligned}
\end{equation}
the connection formula for the $\mathrm{SAdS}_4$ case reads
\begin{equation}\label{eq:conn}
\begin{aligned}
&t^{-\frac{1}{2}+a_0+a_t}(1-t)^{-\frac{1}{2}+a_1}e^{\frac{1}{2}\partial_{a_t}F(t)}\psi_-^{(t)}(z)=\\
&\left(\sum_{\sigma=\pm}\mathcal{M}_{-\sigma}(a_t,a;a_0)\mathcal{M}_{(-\sigma)-}(a,a_1;a_{\infty})t^{\sigma a}e^{-\frac{\sigma}{2}\partial_aF(t)}\right)(1-t)^{-\frac{1}{2}+a_t}e^{i\pi(a_1+a_t)}e^{\frac{1}{2}\partial_{a_1}F(t)}\psi_-^{(1)}(z)+\\
&\left(\sum_{\sigma=\pm}\mathcal{M}_{-\sigma}(a_t,a;a_0)\mathcal{M}_{(-\sigma)+}(a,a_1;a_{\infty})t^{\sigma a}e^{-\frac{\sigma}{2}\partial_aF(t)}\right)(1-t)^{-\frac{1}{2}+a_t}e^{i\pi(-a_1+a_t)}e^{-\frac{1}{2}\partial_{a_1}F(t)}\psi_+^{(1)}(z).
\end{aligned}
\end{equation}
Therefore, the quantization condition for the quasinormal mode frequencies can be written as
\begin{equation}\small\label{quantizationtoexpand}
\begin{aligned}
&\Biggl\{\left[\sum_{\sigma=\pm}\frac{\Gamma(-2\sigma a)\Gamma(2a_1)\Gamma(1-2\sigma a)t^{\sigma a}e^{-\frac{\sigma}{2}\partial_aF(t)+\frac{1}{2}\partial_{a_1}F(t)}e^{2i\pi a_1}}{\Gamma\left(\frac{1}{2}-a_t-\sigma a+a_0\right)\Gamma\left(\frac{1}{2}-a_t-\sigma a-a_0\right)\Gamma\left(\frac{1}{2}-\sigma a+a_1+a_{\infty}\right)\Gamma\left(\frac{1}{2}-\sigma a+a_1-a_{\infty}\right)}\right]\times\\
&\times\biggl(\frac{z-t}{1-t}\biggr)^{-\alpha}\mathrm{Heun}\biggl(t,q+\alpha(\delta-\beta),\alpha,\delta+\gamma-\beta,\delta,\gamma,t\frac{1-z}{t-z}\biggr)+\\
&+\left[\sum_{\sigma=\pm}\frac{\Gamma(-2\sigma a)\Gamma(-2a_1)\Gamma(1-2\sigma a)t^{\sigma a}e^{-\frac{\sigma}{2}\partial_aF(t)-\frac{1}{2}\partial_{a_1}F(t)}}{\Gamma\left(\frac{1}{2}-a_t-\sigma a+a_0\right)\Gamma\left(\frac{1}{2}-a_t-\sigma a-a_0\right)\Gamma\left(\frac{1}{2}-\sigma a-a_1+a_{\infty}\right)\Gamma\left(\frac{1}{2}-\sigma a-a_1-a_{\infty}\right)}\right]\times\\
&\times(z-1)^{1-\delta}\biggl(\frac{z-t}{1-t}\biggr)^{-\alpha-1+\delta}\\
&\times\mathrm{Heun}\biggl(t,q-(\delta-1)\gamma t-(\beta-1)(\alpha-\delta+1),-\beta+\gamma+1,\alpha-\delta+1,2-\delta,\gamma,t\frac{1-z}{t-z}\biggr)\Biggr\}\bigg|_{z=z_{\infty}}=0.
\end{aligned}
\end{equation}

Our goal is to obtain, starting from the connection formula \eqref{quantizationtoexpand}, the analytic expression of the first order correction in $R_h$ of the QNM frequency in a concrete case: $n=0,\, \ell=s=1$. 
Consider the expansion of 
$\omega:=\omega_{0,1,1}$ in powers of $R_h$
\begin{equation}
\omega_{0,1,1}=\sum_{j=0}^{\infty}\omega_jR_h^j,
\end{equation}
where, from the pure AdS case, we take $\omega_0=\ell+2n+2$; in our concrete example, $\omega_0=3$.
We start by expanding at first order in $R_h$ the local solutions for $z\sim 1$, using the three-term recurrence relation for the Heun functions. The first Heun function is
\begin{equation}
\mathrm{Heun}\left(t,q+\alpha(\delta-\beta),\alpha,\delta+\gamma-\beta,\delta,\gamma,t\frac{1-z}{t-z}\right).
\end{equation}
The corresponding series expansion is given by
\begin{equation}
\sum_{j\ge 0}c_j\left(t\frac{1-z}{t-z}\right)^j,
\end{equation}
where the coefficients satisfy the three-terms relation
\begin{equation}
c_{j+1}=\frac{[Q_j+q+\alpha(\delta-\beta)]c_j+P_jc_{j-1}}{R_j},
\end{equation}
with
\begin{equation}
\begin{aligned}
c_0&=1,\\
c_1&=\frac{q+\alpha(\delta-\beta)}{t\delta},\\
P_j&=(j-1+\alpha)(j-1-\beta+\gamma+\delta),\\
Q_j&=j[(j-1+\delta)(1+t)+t\gamma+\alpha-\beta+1],\\
R_j&=t(j+1)(j+\delta).
\end{aligned}
\end{equation}
We solve the recurrence relation to find the leading order and first order in $R_h$ of the coefficients, and then we sum over $j$ to obtain the expansion of the Heun function. The expansion of the first local solution (including both the Heun function and the prefactor) is given by
\begin{equation}\label{expansionfirstlocal}
\frac{1+z}{2z}+\frac{(z-1) \left(\omega_1 z (z-1)^3+2 i z ((z-8) z-2)+6 i\right)+12 i z (2 z-1) \log (z)}{8 (z-1)^3 z^2}R_h+\mathcal{O}(R_h^2).
\end{equation}
Then, we repeat the same for the second solution, starting from the Heun function 
\begin{equation}
\mathrm{Heun}\left(t,q-(\delta-1)\gamma t-(\beta-1)(\alpha-\delta+1),-\beta+\gamma+1,\alpha-\delta+1,2-\delta,\gamma,t\frac{1-z}{t-z}\right). 
\end{equation}
The series defining this Heun function is given by
\begin{equation}
\sum_{j\ge 0}d_j\left(t\frac{1-z}{t-z}\right)^j,
\end{equation}
where the coefficients satisfy the three-terms relation
\begin{equation}\label{recurrence2}
d_{j+1}=\frac{[Q_j+q-(\delta-1)\gamma t-(\beta-1)(\alpha-\delta+1)]d_j-P_jd_{j-1}}{R_j},
\end{equation}
with
\begin{equation}
\begin{aligned}
d_0&=1,\\
d_1&=\frac{q-(\delta-1)\gamma t-(\beta-1)(\alpha-\delta+1)}{t(2-\delta)},\\
P_j&=(j+\alpha-\delta)(j+\gamma-\beta),\\
Q_j&=j[(j+1-\delta)(1+t)+t\gamma+\alpha-\beta+1],\\
R_j&=t(j+1)(j+2-\delta).
\end{aligned}
\end{equation}
This time the procedure is trickier because one can notice that, for $j=2$, the factor $j+2-\delta$ in $R_j$ starts with the first order in $R_h$. We solve the recurrence relation for the leading orders and first orders of $d_j$ for $j\ge 3$, and then we add the contributions from the previous terms by hand.
In this way, the expansion of the second local solution is given by
\begin{equation}\label{expansionsecondlocal}
\begin{aligned}
&\frac{6iz^3-3iz^4+2z\omega_1-\omega_1}{z(z-1)^3(3i+\omega_1)}-\frac{1}{8z^2(z-1)^3(3i+\omega_1)^2}\Bigl\{(z-1)[12i\omega_1(3i+\omega_1)+\\
&+iz^3(-63+174i\omega_1+44\omega_1^2-24\omega_2)-iz^2(63+14\omega_1(3i+2\omega_1)+24\omega_2)+\\
&+z^4(6\omega_1-8i\omega_1^2+3i(-39+8\omega_2))+z(-2\omega_1(-75+2\omega_1(12i+\omega_1))+3i(21+8\omega_2))]+\\
&+24z(3i+\omega_1)[3z^3(z-2)+i(z-1)^3(z+1)\omega_1]\log(1/z)+\\
&+8z(3i+\omega_1)(3i(z-2)z^3+\omega_1-2z\omega_1)[(3i+\omega_1)\log(z-1)-6i\log(z)]\Bigr\}R_h+\mathcal{O}(R_h^2).
\end{aligned}
\end{equation}

We now expand the connection coefficients to the first order in $R_h$.
Since $t$ is proportional to $R_h$ and the leading order of $a$ is given by $\ell+\frac{1}{2}=\frac{3}{2}$,  in both coefficients, the parts proportional to $t^a$ start with the terms of higher order with respect to the parts proportional to $t^{-a}$. So they do not contribute to the first-order expansion of the quantization condition. Then, we remain with the coefficients
\begin{equation}
\begin{aligned}
&\frac{\Gamma(2a)\Gamma(2a_1)\Gamma(1+2a)t^{-a}e^{\frac{1}{2}\partial_aF(t)+\frac{1}{2}\partial_{a_1}F(t)}e^{2i\pi a_1}}{\Gamma\left(\frac{1}{2}-a_t+a+a_0\right)\Gamma\left(\frac{1}{2}-a_t+a-a_0\right)\Gamma\left(\frac{1}{2}+a+a_1+a_{\infty}\right)\Gamma\left(\frac{1}{2}+a+a_1-a_{\infty}\right)},\\
&\frac{\Gamma(2a)\Gamma(-2a_1)\Gamma(1+2a)t^{-a}e^{\frac{1}{2}\partial_aF(t)-\frac{1}{2}\partial_{a_1}F(t)}}{\Gamma\left(\frac{1}{2}-a_t+a+a_0\right)\Gamma\left(\frac{1}{2}-a_t+a-a_0\right)\Gamma\left(\frac{1}{2}+a-a_1+a_{\infty}\right)\Gamma\left(\frac{1}{2}+a-a_1-a_{\infty}\right)}.
\end{aligned}
\end{equation}
Simplifying the common terms, we have
\begin{equation}\label{expansioncoefficients}
\begin{aligned}
&\frac{\Gamma(2a_1)e^{\frac{1}{2}\partial_{a_1}F(t)}e^{2i\pi a_1}}{\Gamma\left(\frac{1}{2}+a+a_1+a_{\infty}\right)\Gamma\left(\frac{1}{2}+a+a_1-a_{\infty}\right)}=-\frac{\omega_1}{6 (\omega_1+3 i)}+\\
&+\frac{24 i \pi\omega_1^3+20 \omega_1^3+126 i \omega_1^2-144 \pi  \omega_1^2-216 i \pi  \omega_1-324 \omega_1-72 i \omega_2-189 i}{144 (\omega_1+3 i)^2}R_h+\mathcal{O}(R_h^2),\\
&\frac{\Gamma(-2a_1)e^{-\frac{1}{2}\partial_{a_1}F(t)}}{\Gamma\left(\frac{1}{2}+a-a_1+a_{\infty}\right)\Gamma\left(\frac{1}{2}+a-a_1-a_{\infty}\right)}=\frac{1}{12}+\frac{1}{144} (-7 \omega_1+9 i)R_h+\mathcal{O}(R_h^2).
\end{aligned}
\end{equation}

Putting together \eqref{expansionfirstlocal}, \eqref{expansionsecondlocal}, and \eqref{expansioncoefficients}, one has the series expansion in $R_h$ of the quantization condition \eqref{quantizationtoexpand}
\begin{equation}
\frac{i(4+\pi\omega_1)}{8}R_h+\mathcal{O}(R_h^2).
\end{equation}
We can therefore conclude that 
\begin{equation}
\omega_1=-\frac{4}{\pi}
\end{equation}
matches the result obtained in \eqref{omega1} for our choice of quantum numbers.

\bibliographystyle{JHEP}
\bibliography{biblio}

\end{document}